\documentclass{emulateapj}
\usepackage{lscape}

\begin{document}

\newcommand{\Wmi}{$W_0^{\lambda2796}$}
\newcommand{\Wf}{$W_0^{\lambda2600}$}
\newcommand{\Wmii}{$W_0^{\lambda2803}$}
\newcommand{\Wmiii}{$W_0^{\lambda2852}$}

\title{Damped Lyman Alpha Systems at $z<1.65$: The Expanded
SDSS HST Sample\altaffilmark{1}}  

\altaffiltext{1}{Based on data obtained from the Sloan Digital
Sky Survey (SDSS) and on observations made with the Hubble Space
Telescope (HST) operated by STScI-AURA for NASA.}

\author{Sandhya M. Rao\altaffilmark{2,4}, David
A. Turnshek\altaffilmark{2}, and Daniel B. Nestor\altaffilmark{2,3}}
                                                                              
\altaffiltext{2}{Department of Physics \& Astronomy, University of Pittsburgh,
Pittsburgh, PA 15260}

\altaffiltext{3}{Astronomy Department, University of Florida, Gainesville, FL 32611}

\altaffiltext{4}{email: rao@everest.phyast.pitt.edu}

\begin{abstract}

We present results of our {\it Hubble Space Telescope} Cycle 11 Survey
for low-redshift ($z<1.65$) damped Ly$\alpha$ systems (DLAs) in the UV
spectra of quasars selected from the Sloan Digital Sky Survey  Early
Data Release.  These quasars have strong intervening MgII-FeII systems
which are known signatures of high column density neutral gas. In
total, including our previous surveys, UV observations of  Ly$\alpha$
absorption in 197 MgII systems with $z<1.65$ and rest equivalent width
(REW) \Wmi\ $\ge 0.3$ \AA\ have now been obtained.  The main results
are: (1) The success rate of identifying DLAs in a MgII sample with
\Wmi\ $\ge 0.5$ \AA\ and FeII \Wf\ $\ge 0.5$ \AA\ is 36($\pm6$)\%, and
increases to 42($\pm7$)\% for systems with \Wmi/\Wf$<2$ and MgI
\Wmiii\ $>0.1$ \AA.  (2) The mean HI column density of MgII systems
with 0.3 \AA\ $\le$ \Wmi $< 0.6$ \AA\ is $\left<N(HI)\right> =
(9.7\pm2.2)\times10^{18}$ cm$^{-2}$. For the larger REW systems in our
sample, $\left<N(HI)\right> = (3.5\pm0.7)\times10^{20}$ cm$^{-2}$.
The mean HI column density remains constant with increasing REW for
\Wmi $\ge 0.6$ \AA, but the fraction of MgII systems containing DLAs
is found to increase with increasing REW.  (3) By combining our
low-redshift results with results at higher  redshift from Prochaska
and Herbert-Fort and at $z=0$ from Ryan-Weber et al.  we find that we
can parameterize the DLA incidence per unit redshift as
$n_{DLA}(z)=n_0(1+z)^\gamma$, where $n_0 = 0.044\pm0.005$ and  $\gamma
= 1.27\pm0.11$. This parameterization is consistent with no evolution
for $z\lesssim 2$ ($\Omega_\Lambda=0.7$, $\Omega_M = 0.3$), but
exhibits significant evolution for $z\gtrsim 2$. (4) The cosmological
neutral gas mass density due to DLAs is constant in the redshift
interval $0.5<z<5.0$ to within the uncertainties, $\Omega_{DLA}
\approx 1\times10^{-3}$.  This is larger than $\Omega_{gas}(z=0)$  by
a factor of $\approx 2$. (5) The slope of the HI column density
distribution does not change significantly with redshift.  However,
the low redshift distribution is marginally flatter due to the higher
fraction of high column density systems in our sample.  (6) Finally,
using the precision of MgII survey statistics, we show that under the
assumption of constant DLA fraction and HI column density suggested by
our current sample, there may be evidence of a decreasing
$\Omega_{DLA}$ from $z=0.5$ to $z=0$. We discuss selection effects
that might affect the results from our survey. We reiterate the
conclusion of Hopkins, Rao, \& Turnshek that very high columns of
neutral gas might be missed by DLA surveys because of their very small
cross sections, and therefore, that $\Omega_{DLA}$ might not include
the bulk of the neutral gas mass in the Universe.

\end{abstract}

                                                                               
\section{Introduction}

Recently, Fukugita \& Peebles (2004) have summarized current
measurements of the local mass-energy inventory. Of the local baryonic
matter, about 6\% is stars or their end states, about 4\% is hot
intracluster x-ray emitting gas, and somewhat less than 2\% is neutral
or molecular gas. The remainder of the baryonic matter is 
assumed to be in the form of a warm-hot intergalactic medium (WHIM),
with properties similar to those discussed by Cen  \& Ostriker
(1999). However, importantly, processes in the neutral  and molecular
gas components most directly influence the  formation of stars in
galaxies. Thus, the determination of empirical  results on the
distribution and cosmic evolution of neutral hydrogen  gas is a key
step in better understanding galaxy formation.  At present, there are
two observational methods to study neutral hydrogen. Locally, the
information is obtained through radio observations of HI 21 cm
emission. But at large distances (redshift $z>0.2$) radio sensitivity
limitations require that the information be obtained through
observations of Ly$\alpha$ absorption in the spectra of background
quasars. Intervening damped Ly$\alpha$ (DLA) absorption-line systems
in quasar spectra provide important non-local probes of the neutral
gas content of the universe since they can, in principle, be tracked
from the present epoch all the way back to the farthest detectable
quasars. Since the first survey for DLAs nearly two decades ago
(Wolfe, Turnshek, Smith, \& Cohen 1986), it has been accepted that
they contain the bulk of the neutral gas content of the universe. This
first survey defined a DLA absorption-line system as an intervening
gaseous HI region with neutral hydrogen column density
$N(HI)\ge 2\times 10^{20}$  cm$^{-2}$.  The damping wings of the
Voigt profile become prominent at column densities near $10^{19}$
 cm$^{-2}$. Thus, even low-resolution spectra which are useful
for the detection of Ly$\alpha$ absorption lines with rest equivalent
widths (REWs) $\ge10$ \AA\ can be adopted to perform DLA searches, and
subsequent studies have used this threshold to describe the statistics
of DLAs (Lanzetta et al.  1991; Rao \& Briggs 1993; Lanzetta, Wolfe,
\& Turnshek 1995;  Rao, Turnshek, \& Briggs 1995, henceforth RTB95;
Rao \& Turnshek 2000, henceforth RT00; Storrie-Lombardi \& Wolfe 2000;
P\'eroux et al. 2003; Prochaska \& Herbert-Fort 2004). The
$N(HI)\gtrsim 10^{20}$  cm$^{-2}$ limit is  believed to be the
threshold above which the gas becomes predominantly neutral and
conducive for future star formation.

The conclusion that DLA surveys identify the bulk of the neutral gas
in the universe is based on three results or assumptions. First,
integration of  the  DLA HI column density distribution   shows that a
relatively small fraction of the neutral gas is contributed by Lyman
limit and sub-DLA absorption systems with $3\times10^{17} < N(HI) <
2\times10^{20}$  cm$^{-2}$, at least  for $z<3.5$ (P\'eroux et
al. 2003, 2005), and perhaps at all redshifts (Prochaska \&
Herbert-Fort 2004). Second, dust obscuration does not cause DLA
surveys to miss a large fraction of the neutral gas (Ellison et
al. 2001, 2004).  Third, the biases introduced by gas cross section
selection are small. However, with regard to this last point, it is
important to emphasize that the interception (or discovery)
probability is the product of gas cross section times comoving
absorber number density, and no DLAs with $N(HI) > 8\times10^{21}$
cm$^{-2}$ have been discovered. Thus, the third assumption requires
that rare systems with relatively low gas cross section and very high
HI column density are either absent or have not been missed to the
extent that the neutral gas mass density will be significantly
underestimated by quasar absorption line surveys. But this assumption
might have to be  reevaluated in order to explain the discrepancy
between the  star formation history (SFH) of DLAs as inferred from
their HI column densities and that determined from galaxies that trace
the optical luminosity function (Hopkins, Rao, \& Turnshek 2005). We
will, therefore, address the validity of this assumption later.

Our main purpose in this paper is to present the results of the most
extensive survey for low-redshift DLAs to date.  Since the Ly$\alpha$
line falls in the UV for redshifts $z<1.65$, Hubble Space Telescope
(HST) spectroscopy is needed to detect and measure DLAs in this
redshift regime that corresponds to the last $\approx70$\% of the age
of the universe. Coupled with the fact that DLAs are rare, the
scarcity of available HST time has meant that a good statistical
description of the  neutral gas content at low redshift is
lacking. Now, with the Space Telescope Imaging Spectrograph (STIS) out
of commission and the installation of the Cosmic Origins Spectrograph
(COS) on HST only a remote possibility, further progress with UV
spectroscopy seems unlikely, at least for the foreseeable
future. 

To implement a low-redshift DLA survey with HST we have used an
approach which differs from the conventional blind searches for quasar
absorption lines.  RTB95 developed a method to determine the
statistical properties of low-redshift DLAs by bootstrapping from
known MgII absorption-line statistics. A similar approach was
originally used by Briggs \& Wolfe (1983) in an attempt to find 21 cm
absorbers towards radio-loud quasars. It has been appreciated for some
time that strong MgII-FeII systems generally have HI column
densities in excess of $10^{19}$  cm$^{-2}$ (e.g., Bergeron \&
Stasi\'nska 1986). Thus, since all high-redshift DLAs are known to be
accompanied by low-ionization metal-line absorption (e.g., Turnshek et
al. 1989, Lu et al.  1993, Wolfe et al. 1993, Lu \& Wolfe 1994,
Prochaska et al. 2003a and references therein), a UV spectroscopic
survey for DLAs can be accomplished efficiently if the search is
restricted to quasars whose spectra have intervening low-ionization
metal-line absorption. Since the MgII$\lambda\lambda2796,2803$
absorption doublet can be studied optically for redshifts $z>0.11$,
MgII turns out to be an ideal tracer for low-redshift DLAs.  If
the incidence of metal lines is known, then the fraction of DLAs in
the metal-line sample gives the incidence of DLAs. We further
developed this method in RT00, and accomplished a three-fold increase
in the number of low-redshift DLAs. We can now confidently use metal
absorption-line properties as a predictor for the presence of DLAs.

In this paper we present results from a sample of nearly 200
MgII systems with UV spectroscopy. Most of these data were
obtained by us during the course of HST Guest Observer programs to
make low-redshift surveys for DLAs.  In principle, once DLAs are
identified, follow-up observations can reveal details of a DLA's
element abundances, kinematic environment, associated galaxy (i.e., a
so-called DLA galaxy), star formation rate, temperature, density,
ionization state, and size. For example, there is now clear evidence
that the neutral gas phase element abundances are increasing from high
to low redshift (e.g., Prochaska et al. 2003b, Rao et al. 2005).
There is a clear trend which indicates that DLAs residing in regions
exhibiting larger kinematic spread have higher element abundances
(Nestor et al. 2003; Turnshek et al. 2005).  At low redshift ($z<1$),
it is now usually possible to identify the DLA galaxy through imaging
(e.g., Rao et al. 2003).  At high redshift, high spectral resolution
observations can be used to test dynamical models for DLA galaxies
(Prochaska \& Wolfe 1998).  When the background quasar is radio loud,
observations of 21 cm absorption provide important results on gas
temperature (e.g., Kanekar \& Chengalur 2003). Observational
constraints on physical conditions (temperature, density, ionization)
also come from high-resolution spectroscopy, and this has led to
estimates of star formation rates in individual objects (Wolfe,
Prochaska, \& Gawiser 2003). Estimates on the contribution of DLAs to
the cosmic SFH have also been made (Hopkins, Rao, \& Turnshek
2005). Finally, observations along multiple closely-spaced sightlines
have led to estimates of DLA region sizes (Monier, Turnshek, \& Rao
2005).  Through such follow-up work, our knowledge of the
characteristic properties of the neutral gas component is steadily
improving.

Thus, our results serve two purposes.  First, they provide a
comprehensive up-to-date list of more than 40 low-redshift ($z<1.65$)
DLAs suitable for follow-up studies.  Second, they provide information
on the distribution and cosmic evolution of neutral gas corresponding
to the last $\approx70$\% of the age of the universe.  We discuss the
MgII sample in \S2. The DLA sample is presented in \S3,
followed by statistical results derived from these systems in
\S4. Notably our study finds no evidence for evolution of the neutral
gas mass of the universe between $0.5<z<5$; at $z=0$ the neutral gas
mass is now estimated to be a factor of $\approx2$  lower. Moreover,
at $z\lesssim 2$ there is no evidence for evolution in the product of
absorber comoving number density and gas cross section, but at
$z\gtrsim 2$ their is clear evidence for an increase in this quantity
in comparison to no evolution models.  A discussion of these and other
new results is presented in \S5. Conclusions are summarized in \S6.

\section{The MgII Sample}

The sample of MgII lines used in our earlier DLA surveys (RTB95; RT00)
was culled from the literature. We observed 36 quasars that have 60 intervening MgII
absorption systems with MgII $\lambda$2796 REWs $W_0^{\lambda2796} \ge
0.3$ \AA\ using HST-FOS in Cycle 6 (PID 6577).  Twenty one of these
MgII systems fell in spectral regions with no flux   because of
intervening Lyman limit systems. Of the remaining 39 systems, 9 were
DLAs. With the addition of UV archival data, the total sample of MgII
systems with UV Ly$\alpha$ information included 82\footnote{Of the 87
systems reported in  Table 4 of RT00, four systems have been
eliminated for reasons noted below, and  one was reobserved in
HST-Cycle 9.  The $z_{abs}=0.1602$ system towards 0151+045 is a biased
system because the galaxy-quasar pair was known prior to the
identification of the MgII system.  The $z_{abs}=0.213$ system towards
1148+386 and the 0.1634 system towards 1704+608 were flagged as
doubtful systems by Boiss\'e et al. (1992). Also, on closer
inspection, the IUE archival spectrum of 1331+170 was inconclusive
with regard to the Ly$\alpha$ line of the $z_{abs}=1.3284$
system. Therefore, these four were eliminated from our current MgII
sample. The $z_{abs}=1.1725$ system towards 1421+330 is the one that
was reobserved in Cycle 9.}   systems of  which 12 were DLAs. We found
that all DLAs in this survey, with the exception of one, had MgII
$W_0^{\lambda2796}$ and  FeII $W_0^{\lambda2600}$ greater than 0.5
\AA. Based on this result, we conducted a similar survey of 54 MgII
systems in 37 quasars with  HST-STIS in  Cycle 9 (PID 8569). Most of
these satisfied the strong MgII-FeII criterion for DLAs. Twenty seven
had useful UV spectra and four of these  were DLAs. The DLA towards
Q1629+120 was discovered in this survey and was reported in Rao et
al. (2003).  Results on the other systems from Cycle 9 are included in
this paper.

Further progress could only be made if the sample size was increased
several fold.  The Sloan Digital Sky Survey (SDSS) sample of quasars,
which numbered in the thousands when this phase of our MgII-DLA
project began, presented an unprecedented leap in the  number of
available survey quasars. The previous largest MgII survey by  Steidel
\& Sargent (1992; SS92) used a sample of 103 quasars. The SDSS  Early
Data Release included nearly 4000 quasars. Nestor (2004) used SDSS-EDR
quasar spectra  to search for MgII systems with the aim of quantifying
the statistical  properties of a large MgII sample (Nestor, Turnshek,
\& Rao 2005, henceforth NTR05) and to conduct follow-up work to search
for DLAs. In Cycle 11 (PID 9382), we targeted a sample of 83 MgII
systems with  $W_0^{\lambda2796}\gtrsim 1$ \AA\ in 75 SDSS quasars
with SDSS magnitude $g\lesssim 19$. There were an additional  16
weaker MgII systems observable in the same set of spectra. Overall,
useful UV information was obtained for 88 systems, 25 of which are
DLAs. Given the large sample from which quasars could be selected for
observation, we were able to minimize the occurrence of intervening
Lyman limit absorption by restricting $z_{em}-z_{abs}$ to be
small. Nevertheless, we were unable to observe the Ly$\alpha$ line
for 11 MgII systems either due to Lyman limits, an intrinsic broad
absorption line trough at the position of Ly$\alpha$ absorption  in
one case,  or due to the demise of STIS.
Table 1 gives the details of the MgII systems that have UV
spectroscopic information. Our entire sample of 197 MgII systems is
included.  Details of the quasar are given in columns 1, 2, and
3. Columns 4-8 give the absorption line information obtained either
from the literature (column 9 gives the reference) or from SDSS quasar
spectral analysis (NTR05). Column 10 is the $N(HI)$ measurement from
the UV spectrum, column 11 is the selection criterion flag used to
determine DLA statistics, and column 12 gives the source of the UV
spectrum.

We now explain the selection criteria used to include MgII systems in
our sample. The total sample is divided into 4 sub-samples which
essentially arose from the process of redefining and improving upon
our selection process. Our first surveys, described in RTB95 and RT00,
included strong MgII systems from the literature; the threshold REWs
were chosen to match the MgII sample of SS92 so that their statistical
results could be used to determine the incidence of  DLA systems. As
demonstrated in RT00, if the incidence of MgII systems  is known, then
the fraction of DLAs in a MgII sample gives the incidence of DLAs.  We
found that  half of the MgII systems in our sample with \Wmi $>0.5$
\AA\ and \Wf $>0.5$ \AA\ were DLAs, and thus, modified  our selection
criteria to include a threshold FeII $\lambda 2600$ REW criterion
as well.  However, we retained the MgII REW thresholds at 0.3 \AA, 0.6
\AA, and 1.0 \AA.

The sub-samples are  defined by the following criteria:

\begin{enumerate}
\item{ \Wmi$\ge0.3$ \AA; }
\item{ \Wmi$\ge0.6$ \AA;}
\item{ \Wmi$\ge0.6$ \AA\ and \Wf$\ge 0.5$ \AA; and }
\item{ \Wmi$\ge1.0$\AA\ and  \Wf$\ge0.5$ \AA. }
\end{enumerate}

Sub-sample 1 includes  all systems surveyed in RT00, as well as
additional systems that happened to fall along  quasar sightlines that
were targeted due to the presence of another stronger system from
sub-samples 2, 3, or 4. Sub-samples 2 and 3 were mainly  targeted for
observation in HST-Cycle 9, and sub-sample 4 includes systems  found
in SDSS-EDR spectra and observed in HST-Cycle 11. A few systems from
the  SDSS-EDR sample have strong MgII and FeII, but have \Wmi$\lesssim
1.0$\AA;  these belong in sub-sample 3.  As we will see in \S 4.2,
this classification is necessary for determining the incidence,
$n_{DLA}(z)$, of the DLAs.

\section{The DLAs}

Here we present Voigt profile fits to the new DLAs.  These 28 systems,
3 of which were observed in HST-Cycle 9, are shown in Figure 1. The
resulting column densities and errors are listed in Table 1.  As is
usually the case for high column density lines,  the Ly$\alpha$ forest
populates DLA troughs making it inappropriate to use an automated
routine such as least squares minimization to fit a Voigt profile to
the data. Therefore, the best fit was estimated using the following
procedure.  Since the continuum fit is the largest source of
uncertainty in determining $N(HI)$, errors were determined by moving
the continuum level by 1$\sigma$ above and below the best-fit
continuum, renormalizing the spectrum, and refitting a Voigt profile
(see RT00).  The differences between these values and $N(HI)$
determined from the best-fit continuum   are listed as the positive
and negative errors in column 10 of Table 1.
\begin{figure}
\plotone{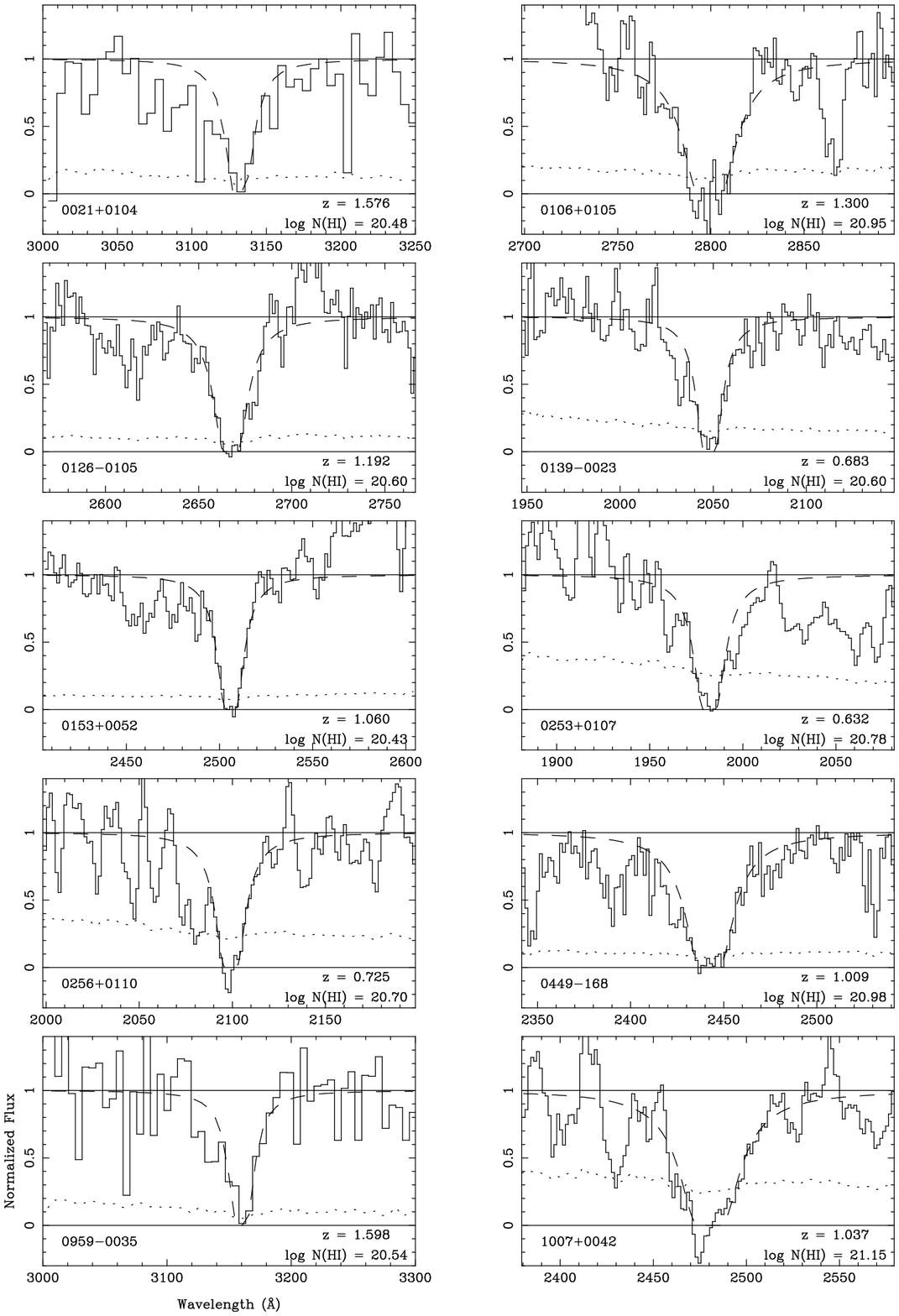}
\caption{Voigt profile fits to the DLA lines. The quasar, MgII
$z_{abs}$,  and $N(HI)$ are given in each panel. The dashed line is
the best-fit Voigt  profile and the dotted line is the 1$\sigma$ error
array. }
\end{figure} 

\begin{figure}
\figurenum{1}
\plotone{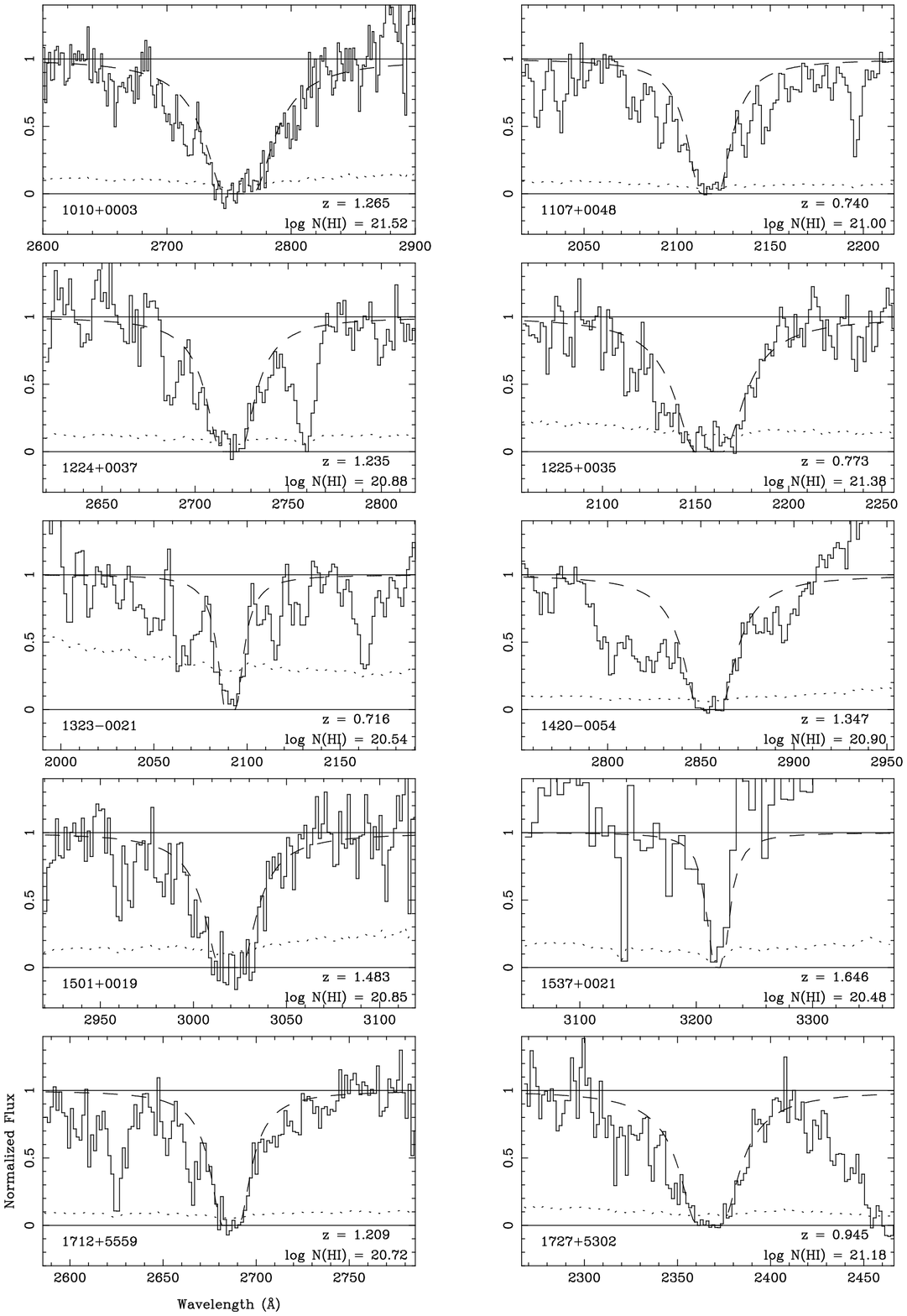}
\caption{cont.}
\end{figure}

\begin{figure}
\figurenum{1}
\plotone{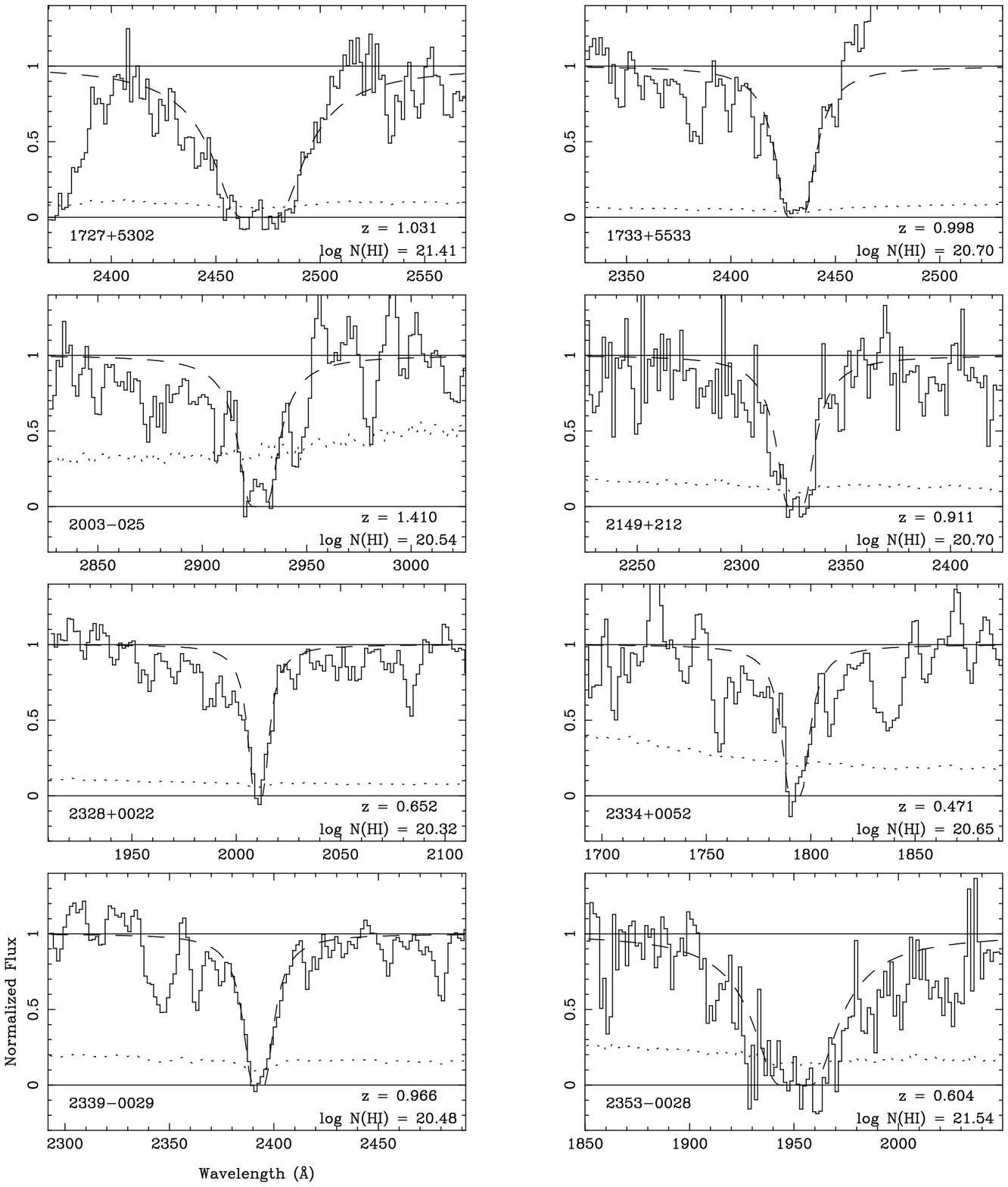}
\caption{cont.}
\end{figure}
\clearpage
\section{Statistical Results}

\subsection{Parameter Distributions and Correlations}
\begin{figure}
\epsscale{1.0}
\plotone{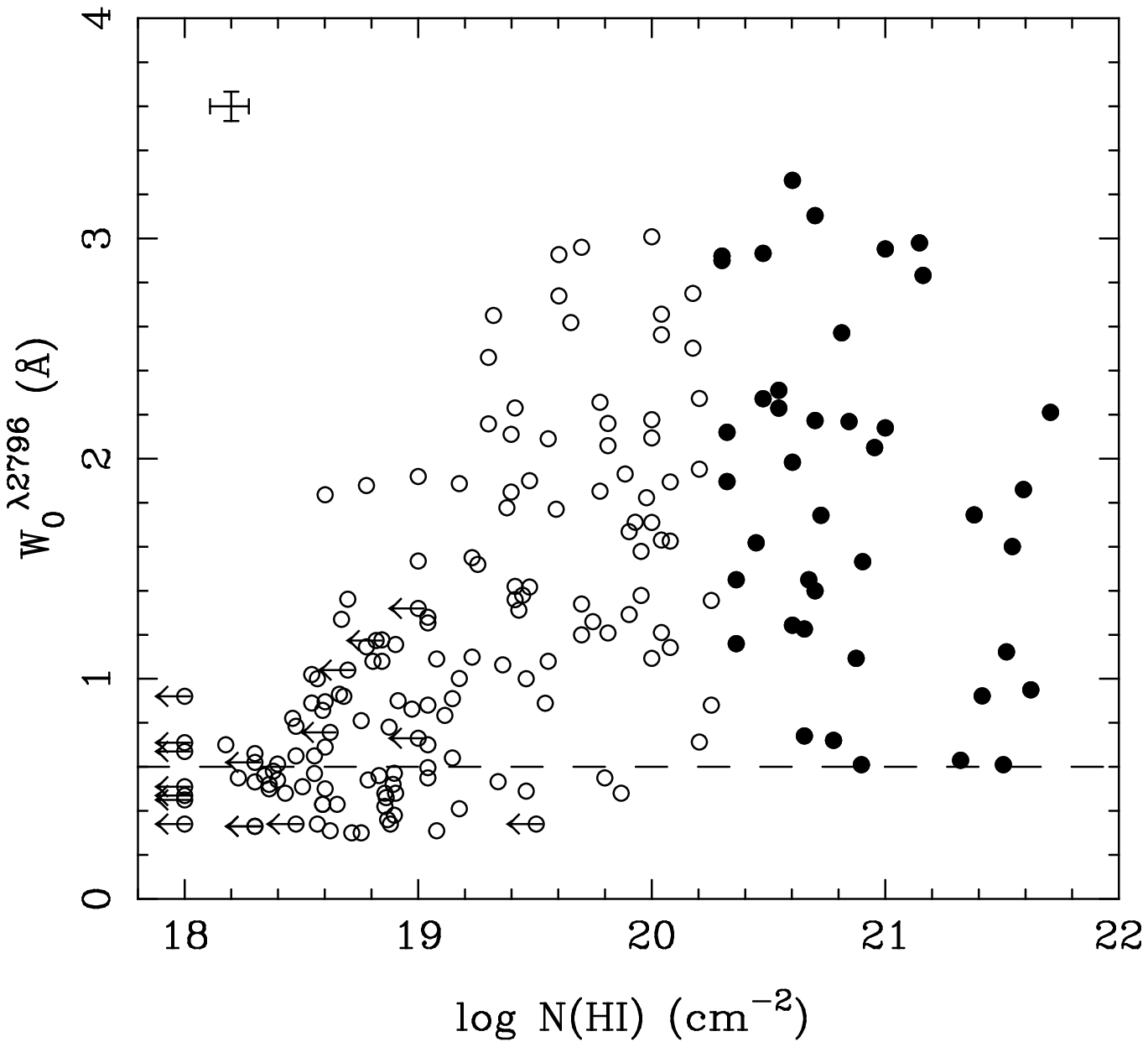}
\caption{Plot of \Wmi\ vs. $\log N(HI)$. Filled circles are DLAs with
$N(HI)\ge 2\times 10^{20}$ atoms cm$^{-2}$. Arrows are upper limits in
$N(HI)$.  Typical uncertainties are given by the error bars in the top
left corner.}
\end{figure}
Since the systems in our sample were selected based on the rest
equivalent  width of MgII $\lambda2796$, measurements of
$W_0^{\lambda2796}$ and $N(HI)$ exist for all 197 systems. MgII
$\lambda2803$, the weaker member of the doublet, was also measured for
all systems; measurements of the FeII $\lambda2600$ and MgI $\lambda
2852$ lines were possible only for a subset.  In this section, we
explore correlations among metal line REWs and HI column density.
Figure 2 is a plot of \Wmi\ versus $\log N(HI)$.  We note that the
upper left region of the figure is not populated, implying that
systems with \Wmi$>2.0$ \AA\ always have HI column densities $N(HI) >
1\times 10^{19}$ cm$^{-2}$. Figure 3 gives the distribution of
MgII \Wmi; the DLAs form the shaded histogram. It is noteworthy that
there are no DLAs with \Wmi$<0.6$ \AA.\footnote{Only one known DLA has
a lower metal-line REW. The 21 cm absorber at $z=0.692$ towards 3C 286
has \Wmi$=0.39$ \AA\ and  \Wf$=0.22$ \AA\ (Cohen et
al. 1994). However, none of the 21 cm absorbers are included in  our
analysis because they are biased systems with respect to the
determination of DLA statistics  (see RT00).}  In addition, the
fraction of systems that are DLAs increases with  increasing
\Wmi. This is shown as a histogram in Figures 4 and 5; the y-axis  on
the left gives the fraction of DLAs as a function of \Wmi.  We also
plot the mean HI column density in each bin as solid circles with the
scale shown on the right. Upper limits are assumed to be
detections.\footnote{The two systems with high $b$ values (the 
$z_{abs}=1.6101$ system towards 1329+412 and the 
$z_{abs}=1.2528$ system towards 1821+107) are included
in the histograms because it is clear that they are not DLAs. However,
since their HI column density is not known, they are not included in
the calculation of the mean column density.}  Figure 4 includes all
observed  MgII systems and Figure 5 includes only the DLAs. The
vertical error bars are standard deviations in the mean and are due to
the spread of $N(HI)$ values in each bin, and the horizontal error
bars indicate bin size. For the MgII systems there is a dramatic
increase of a factor of $\approx 36$ in the mean HI column density
from the first to the second bin, beyond which $\left<N(HI)\right>$
remains constant within the errors. In particular,
Figure 4 shows that for our sample the probability of a MgII system
being a DLA is $P \approx 0$ for \Wmi\ $< 0.6$ \AA\ and, assuming a 
linear dependence, $P \approx
0.16 + 0.18($\Wmi$ - 0.6)$  for $0.6 \le$ \Wmi\ $< 3.3$ \AA.
For systems with 0.3 \AA\ $\le$
\Wmi $<$ 0.6 \AA,  $\left<N(HI)\right> = (9.7\pm 2.2)\times 10^{18}$
cm$^{-2}$, and $\left<N(HI)\right> = (3.5 \pm 0.7) \times 10^{20}$
cm$^{-2}$  for systems with \Wmi $\ge$  0.6 \AA. Figure 5 shows a
trend for decreasing DLA column density with \Wmi. The reasons for
this are not obvious, but are likely to  be due to small number
statistics (see Figure 2), a real physical effect, or a selection
effect that is not yet understood (see Turnshek et al. 2005, \S 5.1).
\begin{figure}
\plotone{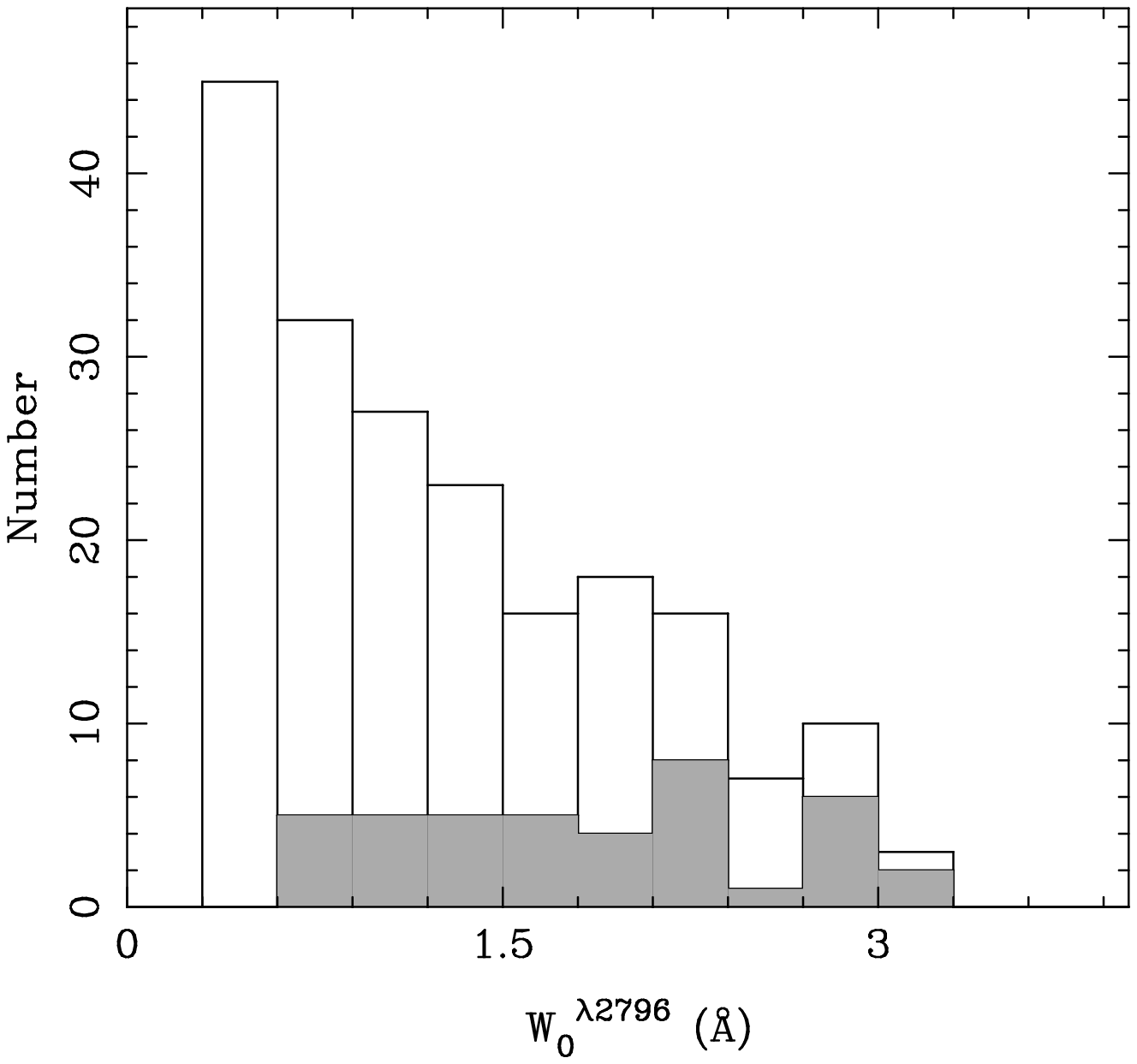}
\caption{Distribution of MgII$\lambda 2796$ rest equivalent widths,
\Wmi. The shaded histogram represents systems that are DLAs. Note that
there are no DLAs in the first bin, i.e., for MgII \Wmi$<0.6$ \AA. The
fraction of DLAs increases with increasing \Wmi\ for \Wmi$\ge 0.6$ \AA.}
\end{figure}

\begin{figure}
\plotone{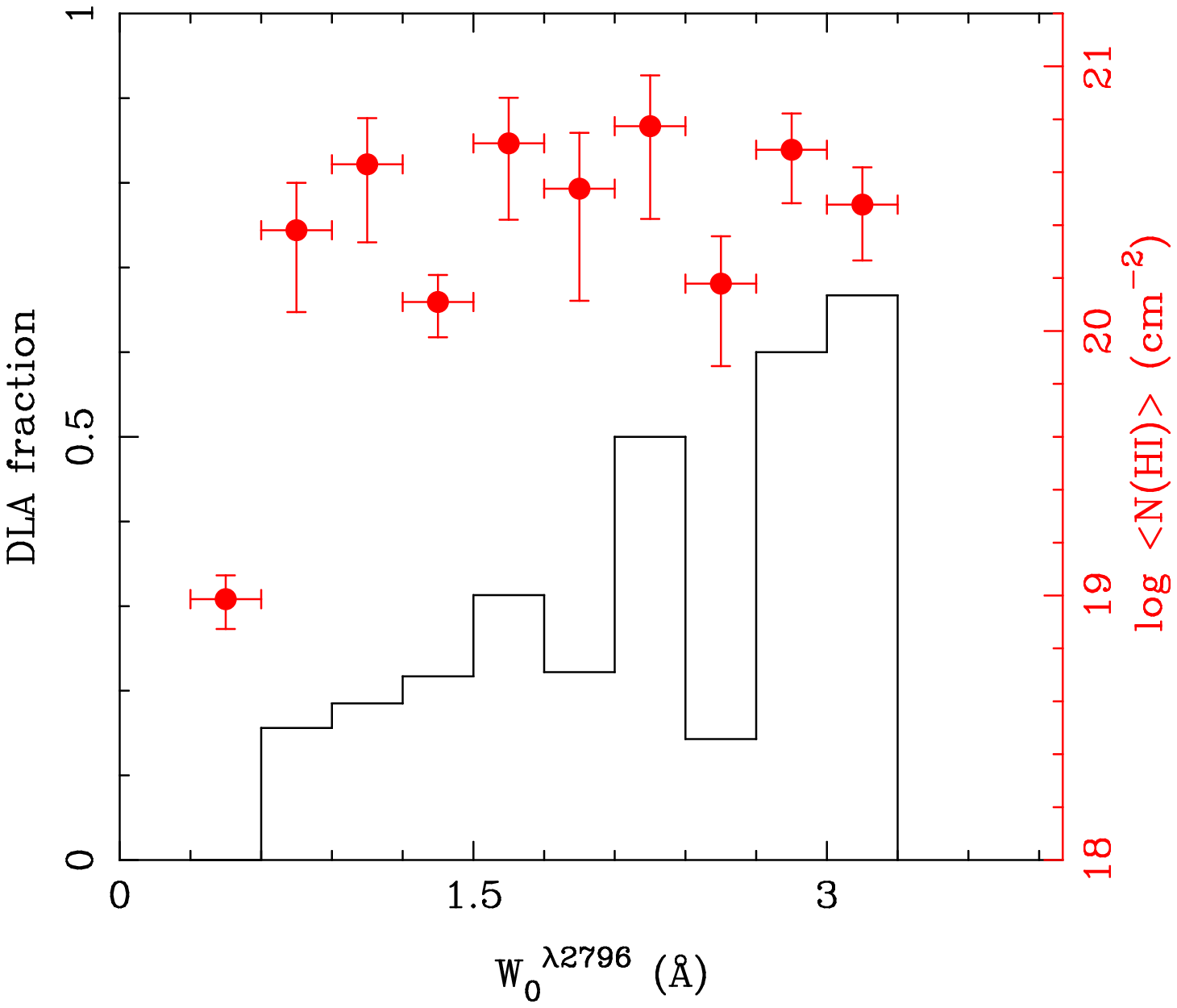}
\caption{The histogram shows the fraction of MgII systems that are
DLAs as a function of MgII \Wmi, with the scale shown on the left
axis. The  solid circles are the logarithm of the mean HI column
density in each bin. The scale is shown on the right.}
\end{figure}

\begin{figure}
\plotone{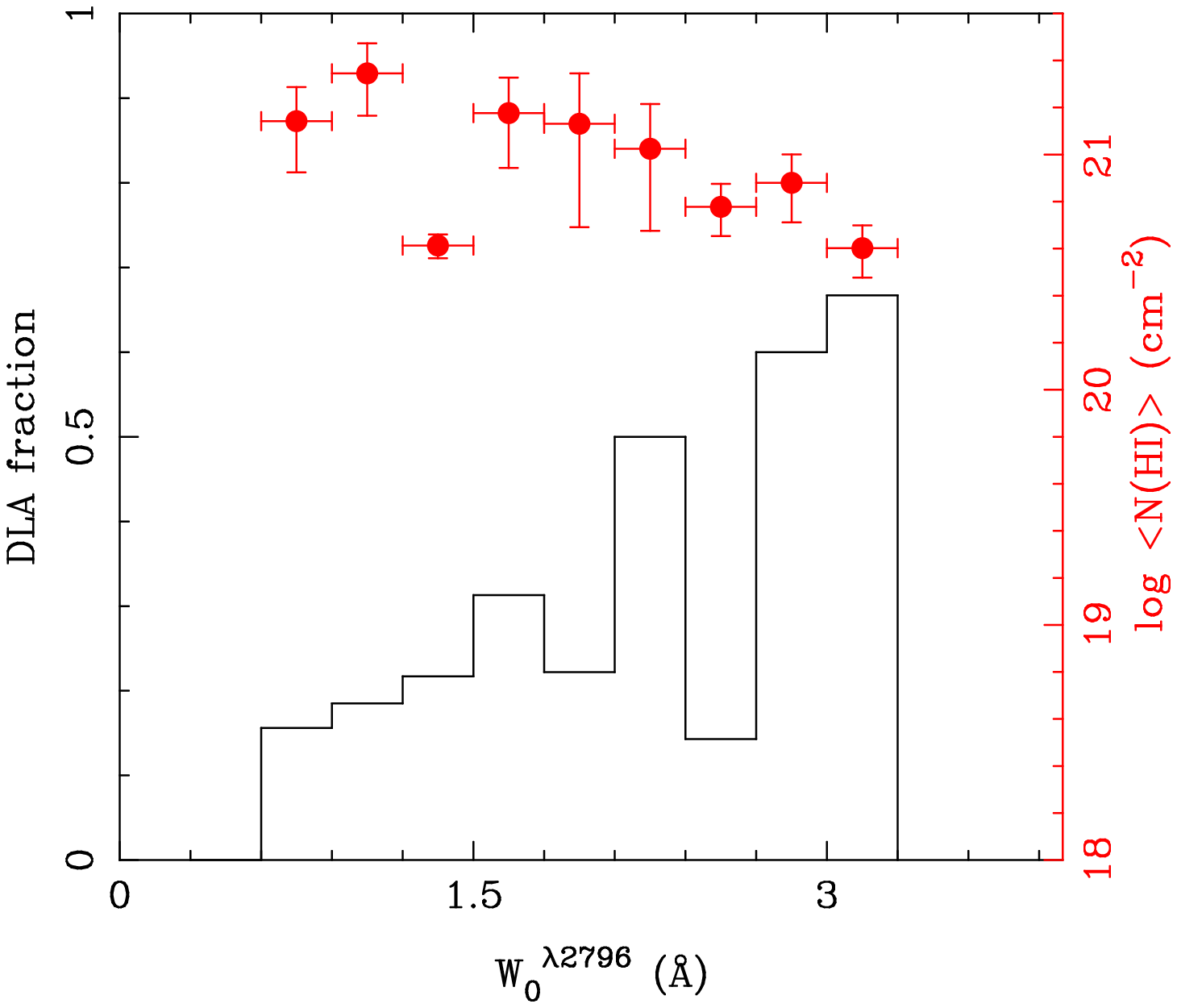}
\caption{Same as Figure 4, but points are for DLAs only.}
\end{figure}
Since some of the higher \Wmi\ systems in the sample have an
FeII$\lambda 2600$ selection criterion  folded in,  we  repeat the
above analysis here for the non-FeII selected part of the
sample. Figure 6 is a \Wmi\ distribution for systems that do not
include the FeII selection criterion. These are systems that belong to
sub-samples 1 and 2. The DLAs form the shaded histogram. The first bin
contains the same systems as in Figure 3. Even with this smaller
sample, 111 systems compared to 197, we find that the fraction of DLAs
increases with \Wmi.  The mean HI column density for this sample is
shown in Figure 7. Since the number of  systems in the higher \Wmi\
bins is small, we bin the data differently from Figure 4 and, for
comparison, show the Figure 4 sample rebinned as well. We find that
the FeII selection has no effect on the mean column density as a
function of \Wmi. Consistent with our  larger sample, for systems with
\Wmi $\ge$  0.6 \AA\ we find $\left<N(HI)\right> =  (3.40 \pm 1.25)
\times 10^{20}$ cm$^{-2}$. One might expect that since the fraction of
DLAs increases with increasing \Wmi\ and that the FeII selection
primarily affects higher \Wmi\ systems, the mean HI column density
should be higher in the FeII selected sample. However, again, this may
be offset by the fact that the mean DLA HI column density decreases
with increasing  \Wmi, thus keeping the mean HI column density of FeII
and non-FeII selected samples indistinguishable.
\begin{figure}
\plotone{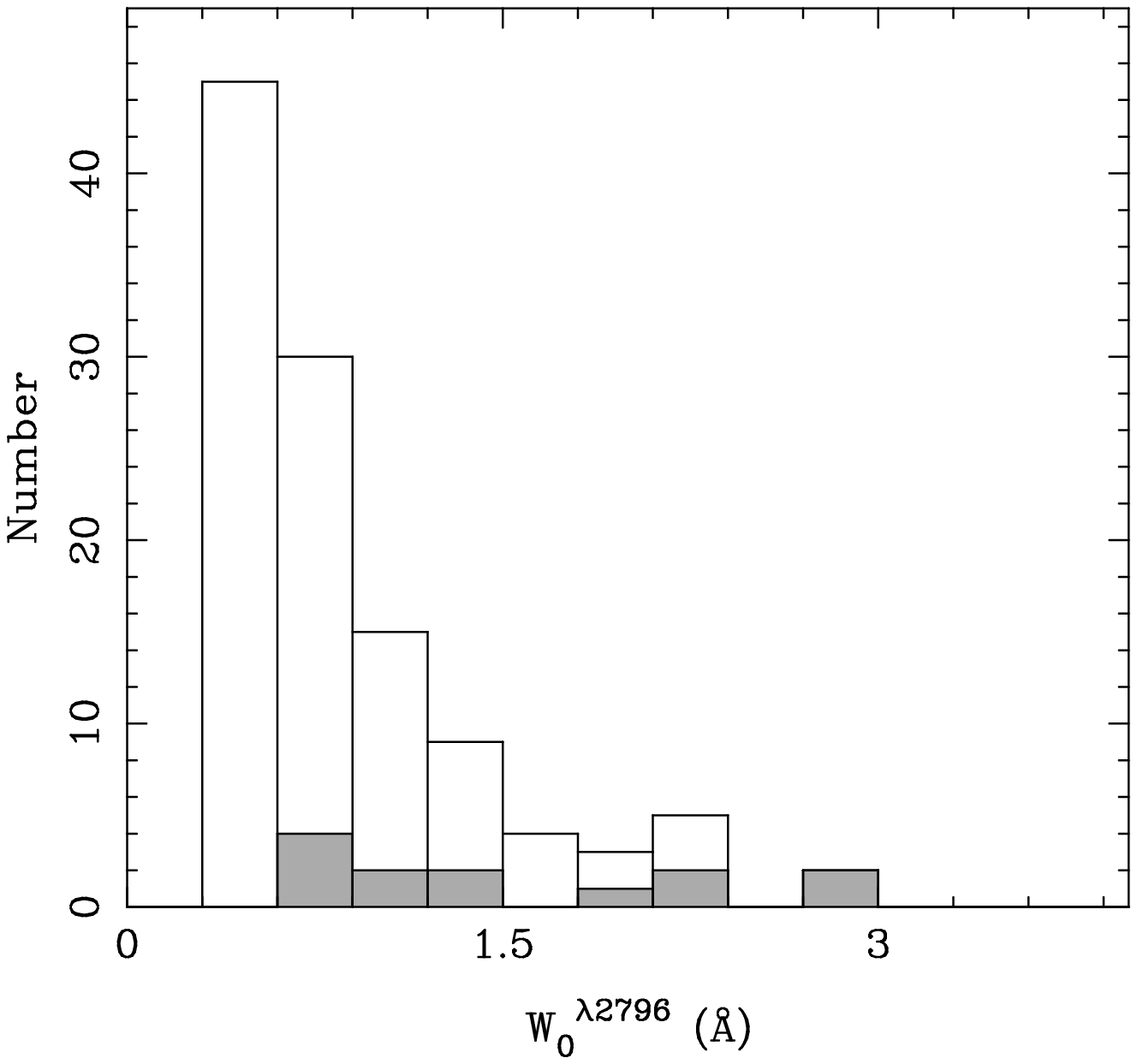}
\caption{Same as Figure 3 but for non-FeII selected systems. These
form sub-samples 1 and 2. The shaded histogram shows the
DLAs. As in Figure 3, the fraction of DLAs in bins with MgII \Wmi$\ge
0.6$ \AA\  increases with  increasing \Wmi.}
\end{figure}

\begin{figure}
\plotone{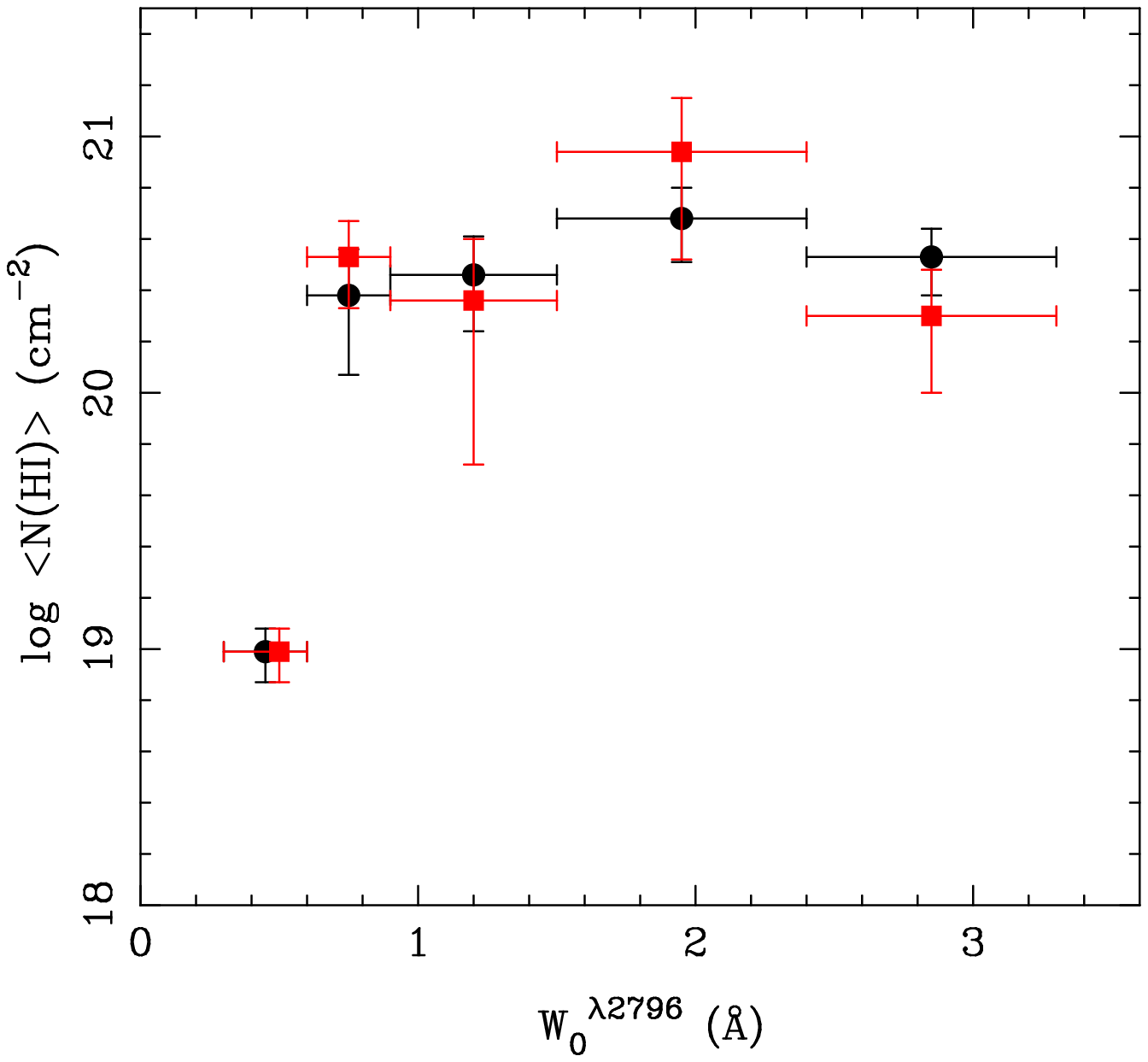}
\caption{Logarithm of the mean HI column density of absorbers as a
function of MgII \Wmi.  The red solid squares are for non-FeII
selected systems, i.e., for sub-samples 1 and 2. The black solid
circles are the same data shown in Fig. 4,  but rebinned to match the
binning of the sub-sample without FeII selection. The data points in
the first bin are identical but have been displaced for clarity.  The
FeII selection has no effect on the mean HI column density as a
function of MgII \Wmi.  }
\end{figure}
 
\begin{figure}
\plotone{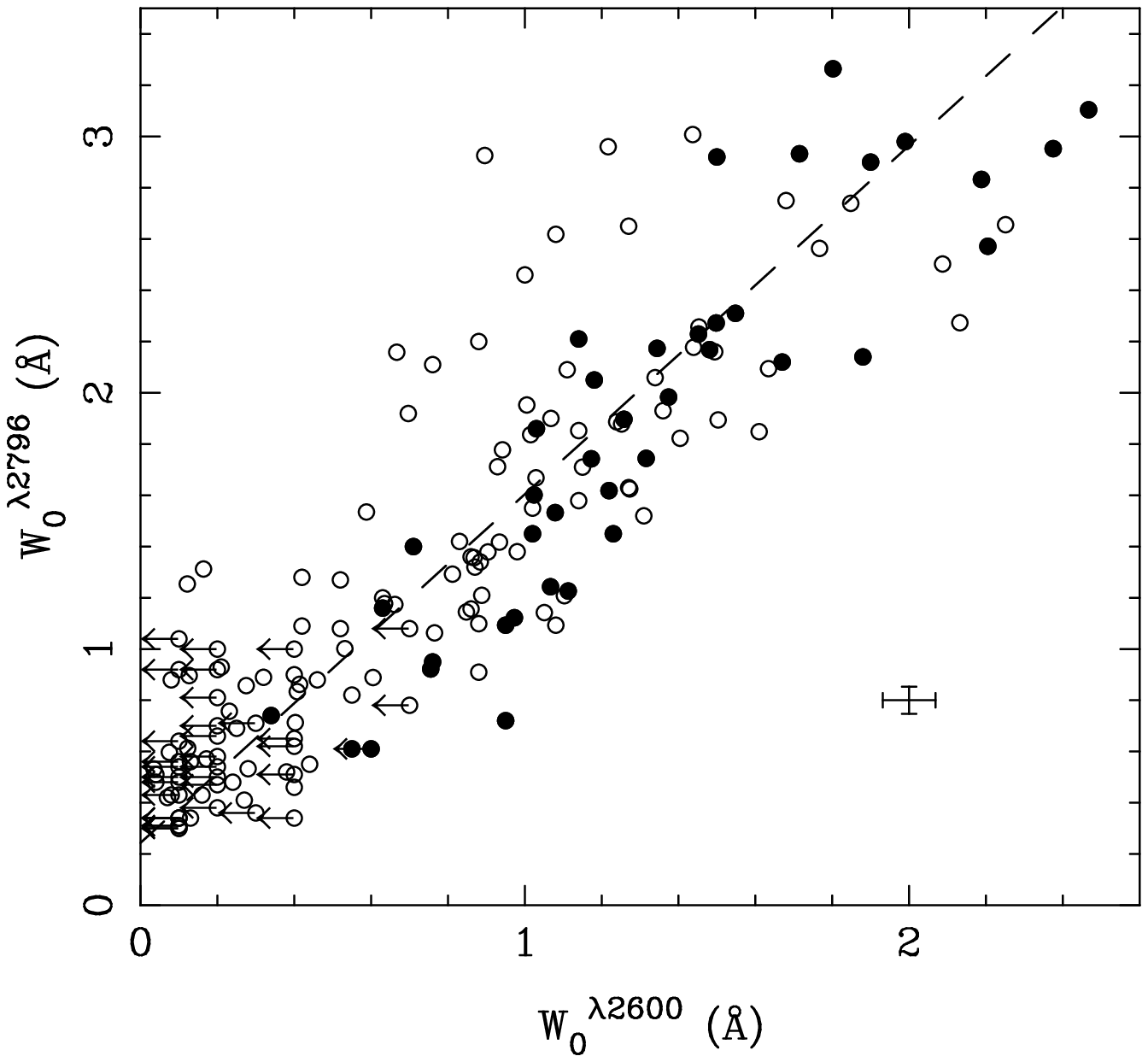}
\caption{Plot of \Wmi\ vs. \Wf\ for all MgII systems that have
measured values of \Wf. Filled circles are DLAs. Typical error bars
are shown at lower right.  The dashed line is the best fit linear
correlation described in the text with slope $b=1.36$. }
\end{figure}

\begin{figure}
\plotone{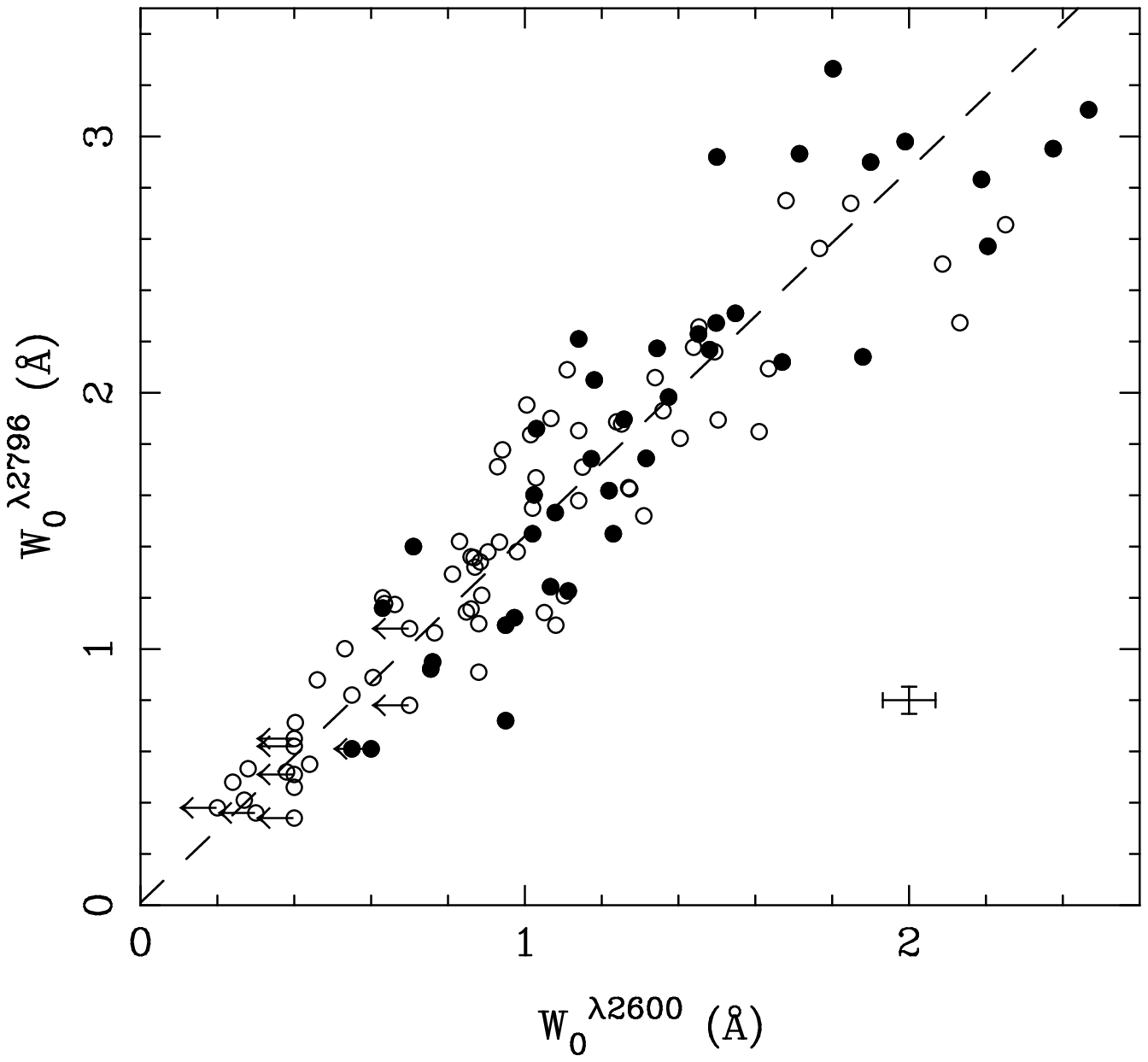}
\caption{Plot of \Wmi\ vs. \Wf\ for systems with \Wmi/\Wf$<2$. Filled
circles are DLAs. Non-DLAs in the upper left and  lower left region of
the diagram have been eliminated and the correlation is tighter than
what is seen in Figure 8. Also, 38\% of all systems are DLAs
regardless of  the values of \Wmi\ and \Wf; we believe this to be a
more robust predictor of the presence of DLAs in MgII-FeII
systems. The slope of the best fit linear correlation is $b=1.43$.}
\end{figure}
Figure 8 is a plot of  \Wmi\ vs. \Wf\ for systems with measured values
of \Wf, including upper limits.  In RT00 we found that 50($\pm 16$)\%
of the 20 systems  (excluding  upper limits and 21 cm absorbers) with
\Wmi\ $> 0.5$ \AA\ and \Wf\ $> 0.5$ \AA\ are DLAs. Now, with the
expanded sample that includes 106 systems in this   regime, we find
that 36($\pm 6$)\% are DLAs. The dashed line is a least-squares fit
with slope $b=1.36 \pm 0.08$ and intercept $a=0.24 \pm 0.06$. It was
determined using the BCES estimator of Akritas \& Bershady (1996),
assuming intrinsic scatter but uncorrelated errors in \Wmi\ and
\Wf. Upper limits were not used for the fit.  We note that DLAs do not
populate the  top left region of the diagram where the \Wmi\ to \Wf\
ratio is  $\gtrsim 2$. In fact, if the sample is restricted to systems
with \Wmi/\Wf$<2$, all but one of the DLAs in  Figure 8 are retained,
the outliers in the top left region are excluded as are most systems
in the lower left corner of the plot. Figure 9 shows this truncated
sample; the slope of the least-squares fit does not change
significantly. We find $b=1.43 \pm 0.08$ and $a=0.01 \pm 0.08$ for
this definition of the sample. The only DLA that has been eliminated
is the one with the smallest value of  \Wf. However, given the
measurement errors for this system, its  \Wmi/\Wf ratio is within
1$\sigma$ of 2. The implication is that a system with metal line ratio
\Wmi/\Wf $>2$ has nearly zero probability of being a DLA.  For this
truncated sample with \Wmi/\Wf$<2$, but no restrictions on the
individual values of \Wmi\ or \Wf,  38($\pm 6$)\% are DLAs.  In
addition, all known 21 cm absorbers, including the $z=0.692$ system
towards 3C 286 mentioned above, have \Wmi/\Wf$<2$. Thus, the \Wmi/\Wf\
ratio provides a more robust predictor of the presence of a DLA.
\begin{figure}
\plotone{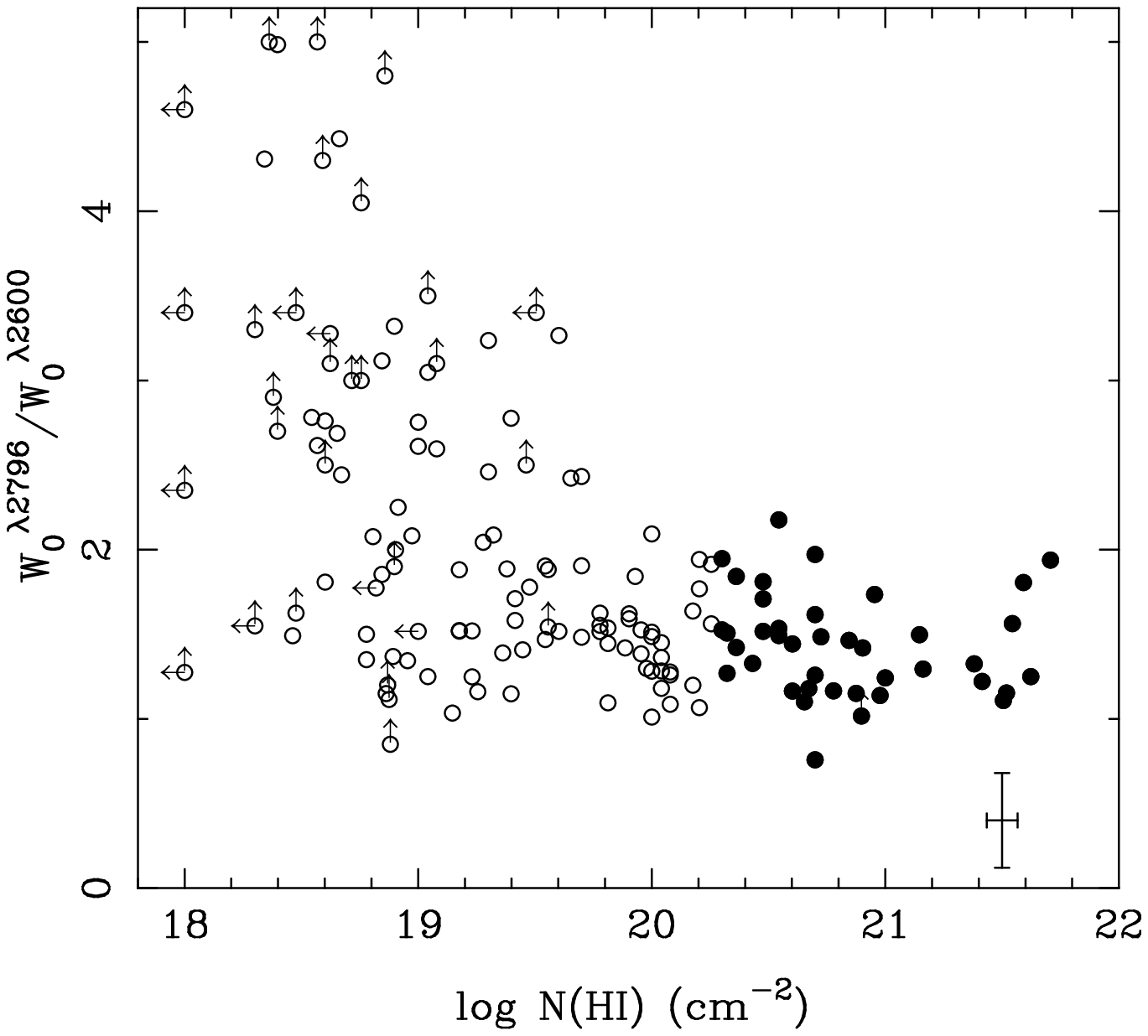}
\caption{Plot of \Wmi/\Wf\ vs. $N(HI)$. Systems with \Wmi/\Wf$>5$ are
not shown for clarity; all of these have $\log N(HI)<19.6$. The DLAs
(filled circles) are confined to the region of the plot where
$1\lesssim$\Wmi/\Wf$\lesssim 2$.}
\end{figure}
 
\begin{figure}
\plotone{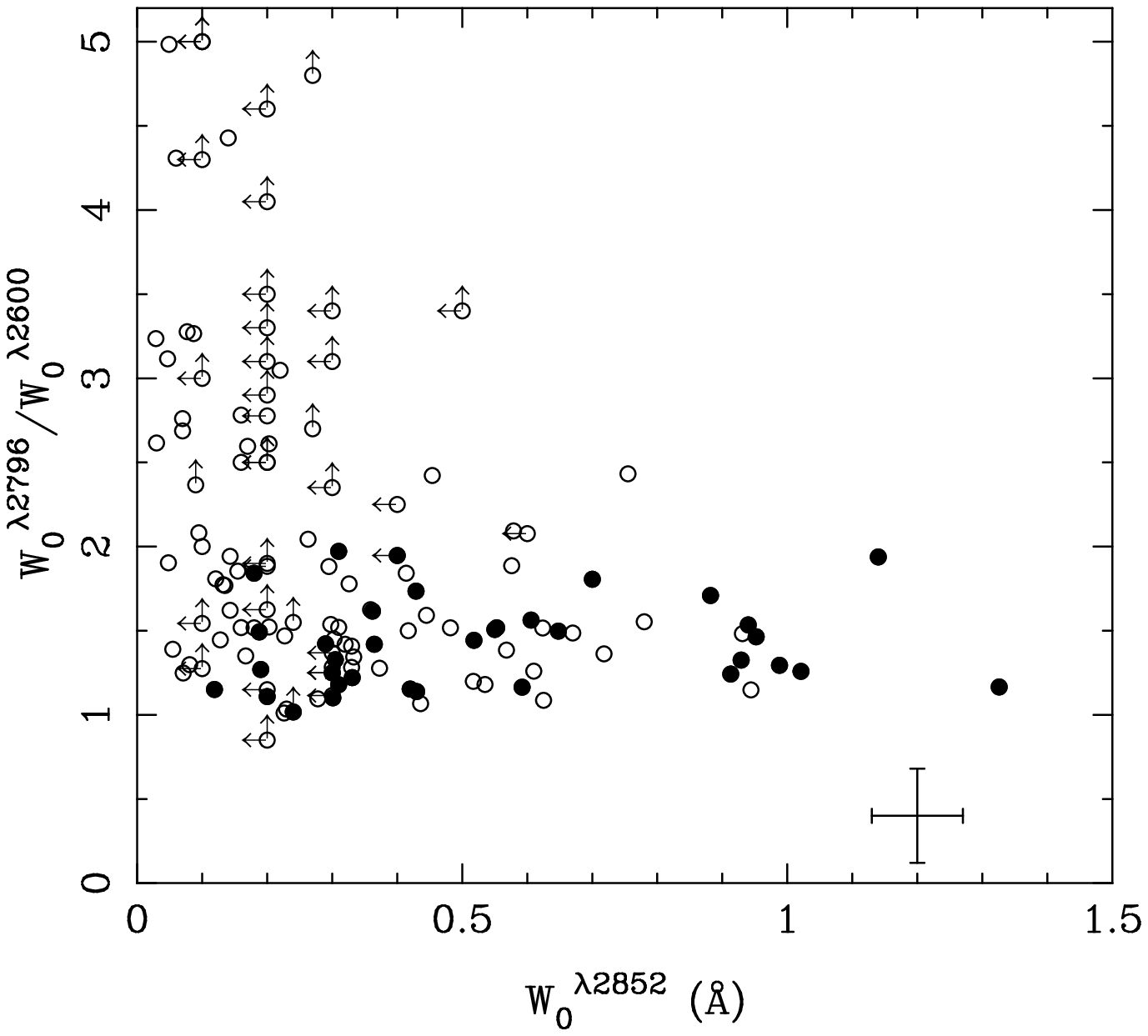}
\caption{Plot of \Wmi/\Wf\ vs. MgI \Wmiii. Filled circles are
DLAs. Typical error bars are shown in the lower right corner. Again,
the DLAs are  confined to the region where
$1\lesssim$\Wmi/\Wf$\lesssim 2$, but span almost the entire range of 
\Wmiii\ (\Wmiii $>0.1$ \AA).}
\end{figure}
This result is shown more dramatically in Figures 10 and 11. The ratio
\Wmi/\Wf\ is plotted as a function of $N(HI)$ in Figure 10. Ratios
above 5 are not shown for clarity. These are mainly confined to $\log
N(HI) < 19.2$ with  only one system above this column density at $\log
N(HI) = 19.6$.  The DLAs populate the region of the plot where
$1\lesssim$\Wmi/\Wf$\lesssim 2$; the two outliers lie within $1\sigma$
of this range.  A plot of the ratio \Wmi/\Wf\ vs. \Wmiii\ for systems
with  measured values of  \Wmiii, including upper limits, is shown in
Figure 11. Again, the DLAs are  confined to the region where
$1\lesssim$\Wmi/\Wf$\lesssim 2$, but span almost the entire range of 
\Wmiii\ (\Wmiii $>0.1$ \AA).
The two systems outside the range $1\lesssim$\Wmi/\Wf$\lesssim 2$ from
Figure 10 do not have information on \Wmiii. Of the systems with
measured values of \Wmiii, 32 of the 77 systems  with
\Wmi/\Wf$\lesssim 2$ and \Wmiii$\ge0.1$, i.e, 42($\pm 7$)\%, are DLAs.  The
other 9 DLAs either do not have measured values of \Wmiii, or have
high upper limits due to poor data quality.  We also find that 9 out
of the 11 systems with \Wmiii$>0.8$\AA\  are DLAs.  We note that
systems with \Wmi/\Wf$\gtrsim2$ are likely to have low values of
\Wmiii.
\begin{figure}
\plotone{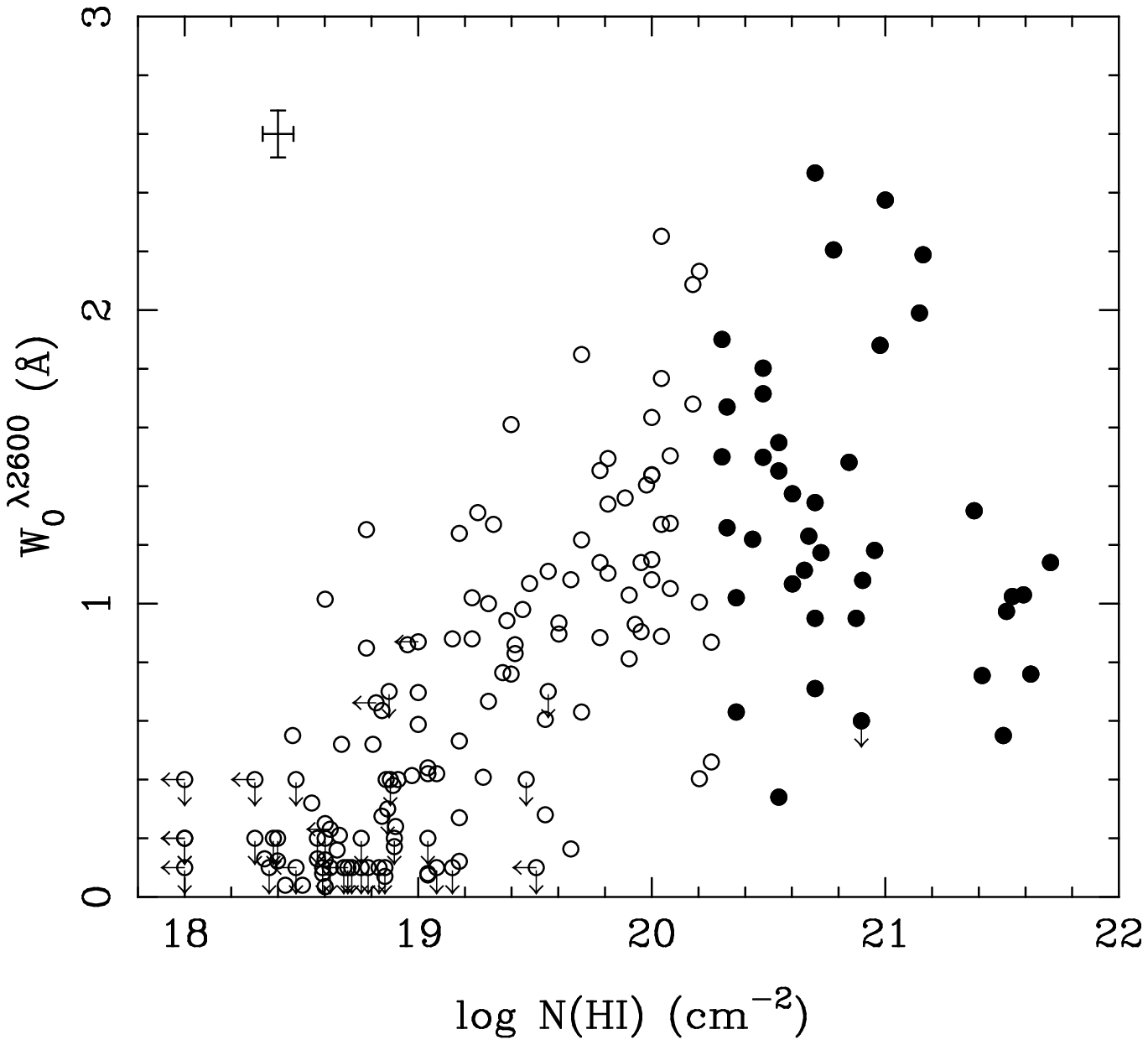}
\caption{ Plot of \Wf\ vs. $\log N(HI)$. Arrows indicate upper limits.
Filled circles are DLAs. Typical uncertainties are given by the error 
bars in the top left corner.}
\end{figure}

\begin{figure}
\plotone{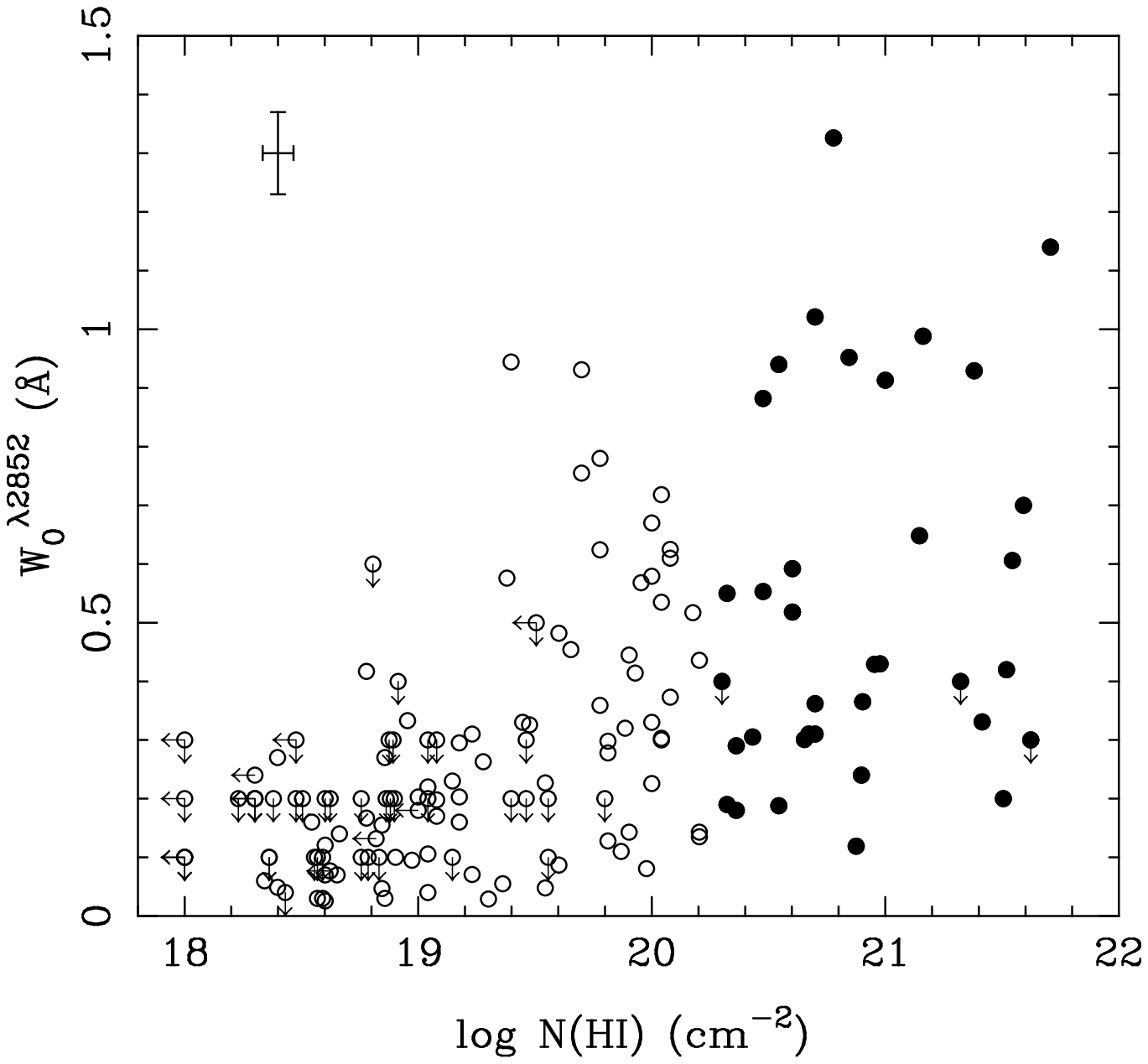}
\caption{Plot of \Wmiii\ vs. $\log N(HI)$. Arrows indicate upper
limits.  Filled circles are DLAs. Typical uncertainties are given by 
the error bars in the top left corner.}
\end{figure}
For completeness, we also plot \Wf\ vs. $\log N(HI)$ in Figure 12 and
\Wmiii\ vs. $\log N(HI)$ in Figure 13. There is no obvious trend in
these distributions except for the fact that the upper left regions of
the plots are not populated. There are no high REW, low HI column
density systems. This is not a selection effect since column densities
as low as $10^{18}$ cm$^{-2}$ can often be easily measured. This
implies  that systems  with \Wf $ \gtrsim 1$ \AA\ or \Wmiii $\gtrsim
0.5$ \AA\  generally have HI column densities $N(HI) >10^{19.0}$
cm$^{-2}$.  Below this fairly sharp boundary, metal-line REWs  span
all values of HI column density.

\subsubsection{Discussion}

How can these trends be interpreted? Apart from the upper envelopes in
Figures 2, 12, and 13, there is no other simple correlation between
metal-line rest equivalent width and HI column density. Since the
metal lines are saturated, the rest equivalent width is more a measure
of  velocity spread, not column density. High resolution observations
of MgII absorption lines have shown that the stronger systems break up
into many components (e.g., Churchill, Vogt, \& Charlton 2003), and
span velocity intervals of up to 400 km/s. Turnshek et al. (2005) show
line equivalent widths in velocity units of $\gtrsim800$ km/s  in the
strongest systems found in the SDSS.  These highest equivalent width
systems  may arise in galaxy groups; however, the more common systems
like those in our DLA survey are more likely to arise in clouds that
are  bound in galaxy-sized potentials. A DLA is observed if at least 
one of the
clouds along the sightline happens to be cold (less than a  100 K),
and with a velocity dispersion of a few 10s of km/s.  A simple
interpretation is that the greater the  number of clouds along the
sightline, the higher the probability of encountering a DLA. This
would explain the higher fraction of DLAs among large \Wmi\ systems
and the lack of a correlation between  \Wmi\ and $N(HI)$ other than
the upper envelopes in Figures 2, 12, and 13. Only rarely would a
sightline intersect a single cloud resulting in small \Wmi\ {\it and}
high $N(HI)$, as in the 3C 286 system  described in \S4.1. This
probabilistic approach to explain metal-line and  HI strengths in
high-$N(HI)$ absorbers was also proposed by Briggs \& Wolfe (1983) to
explain their MgII survey for 21 cm absorbers. They proposed a
two-phase model  where the 21 cm absorption is produced in galaxy
disks, and the metal-line  components that do not produce 21 cm
absorption are produced in galactic halos.  However, this
multi-component/cloud model is likely to be valid in any gas-rich
galaxy, as is evidenced by DLA galaxy imaging studies (Le Brun et
al. 1997; Rao \& Turnshek 1998; Turnshek et al. 2001; Rao et al. 2003;
Turnshek et al. 2004). The disk models of Prochaska \& Wolfe (1997)
and the Haehnelt, Steinmetz, \& Rauch (1998) models of infalling and
merging clouds could reproduce these observations equally well. In
other words, DLAs arise in pockets of cold gas   embedded within warm
diffuse gas or gas clouds in any bound system.
 
Twenty one cm observations of low-redshift DLAs also reveal some cloud
structure. For example, the $z=0.313$ system towards PKS 1127$-$145
shows 5 components and the $z=0.394$ system towards  B0248+430 is
resolved into 3 components (Lane 2000; Lane \& Briggs 2001; Kanekar \&
Chengalur 2001). Since the MgII line for these systems has not been
observed at a resolution as high as the 21 cm observations, a
one-to-one correspondence between the metal-line and 21 cm  clouds
cannot be drawn. In other instances, both warm and cold gas have been
detected in a 21 cm DLA; Lane, Briggs, \& Smette  (2000) find that
two-thirds of the column density in the $z=0.0912$ DLA towards
B0738+313  is contained in warm phase gas, and the rest is contained
in two narrow components. The $z=0.2212$ absorber towards the  same
quasar was also found to exhibit similar characteristics (Kanekar,
Ghosh, \& Chengalur 2001). In each of these cases, the line of sight
probably intersects two cold clouds in addition to warm diffuse gas
spread over a wider range of velocities that can be detected only in
21 cm  observations of very high sensitivity. There are also several
instances of DLAs not being detected at 21 cm (Kanekar \& Chengalur
2003). High spin temperatures ($T_s\gtrsim 1000$ K) corresponding to
warm diffuse gas and/or covering factors less than unity towards
extended quasar radio components have been  suggested as possible
explanations (Kanekar \& Chengalur 2003; Curran et al. 2005).

Clearly, a wide variety of cloud properties and their combinations are
responsible for the observed properties of MgII, DLA, and 21 cm
absorption lines.   Large simulations of galaxy sightlines with
varying cloud properties that reproduce the metal-line versus DLA
correlations shown in Figures 2-13 would be an important next step
towards improving our understanding of these absorption line
systems. The simulations should not only be able to reproduce the
frequency of occurrence of DLAs in MgII systems, but also the  number
density evolution of MgII systems and DLAs. Moreover, further analysis
on large data sets  might enable us to predict the occurrence of DLAs
among metal-line systems and determine their HI column densities to
some degree of  accuracy, but this is a project for future study.  For
the remainder of this paper we discuss the statistical properties of
neutral gas in the low redshift universe as derived from the expanded
HST sample.

\subsection{Redshift number density $n_{DLA}$}

The redshift number density of DLAs, $n_{DLA}$, sometimes written as
$dn/dz$,  can be determined using the equation
\begin{equation}
n_{DLA}(z) = \eta(z)\, n_{MgII}(z),
\end{equation}
where $\eta(z)$ is the fraction of DLAs in a MgII sample as a function
of  redshift and $n_{MgII}(z)$ is the redshift number density of MgII
systems.  Since our MgII sample was assembled under various selection
criteria  (see \S2), $n_{MgII}(z)$ needs to be evaluated
carefully. We can express $n_{MgII}(z)$ for our sample as
\begin{equation}
n_{MgII}(z)=\frac{1}{197}\sum_i w_i\, n_{{MgII}_i}(z),
\end{equation}
where the sum is over all 197 systems, $w_i$ is a weighting factor
that depends on the $i^{th}$ system's selection criterion for being
included in the survey, and $n_{{MgII}_i}(z)$ is the  $i^{th}$
system's  $dn/dz$ value calculated using the parametrization derived
in the Appendix of NTR05:
\begin{equation}
dn/dz=N^*\,(1+z)^\alpha\,e^{-\frac{W_0}{W^*}(1+z)^{-\beta}},
\end{equation}
where we have retained the notation given in NTR05 and $W\equiv$ \Wmi.
$N^*$, $W^*$, $\alpha$, and $\beta$  are constants.  This expression
is an integral over all \Wmi\ greater than $W_0$. For our calculation,
$W_0$ is different for each of four sub-samples that comprise our
total sample (see \S2 and Table 1). Thus, for example, a system that
belongs  to sub-sample 1 has a REW threshold $W_0 = 0.3$ \AA\ in
Equation 3  and weight $w_i=1$ in Equation 2, while a system in
sub-sample  2 has  $W_0=0.6$ and $w_i=1$. This is because sub-samples
1 and 2 are purely  MgII-selected samples with no regard to the
strength or presence of the FeII $\lambda 2600$ line. On the other
hand, a system that  belongs to sub-sample 3 has $W_0=0.6$  in
Equation 3 and weight $w_i=0.54$  in Equation 2. This is because an
FeII $\lambda 2600$ criterion was  used to select the system in
addition to \Wmi, and  54\% of the 1,130  \Wmi$\ge 0.6$ \AA\ systems
in the MgII survey of NTR05 have \Wf$\ge 0.5$  \AA. Similarly, for
systems in sub-sample 4, $W_0=1.0$ and $w_i=0.72$.  In this case, 72\%
of the 781 \Wmi$\ge  1.0$ \AA\ systems in NTR05 have  \Wf$\ge 0.5$
\AA. For sub-samples 3 and 4  we have assumed that  the  fraction of
MgII systems that are also strong FeII systems is independent  of
redshift.

The MgII doublet moves out of the SDSS spectroscopic range for
redshifts $z<0.36$.  In order to extend MgII statistics to lower
redshifts, we conducted a survey  of quasars with the Multiple Mirror
Telescope (MMT) on Mount Hopkins, AZ (Nestor, 2004). These  results, 
which will be
presented in a forthcoming paper (Nestor, Turnshek, \& Rao 2006, in
preparation), were used to determine DLA statistics for the redshift
range $0.11\le z\le0.36$ using the same procedure described above. Of the
11 systems from our sample in this redshift range, nine belong in
sub-sample 1 and two are in  sub-sample 2.

Figure 14 shows the results for $n_{DLA}(z)$ at low redshift split
into two redshift  bins (solid squares). We find 18 DLAs in 104 MgII
systems in the redshift interval $0.11<z\le0.9$ with $n_{DLA}(z=0.609)
= 0.079 \pm 0.019$ and 23 DLAs in 94 MgII systems in the redshift
interval $0.9<z\le1.65$ with $n_{DLA}(z=1.219) = 0.120  \pm
0.025$. We did not find it necessary to apply a Malmquist bias correction 
for the number of systems with $N(HI)\ge 2 \times 10^{20}$ cm$^{-2}$
(as was done in RT00) because the sample contains an equal number of systems
within $1\sigma$ above and below this threshold value.
Standard error propagation procedures were used to determine
uncertainties. The points are plotted at the mean redshift of the MgII
samples.  The high-redshift data points are from Prochaska \&
Herbert-Fort (2004)  and the $z=0$ point was estimated by Zwaan et
al. (2005a)  from a WSRT survey of HI in the local
universe. The solid curve is a no-evolution curve in the standard
$\Lambda$CDM cosmology that we refer to as the ``737'' cosmology where
($h$,$\Omega_M$,$\Omega_\Lambda$) = (0.7, 0.3, 0.7).  This curve,
which  has been normalized at the $z=0$ data point, shows that the
comoving  cross section for DLA absorption declined rapidly by a
factor of $\approx 2$ until $z\approx 2$ and has remained constant
since then. This behavior might be a consequence of what has been
observed in other studies of galaxy evolution, namely, that today's
galaxies were in place by $z\approx 1$ and are a consequence of  rapid
merger and/or collapse events that occurred prior to this epoch.
\begin{figure}
\plotone{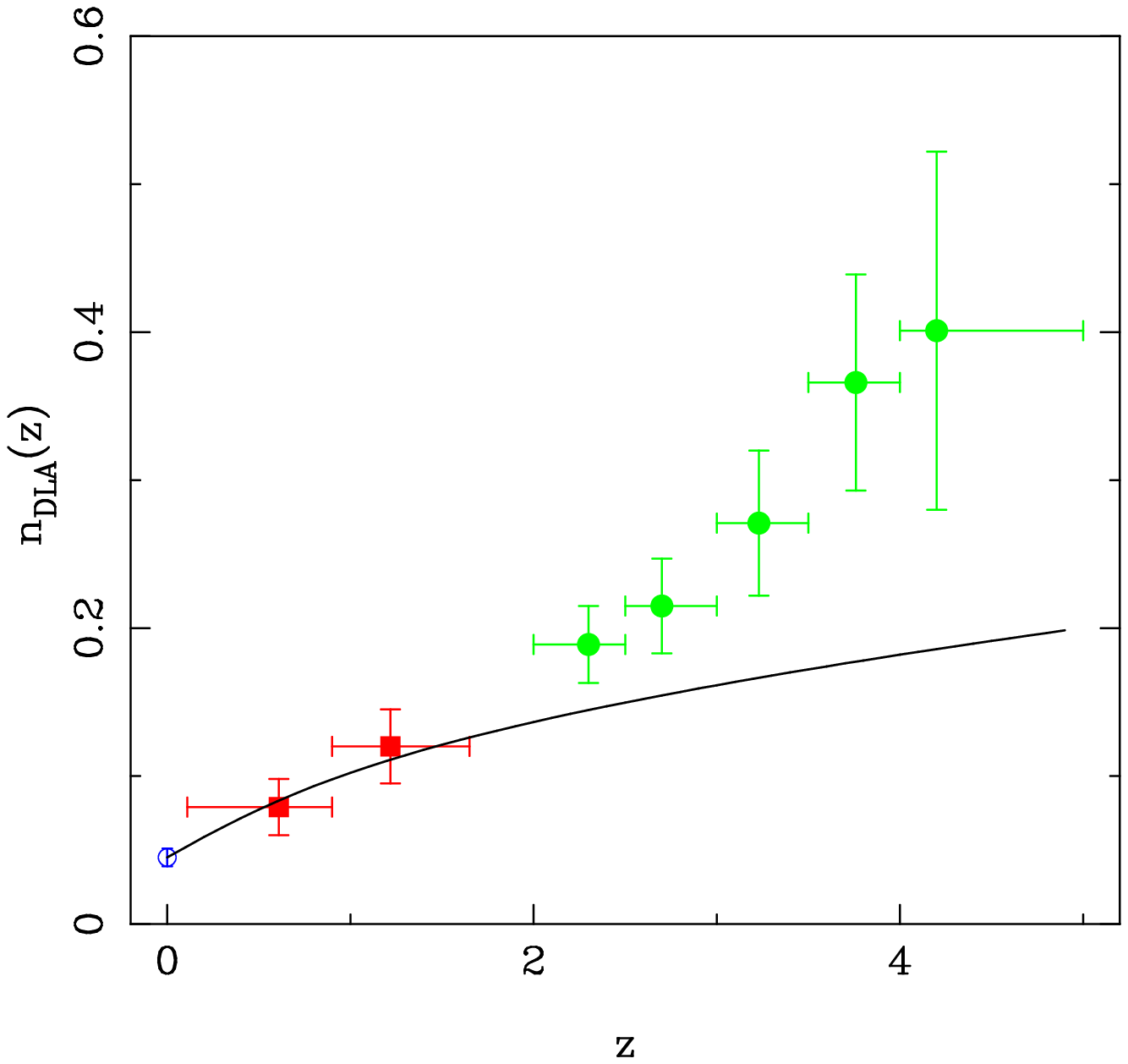}
\caption{Plot of $n_{DLA}(z)$ versus redshift. The new low-redshift
data points are shown as filled squares. The high-redshift points
(filled circles) are from Prochaska \& Herbert-Fort (2004) and the
$z=0$ data point is from an analysis of local HI using the WSRT (Zwaan
et al. 2005a). The solid line is a no-evolution curve in a
standard ``737'' $\Lambda$CDM cosmology with
($h$,$\Omega_M$,$\Omega_\Lambda$) = (0.7, 0.3, 0.7)  normalized at the
$z=0$ data point. This curve implies that the comoving  cross section
for absorption declined rapidly by a factor of $\approx 2$ until
$z\approx 2$ and has remained constant since then. This is consistent
with the idea that today's structures have been in place since
$z\approx 1$ and are a consequence of merger events that occurred
prior to this epoch.}
\end{figure}

\begin{figure}
\plotone{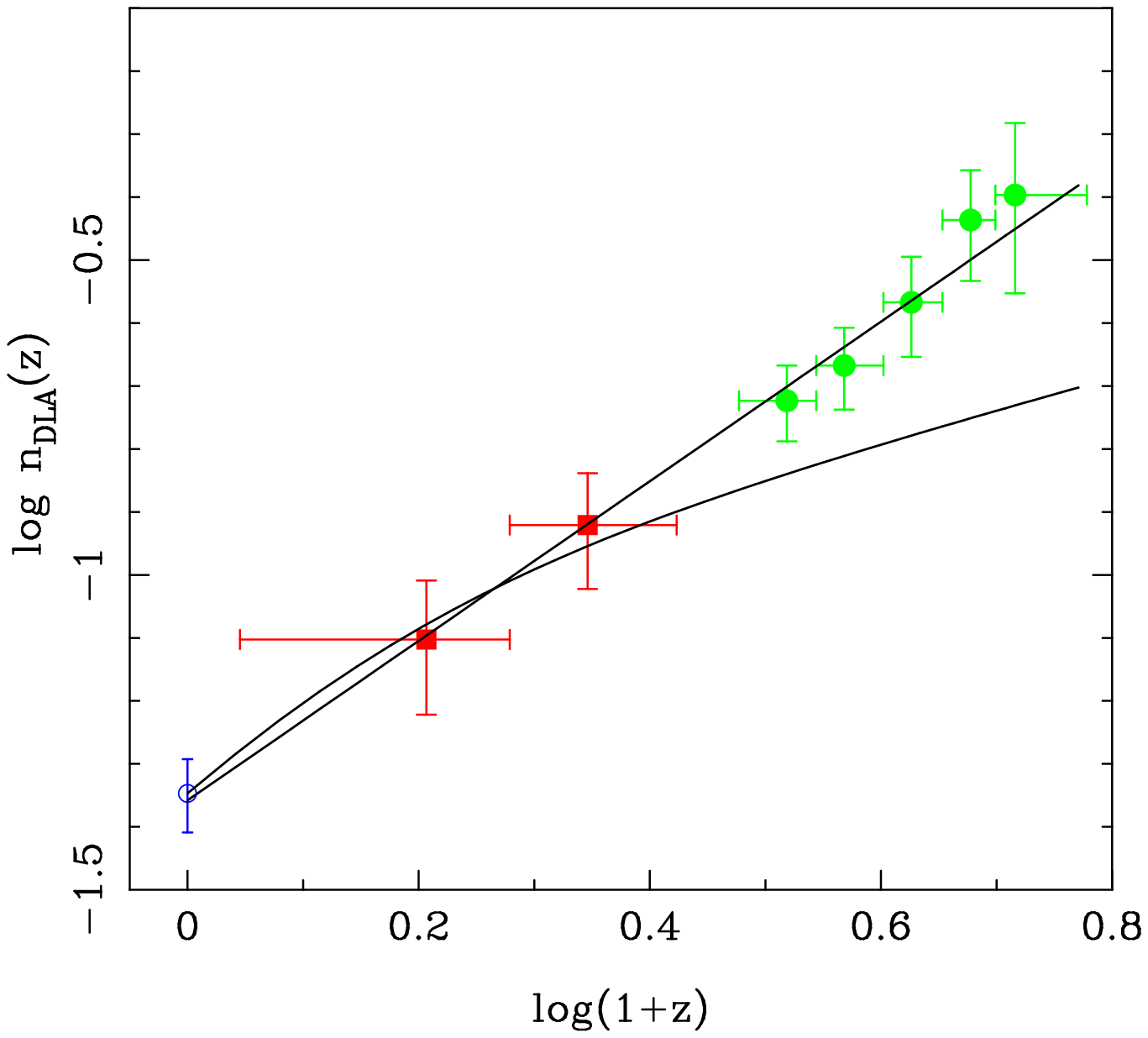}
\caption{Plot of $\log n_{DLA}(z)$ as a function of $\log (1+z)$. The
straight line is the power law fit to the data points with slope
$\gamma=1.27\pm 0.11$, and the  curve is the no-evolution function
shown in Figure 14.}
\end{figure} 
It has been customary in quasar absorption line studies to plot the
logarithm of the redshift number density in order to illustrate its
power law dependence with redshift, i.e.,  $n(z)=n_0(1+z)^\gamma$. In
$\Lambda=0$ cosmologies, the exponent is a  measure of evolution.  For
example,  $\gamma=1$ for $q_0=0$ or $\gamma =0.5$ for $q_0=0.5$
implies no intrinsic evolution of the absorbers. Any significant
departure from these values for $\gamma$ was considered as evidence
for evolution in the product of the comoving number density and cross
section of absorbers. We plot $\log n_{DLA}(z)$  as a function of
$\log (1+z)$ in Figure 15. The straight line is the power law fit to
the data points with slope $\gamma=1.27\pm 0.11$, and the  curve is
the same no-evolution function  shown in Figure 14. Thus, in the past,
the observations would have been interpreted as being consistent with
the DLA absorbers undergoing no intrinsic evolution in a $q_0=0$
universe, and marginally consistent  with evolution in a $q_0=0.5$
universe. With the now widely accepted concordance cosmology, the
interpretation has changed quite dramatically; as noted above, the
nature of the evolution is redshift dependent.

Further implications of this evolution are discussed in \S5 along with
inferences drawn from the evolution in $\Omega_{DLA}$ (\S4.3) and the
HI column density distribution (\S4.4).

\subsection{Cosmological mass density $\Omega_{DLA}$}

We can determine $\Omega_{DLA}$ from the DLA column densities listed
in Table 1 and $n_{DLA}(z)$ via the expression
\begin{equation}
\Omega_{DLA}(z)= \frac{\mu m_H H_0}{c \rho_c} n_{DLA}(z)
\left<N(HI)\right>  \frac{E(z)}{(1+z)^2},
\end{equation}
where
\begin{equation}
E(z)=\frac{H(z)}{H_0}=[\Omega_M(1+z)^3 + (1-\Omega_M-\Omega_\Lambda)(1+z)^2 +
\Omega_\Lambda]^{1/2}.
\end{equation}
Again, the ``737'' cosmology has been used in the calculation of
$\Omega_{DLA}$.  Also, $\mu=1.3$ corrects for a neutral gas
composition of 75\% H and 25\% He by mass, $m_H$ is the mass of the
hydrogen atom,  $\rho_c$ is the critical mass density  of the
universe, and $\left<N(HI)\right>$ is the mean HI column density of
DLAs in each bin.

In contrast to the redshift number density evolution shown in Figure
14, we find  that $\Omega_{DLA}$ has remained constant from $z=5$ to
$z=0.5$ to within the uncertainties. Figure 16 shows the new results
as solid squares.  Specifically, for the redshift range $0.11<z\le
0.90$, we find $\left<N(HI)\right> = (1.27 \pm 0.36) \times 10^{21}$
cm$^{-2}$ and $\Omega_{DLA}(z=0.609) = (9.7 \pm 3.6) \times 10^{-4}$,
and for the range $0.90< z<1.65$ we get $\left<N(HI)\right> =  (1.07
\pm 0.23) \times 10^{21}$  cm$^{-2}$ and $\Omega_{DLA}(z=1.219)=(9.4
\pm 2.8) \times 10^{-4}$.  The uncertainties have been reduced
considerably in comparison to our results in RT00.  The reasons for
this are two fold. First, the uncertainty in $n_{MgII}$ has been
significantly reduced due to the fact that the MgII sample size was
increased 10-fold. Second, the number of DLAs in each bin has
increased by more than a factor of 3. Thus, the uncertainties in the
low- and high-redshift data points are now comparable. Note  that the
statistics of the high-redshift data are also improved due to the
inclusion  of an SDSS DLA sample (Prochaska \& Herbert-Fort
2004). Nevertheless, our basic conclusion from RT00 has remained
unchanged, namely, that the cosmological mass density of neutral gas
remains roughly constant from $z\approx 5$ to $z \approx 0.5$.
\begin{figure}
\plotone{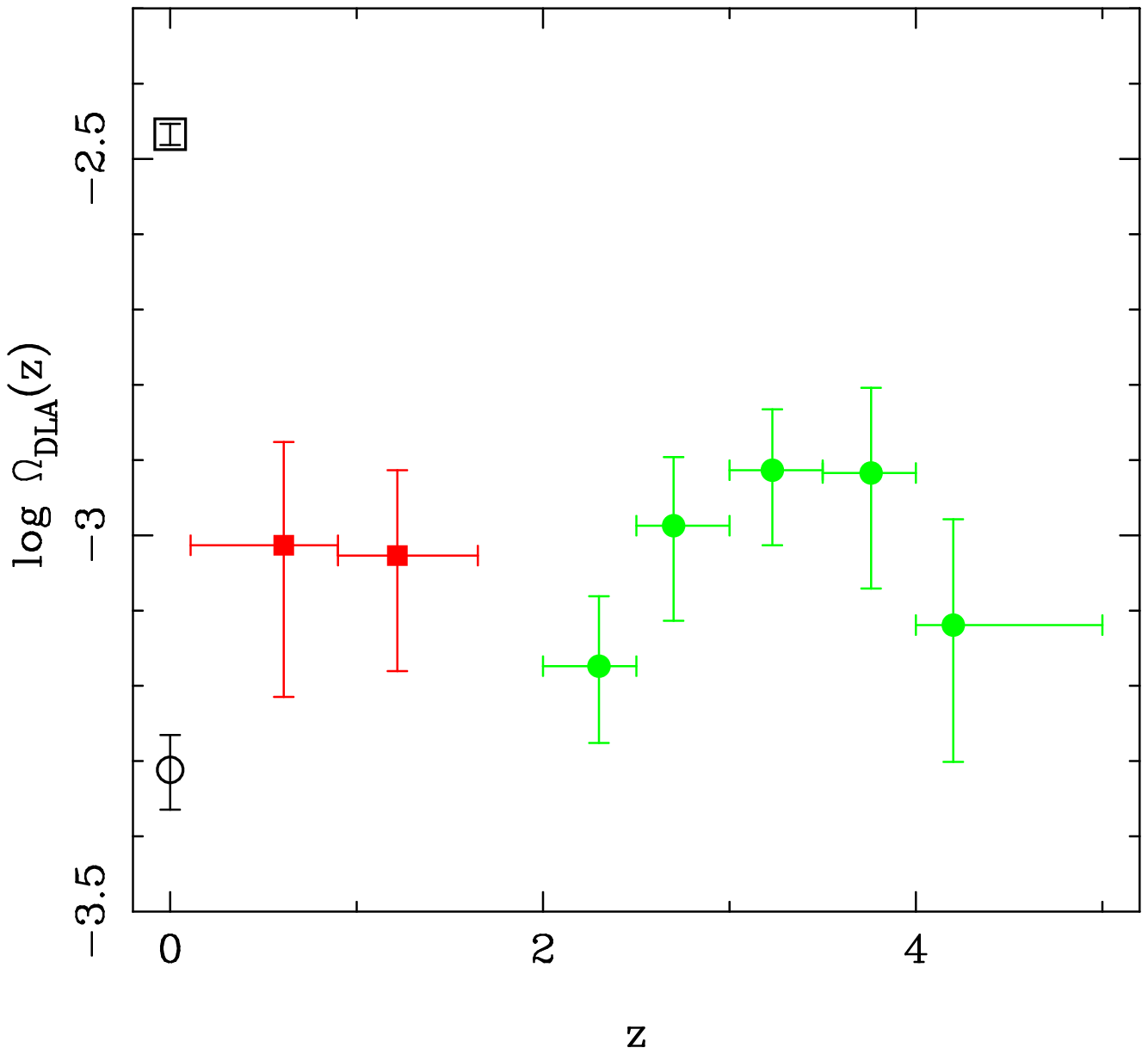}
\caption{Cosmological mass density of neutral gas, $\Omega_{DLA}$, as
a  function of redshift. The filled squares are the new low-redshift
data points. The high-redshift points (filled circles) are from
Prochaska   \& Herbert-Fort (2004) and the open circle at $z=0$ is
from Zwaan et al.  (2005b). The open square at $z=0$ is the mass
density in stars estimated by Panter, Heavens, \& Jimenez (2004) from
the SDSS. While the statistics have improved considerably, our basic
conclusion from RT00 has remained unchanged, namely, that the
cosmological mass density of neutral gas has remained constant from
$z\approx 5$ to $z \approx 0.5$.}
\end{figure}
The drop in redshift number density from $z=5$ to $z=2$ along with a
constant mass density in this range indicates that while the product
of galaxy cross section and comoving number density  is declining,
the mean column density per absorber is increasing. This is, again,
consistent with the assembly of higher density clouds as galaxy
formation  proceeds.

On the other hand, a constant cross section from $z\approx 1$ to $z=0$
along with  a drop in mass density from $z\approx 0.5$ to $z=0$ is
indicative of star formation  that depletes the highest column density
gas while keeping the absorption  cross section constant. This would
in turn require that the column density distribution of DLAs change
such that the ratio of high to low column densities decreases from
low-redshift to $z=0$. As we will see in the next section, the column
density distribution does show some evidence for this.

\subsection{Column density distribution $f(N)$}

Figure 17 shows the normalized cumulative column density distribution
(CDD) for the three redshift regimes. The dashed curve is the $z=0$
CDD from an analysis of an HI diameter-limited sample of local
galaxies from Rao \& Briggs (1993) while  the dotted curve is the
$z=0$ distribution derived by Ryan-Weber et al. (2003, henceforth
R-W03; 2005) from HIPASS data. The thick, solid curve is derived from
the DLAs in Table 1 and the thin, solid curve  is from the ``total''
sample of Prochaska \& Herbert-Fort (2004). The  change in the three
CDDs with redshift  is exactly what is expected based on the $n_{DLA}$
and $\Omega_{DLA}$ results.  Namely, that the low-redshift CDD shows a
higher incidence of high column density systems than at high redshift
presumably due to the assembly of gas as galaxy formation proceeds,
followed by a decrease in  the fraction of high column density systems
to $z=0$, presumably due to the  depletion of gas during star
formation. Thus, at least qualitatively, the evolutionary behavior of
$n_{DLA}$, $\Omega_{DLA}$, and the CDD are entirely consistent with
one another. A KS test shows that there is a 25\% probability that the
high- and low-redshift curves are drawn from the same population; this
is significantly higher than what we observed in RT00, where the  two
samples had only a 2.8\% probability of being drawn from the same
population. However, the general trend that the low-redshift sample
has a  higher fraction of high column density system still remains.
\begin{figure}
\plotone{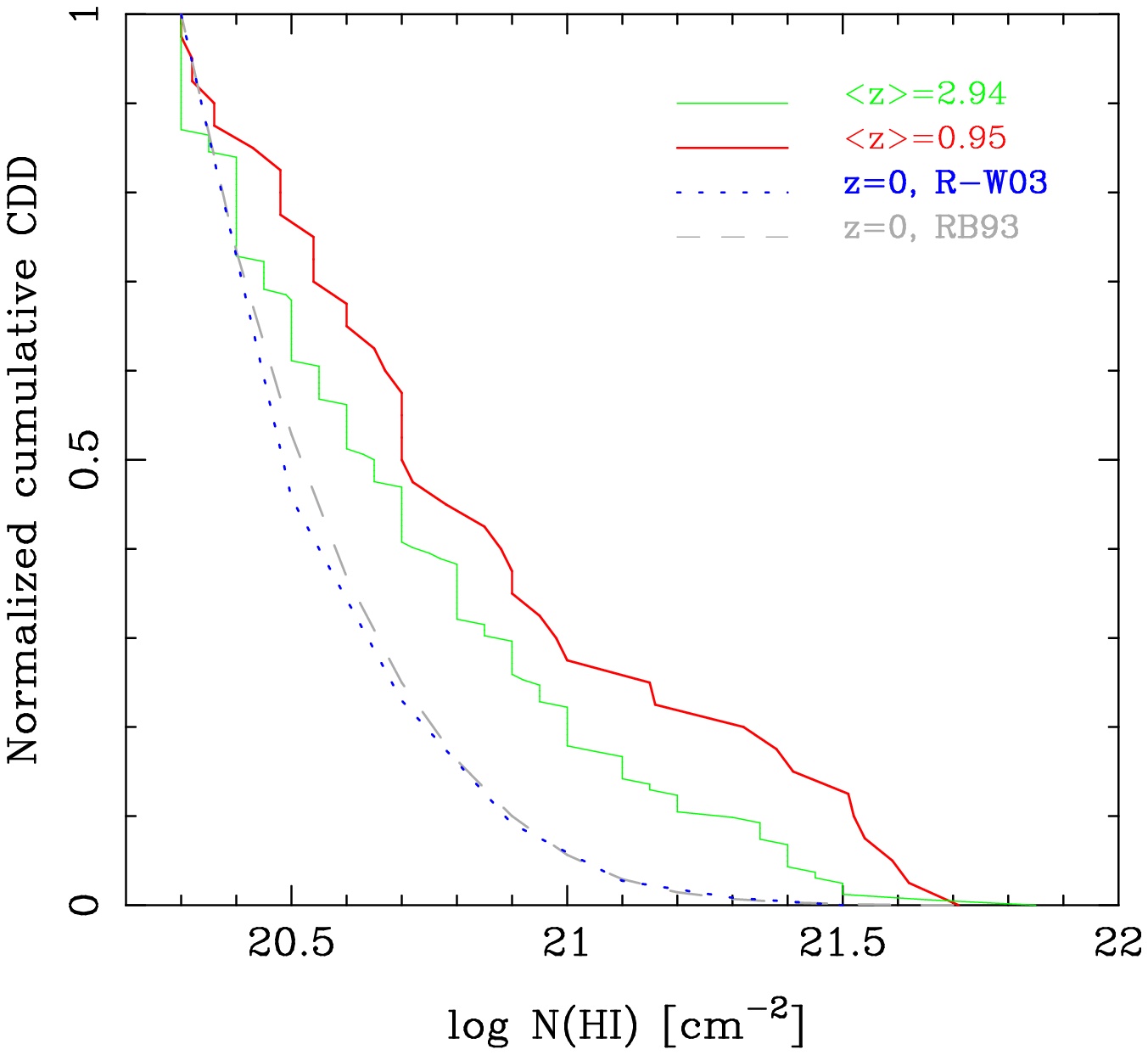}
\caption{The normalized cumulative column density distribution of DLAs
for three redshift regimes. The top curve (red, thick solid line)
includes the 41 low-redshift DLAs from Table 1 at a median redshift of
0.95.  The middle curve (blue, thin solid line) includes 163
high-redshift systems with mean redshift 2.94 (Prochaska \&
Herbert-Fort 2004),  and the bottom two curves are estimates at
$z=0$. The dashed curve is from Rao \& Briggs (1993) and the dotted
curve is from R-W03.}
\end{figure}

\begin{figure}
\plotone{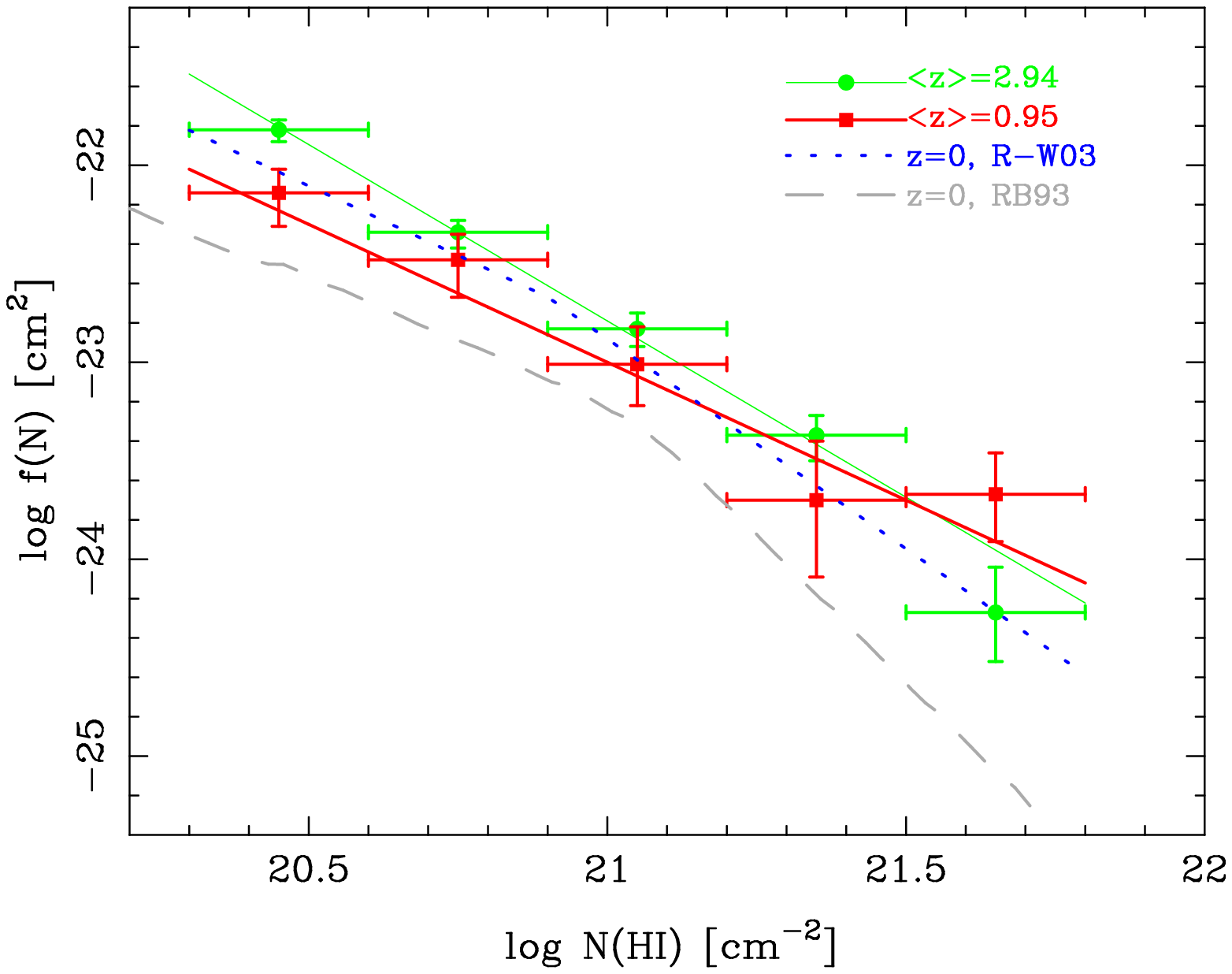}
\caption{Absolute column density distribution (CDD) function for the
three redshift regimes. The red, thick solid line is a least squares
fit to the low redshift data points and has slope $\beta=1.4$; the
blue, thin solid line is a least squares fit to the high redshift data
points with a slope of $\beta=1.8$.  The dotted line is the column
density distribution derived by R-W03 from 21 cm HIPASS data of local
galaxies. It has  slope $\beta = 1.4$ for $\log N(HI)<20.9$ and
$\beta=2.1$ for $\log N(HI)\ge 20.9$.  The dashed line is from HI
measurements of an optically-selected sample of local galaxies (Rao \&
Briggs 1993).  The offset between the two $z=0$ curves arises from the
different normalizations of the HI-mass and optical luminosity
functions, respectively. See text.}
\end{figure}
The absolute CDD can be determined using the equation
\begin{equation}
f(N,z) = n_{DLA}(z) \frac{E(z)}{(1+z)^2}  \frac{y(N,z)}{\Delta N}
\end{equation}
where $y(N,z)$ is the fraction of DLAs with column densities between
$N$ and $N+\Delta N$ at redshift $z$, and $E(z)$ is as given  in
Equation 5.  Figure 18 is a plot of the log of the absolute CDD
function, $\log f(N)$, as a  function of $\log N(HI)$. The turnover
with redshift is most apparent in  the lowest and highest column
density bins.  We derive $\beta = 1.4\pm 0.2$ and $\beta=1.8 \pm 0.1$
at low and high redshift respectively, where the CDD is expressed as
$f(N)=BN^{-\beta}$. At $z=0$, R-W03 derive $\beta=1.4\pm0.2$ for $\log
N(HI)<20.9$ and $\beta=2.1\pm0.9$ for $\log N(HI) \ge 20.9$. The
general form of the absolute CDD does not vary considerably  with
redshift, which in turn explains the roughly constant value of
$\Omega_{DLA}$.  The differences in the $f(N)$ distributions are
subtle, implying that the gas content in DLAs is not changing
drastically. This is strong evidence that DLAs do not have high SFRs
and are, therefore, a different population of objects than those
responsible for much of the observed luminosity in the high redshift
universe (see also \S 5.1 and Hopkins et al. 2005).  On the other 
hand, a non-evolving DLA population might be
observed if the gas that is used up in star formation is replenished
from the inter-galactic medium at a comparable rate. This possibility
seems rather contrived, and requires more proof than the current
observational evidence can provide. In \S5, we discuss further
evidence that  suggests that DLAs and high SFR galaxies (e.g., Lyman
break galaxies) are mutually exclusive populations.

We'd like to draw the reader's attention to the $z=0$ curves in
Figures 17 and 18. The general shape of the CDD at the present epoch
is reproduced in both the RB93 and R-W03 estimates (Figure 17), with
only a minor difference in the column density where the slope of the
distribution changes (Figure 18). This break in the CDD is a direct
consequence of the exponential distribution of $N(HI)$ in disks, with
the position of the break depending on the maximum column density in a
face-on galaxy.  The RB93 CDD was determined using HI 21 cm maps of an
optical diameter-limited sample of local galaxies normalized by the
luminosity function of late-type galaxies locally (Rao 1994). On the
other hand, the R-W03 estimate was derived from the HIPASS survey of
HI in the local universe and the HI mass function of galaxies locally
(Zwaan et al. 2003).   The offset between the two curves (Figure 18)
is a result of different normalizations in the luminosity and HI mass
functions of the two estimates, respectively. We now know that the
local galaxy luminosity function (as determined prior to 1993) did not
include gas-rich galaxies that  occupy the low-luminosity tail of the
galaxy luminosity function. The overall normalization  might have been
underestimated as well. This offset is also manifested in the values
derived for $n(z)$, since $n(z)$ depends on the volume number density
of  absorbers. In Figure 14, we used the most recent estimate of
$n(z=0) = 0.045\pm0.006$  derived by Zwaan et al. (2005a).
 Ryan-Weber et al. (2005) derive a  similar value; these
are a factor of 3 higher than the RB93 estimate. It is also of
interest to note that the local value of $\Omega_{HI}$ derived by RB93
is only $\sim 30$\% smaller than more recent estimates. This means
that while the bulk of the neutral gas at $z=0$ is in the more
luminous galaxies, a significant fraction of the HI cross section is
contributed by optically low-surface-brightness and dwarf
galaxies. This conclusion is indeed borne out by the recent comparison
of  HI in low- and high-surface brightness galaxies by Minchin et
al. (2004).  In any case, it is now clear that the recent deep, large
scale surveys of HI gas in  the local universe have provided a better
understanding of the distribution of HI at $z=0$, allowing for more
precise determinations of its statistical properties for comparison
with quasar absorption line studies.

\section{Discussion}

\subsection{Selection Effects and Biases}

As with any survey, selection effects and biases need to be well
understood in order to correctly interpret results. Here we raise some
of the important ones that may affect our survey.

{\bf 1.} We have determined DLA statistics under the assumption that
all DLAs exhibit MgII absorption, and therefore, that DLAs form a
subset of MgII absorbers. At the redshifts probed by our UV surveys,
$0.11<z<1.65$, we find that there is little chance of encountering a
DLA unless \Wmi $\ge 0.6$ \AA. Since our MgII sample includes  systems
with \Wmi $\ge 0.3$ \AA, we believe this result to be  fairly
secure. An exception may occur in the rare case where the DLA
sightline passes through a single cloud. Its velocity width, i.e., $b$
parameter  and therefore, \Wmi, would then be small, perhaps even
smaller than in the  DLA towards 3C 286 (see \S 4.1.1). The DLA
towards the {\it D} component of the Cloverleaf gravitationally lensed
quasar might be an example of such a case (Monier et al. 2005). The
$N(HI)=2\times 10^{20}$ cm$^{-2}$, $z=1.49$ DLA is not detected in the
three brighter components of this quadruply lensed quasar. A composite
Keck spectrum of all four components combined does not detect MgII
down to a 3$\sigma$ limit of \Wmi$=0.06$ \AA.  Given the relative
brightnesses of the four components, any MgII absorption towards
component {\it D} could be diluted by a factor of 5 - 10 in the
composite spectrum. Thus, the absence of metal lines in this DLA need
not be an indication of unpolluted gas, but instead,  of low velocity
dispersion gas that might only be detected with high resolution
spectra of component  {\it D}.

Although the highest redshift DLAs have not been shown to have MgII
absorption because the MgII doublet is shifted into the infrared,
metals have been detected in DLAs at redshifts as high as $z=3.9$
(Prochaska et al. 2003a) and are, therefore, expected to also include
MgII. Apart from testing the DLA-MgII connection at high redshift and
exploring any evolution, assembling a near-infrared MgII sample with
follow-up optical spectroscopy to search for DLAs would be important
for comparison with blind optical DLA surveys.  In addition, Nestor et
al. (2003) and Turnshek et al. (2005) have shown that a positive
correlation exists between \Wmi\ and neutral gas metallicity when an
ensemble average of strong (\Wmi $\ge 0.6$ \AA) SDSS MgII absorbers is
considered. Since the REW of saturated MgII lines is an indication of
gas velocity spread, this correlation indicates a
metallicity-kinematics relation for the average MgII absorber. The
evolution of this relationship to higher redshift will also provide
important constraints on CDM simulations of galaxy and structure
formation.

Ultimately, our study is based on the premise that all DLAs have MgII
absorption, and unless significant counter examples are found, this
assumption is now on fairly firm ground.

{\bf 2.} Figure 4 shows that the fraction of DLAs in a MgII sample is
a function of \Wmi, rising from a fraction near 16\% just above  the
threshold value of \Wmi$= 0.6$ \AA\ to about 65\% at the highest
values near \Wmi$= 3$ \AA. At present these should be considered
approximate fractions since the presence of FeII$\lambda2600$ has been
used to increase the probability of finding a DLA. However, for
reasons that are not yet clear, the FeII criterion does not affect the
mean HI column density as a function of \Wmi\ (see Figure 7),  and so
the FeII inclusion effect is probably not significant for our
sample. Nevertheless, the \Wmi\ dependence of the DLA fraction will
introduce a bias in $n_{DLA}$ unless the observed sample's \Wmi\
distribution matches the true distribution. We have accounted for this
by carefully defining our samples as described in \S2, and by making
use of Equations 1, 2, and 3 to calculate $n_{DLA}$.

{\bf 3.} The degree to which $N(HI)$ may be biased by \Wmi\ can be
seen  by examining Figure 5. The mean HI column density of identified
DLAs is $N(HI) \approx 2\times10^{21}$ cm$^{-2}$ when  $0.6$
\AA\ $\le$ \Wmi $< 1.2$ \AA, but it seems to decrease by a factor of
$\approx4$ at \Wmi $\approx3$ \AA.  However, inspection of Figure 2
suggests that this trend is not particularly tight nor is it well
established for DLAs by themselves. On the other hand, if one
considers all the points in Figure 2, it is clear that the $N(HI)$
distribution changes for  different \Wmi\ intervals.  It is
interesting that in MgII-selected surveys for DLAs,  the determination
of the cosmological mass density of neutral  gas, $\Omega_{DLA}$, has
(so far) not revealed any dependency  on \Wmi\ selection. This is
because in our sample, the increased probability (by a factor of
$\approx 4$) of finding a DLA  at the largest \Wmi\ values is
approximately compensated for by the  corresponding decrease in mean
HI column density (by a factor of $\approx 4$) at the largest \Wmi\
values.  It is worth pointing out that although the MgII selection
criteria lead to reasonably complete samples of DLAs, incompleteness
must set in at HI column densities in the  sub-DLA regime because
systems with \Wmi $<0.3$ \AA\ can have sub-DLA HI column
densities. Therefore, only the $N(HI)$ distribution in the DLA regime
can be reliably considered with the available data.

{\bf 4.}   Hopkins et al. (2005; see also Rao 2005) have discussed the
question of whether the observed population of DLAs  can account for
the observed SFH of the universe from low to high redshift.   By
applying the Kennicutt (1998) formulation of the Schmidt law to  the
properties of the currently observed population of DLAs they find that
the DLAs cannot account for the cosmic SFR density inferred from the
luminosity density of high-redshift galaxies (see figure 2 in Hopkins
et al. 2005). An even larger discrepancy occurs when one compares DLA
metallicities to the metallicities expected on the basis of the cosmic
SFR (see figure 4 in Hopkins et al. 2005 and figure 13 in Rao
2005). One way to avoid  this discrepancy is to postulate that the
MgII and blind DLA surveys are not yet large enough to include
absorbers with  very small individual cross sections that nevertheless
may dominate the cosmic SFR and be the main reservoirs for the metals
as well. Indeed, these star forming regions will be rich in molecular
gas, the direct fuel for star formation, but with HI column densities
that may exceed the observed DLA regime.  Kennicutt (1998) points out
that in normal disks star formation generally takes place in regions
that contain $1-100$ M$_{\odot}$ pc$^{-2}$ (i.e., $\approx 10^{20} -
10^{22}$ atoms or molecules cm$^{-2}$), whereas the more rare (and
smaller) star burst regions contain $10^2 - 10^5$ M$_{\odot}$
pc$^{-2}$ (i.e., $\approx 10^{22} - 10^{25}$ atoms or molecules
cm$^{-2}$).   For example, an absorber with a size of about 100 pc,
comparable to giant molecular clouds (GMCs), has a cross section that
is $\approx 10^4$  times smaller than known DLAs, which typically have
effective radii of  $\approx 10$ kpc (Monier et al. 2005). Assuming
that there are on the order of 10 GMCs per galaxy, the total cross
section per unit volume, i.e., interception probability, for these
very high column density gas systems would be on the order of $10^3$
times smaller.  This means that $10^3$ DLAs need to be detected in
order to find one very high column density system.  With the SDSS, we
are getting close, but are not quite there yet. A one in a thousand
system with $N(HI+H_2)=10^{24}$ cm$^{-2}$ would increase the  SFR
density of DLAs by more than a factor of 2, and bring the DLA SFR
density into agreement with the luminous SFR density. Searches for
molecular gas in DLAs have resulted in only a handful of detections.
Moreover, the molecular gas fraction in the few DLAs with H$_2$
detections  is very small (e.g. Ledoux et al. 2003), and  is
consistent with the  idea that the known sample of DLAs does not trace
the majority of the star forming gas in the universe.  It therefore
seems reasonable to conclude that most of the neutral and molecular
gas mass has so far been missed in DLA surveys.

However, the possibility that these very high column density gas
systems are being missed  by DLA surveys may not only be due to their
small gas cross sections, but also because they are likely to be  very
dusty. Ledoux et al. (2003) find that the DLAs in which H$_2$ is
detected  have among the highest metallicities and the highest
depletion factors, hinting at the possibility of much higher
depletions in much higher column density molecular  gas clouds.
Although radio loud quasar surveys for DLAs have not revealed any
significant dust bias in optical surveys (Ellison et al. 2001; 2004;
Akerman et al. 2005), the radio loud quasar surveys for DLAs may
themselves suffer from the small cross section selection effect.  Not
enough radio loud quasars have yet been surveyed to find the putative
one in a thousand very high column density system.  But if significant
mass has been missed due to small total cross section for star forming
regions, whether or not these high-gas-mass regions will be found once
sample sizes are much larger is unclear. Substantial dust-induced
reddening may prevent complete samples from ever being discovered via
optical quasar absorption-line spectroscopy.

In this regard it is interesting that Gardner et al. (1997)  found in
their CDM simulations that depletion of the gas supply by star
formation only affected the DLA statistics at $z>2$ for $N(HI) >
10^{22}$ atoms cm$^{-2}$ (i.e., in a regime where DLAs have not been
found), even though roughly half of the cold collapsed gas was
converted to stars by $z=2$.

{\bf 5.} Gravitational lensing has the opposite effect on DLA surveys.
Magnification by DLA galaxies could brighten background quasars, and
preferentially include them in magnitude-limited samples. Le Brun et
al. (2000), with HST imaging observations, showed no evidence for
multiple images of background quasars and concluded that the quasars
were magnified by at most 0.3 magnitudes. In addition, Ellison et
al. (2004) and P\'eroux et al. (2004) using statistical tests on low
redshift MgII and DLA  samples, showed that lensing bias is a minor
effect. More recently, using the SDSS MgII survey results of Nestor
(2004), M\'enard et al. (2005) show that quasars behind strong MgII
absorbers, of which DLAs are a subset, show little magnification bias,
and that its effect on $\Omega_{DLA}$ at low redshift is negligible
(see also M\'enard 2005).  It is also unlikely that the  lowest
redshift points that we derived from our HST-UV data (Figure 16) are
affected by lensing bias. This is because the DLAs with the highest HI
column densities at $z\approx 0.5$ arise in dwarf galaxies (Rao et
al. 2003), and consequently, do  not have the mass required to produce
significant magnification.

\subsection{Interpretation of the Statistical Results on DLAs}

As discussed in \S4, the evolutionary behavior of $n_{DLA}$,
$\Omega_{DLA}$, and the CDD are, at least qualitatively, consistent
with one another in terms of a simple galaxy formation scenario. We
find evidence for a rapid decline of $n_{DLA}$ (by a factor of 2) from
$z=5$ to $z\approx 2$ followed by no evolution down to $z=0$.  For
comparison, the evolution of Ly$\alpha$ forest lines with  $\log N(HI)
\gtrsim 14$, which have been shown to be associated with the same
large scale structure that traces galaxies (Tripp, Lu,  \& Savage
1998; Dav\'e et al. 1999; Penton, Stocke, \& Shull 2002), also slows
down dramatically near $z\approx 1.5$ (e.g.,  Kim et al. 2002, Weymann
et al. 1998).  Thus, both the DLAs and the higher $N(HI)$ Ly$\alpha$
forest appear to follow similar evolutionary histories consistent with
the collapse and assembly of baryonic structures near $z\sim 1.5$ or
2. The near constant value of  $\Omega_{DLA}$ during the phase of rapid
evolution in $n_{DLA}$ implies an increase in the HI column densities
of individual clouds. The observed evolution in the CDD of
the DLAs, although mild, is evidence  for this.  The subsequent drop
in $\Omega_{DLA}$ down to $z=0$ along with  an unevolving $n_{DLA}$ is
indicative of star formation that depletes gas  while keeping the
absorption cross section constant. The change in slope of the CDD from
low redshift to $z=0$, i.e., the decrease in the  ratio of high to low
column densities,   is again consistent with this scenario. Further
details on the evolution of HI from low redshift to the present epoch
can be studied only when the sample of low-redshift DLAs becomes large
enough to split the $0<z<1.65$ redshift interval, without compromising
on the  uncertainties, into finer than the current two bins, and
now we consider this possibility by adopting some reasonable 
assumptions.
\begin{figure}
\plotone{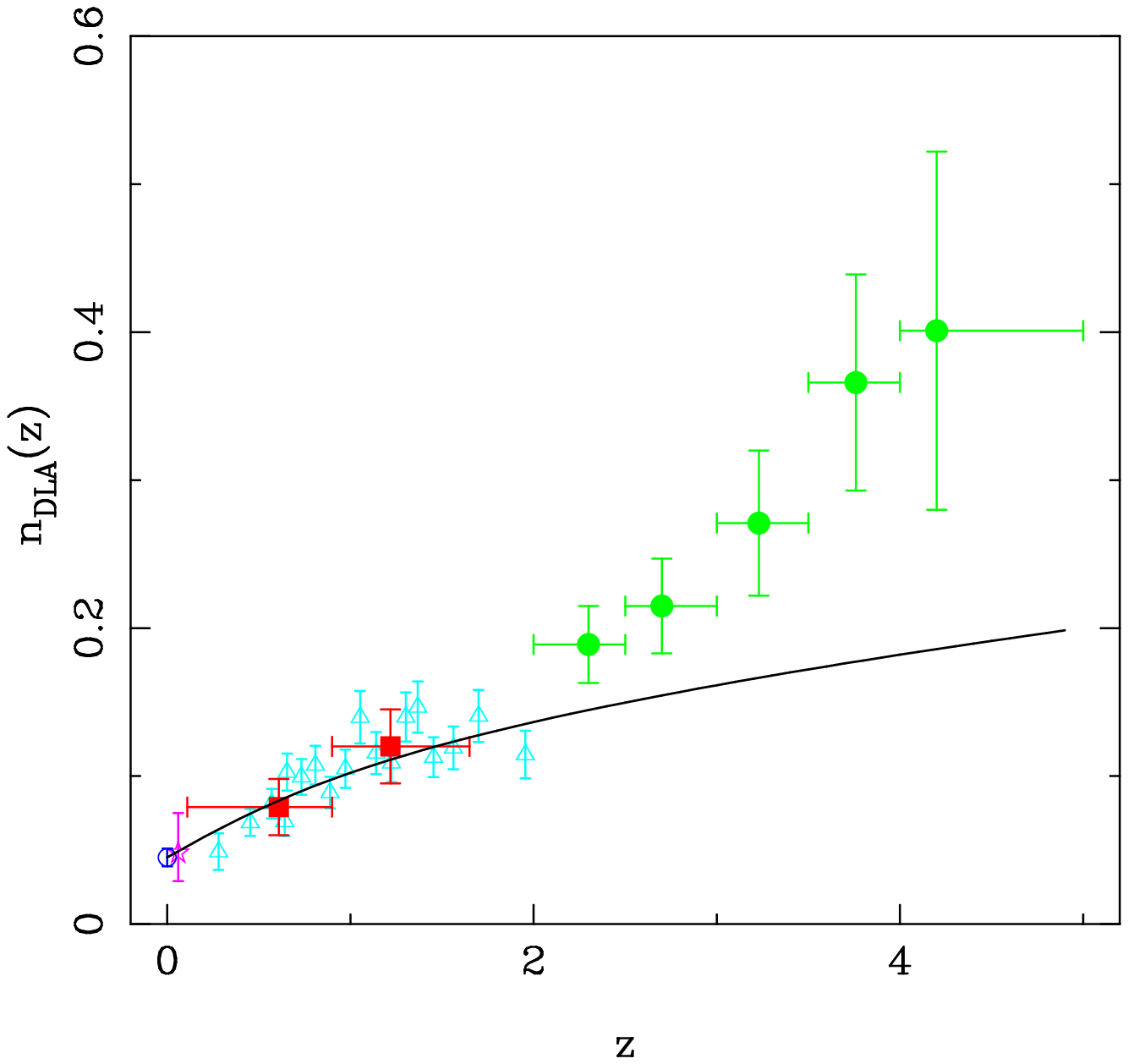}
\caption{Redshift number density of DLAs as a function of
redshift. Symbols are the same as in Figure 14 with the addition of
open triangles, which are derived from the \Wmi $\ge 0.6$ \AA\ MgII
redshift number density and assuming that the  fraction of DLAs in
these MgII systems is constant at 22\%. The errors are therefore
indicative of statistical errors in the MgII sample alone. The open
star at $z=0.06$ is similarly derived from the HST MgII sample of
Churchill (2001).  Including errors in the DLA fraction will
systematically move the data up or down by $\sim 25$\%.}
\end{figure}
\begin{figure}
\plotone{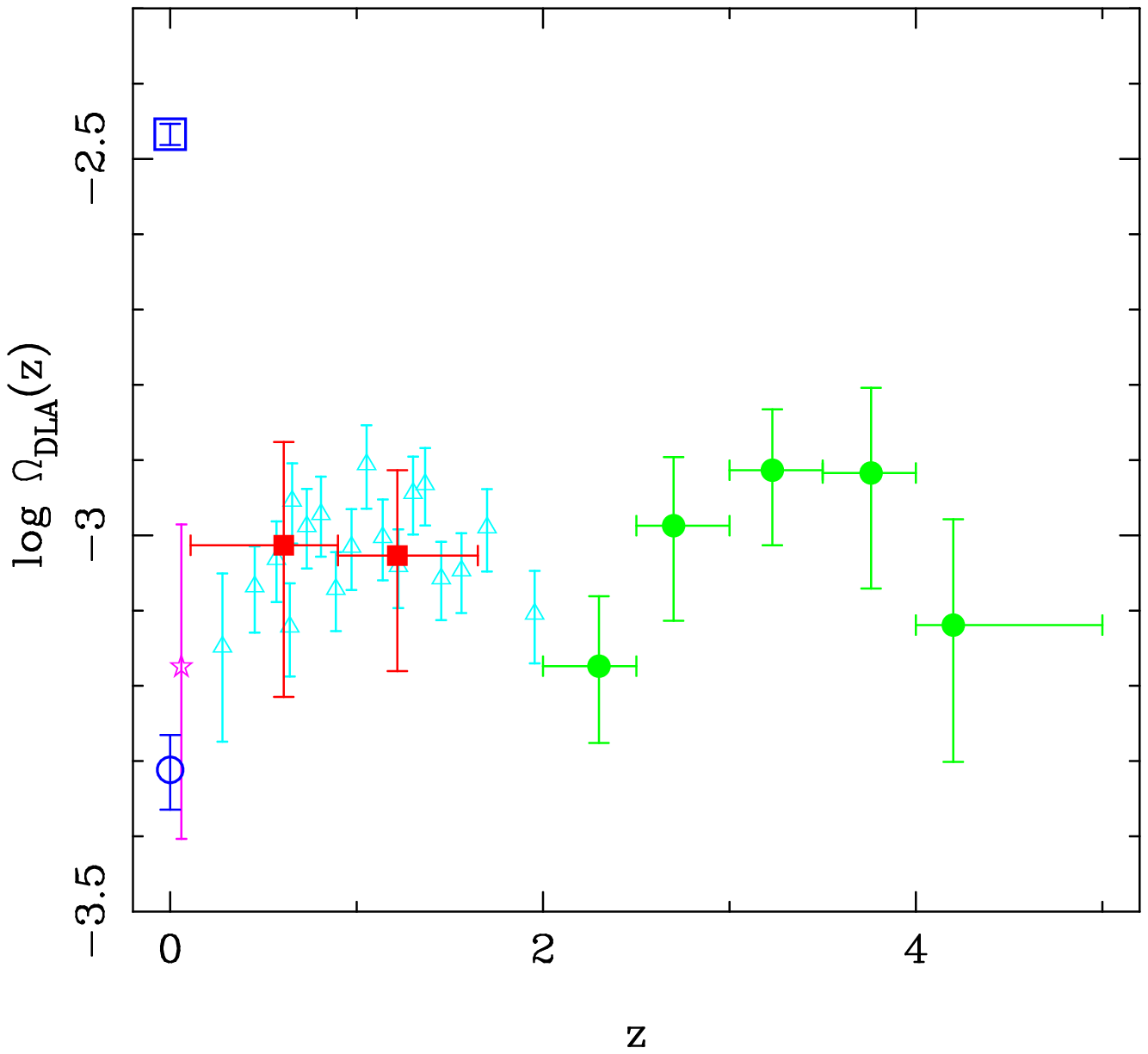}
\caption{Cosmological mass density of neutral gas in DLAs as a
function of redshift.  Symbols are the same as in Figure 16 with the
addition of open triangles, which are derived from the \Wmi $\ge 0.6$
\AA\ MgII redshift number density, and assuming that the  fraction of
DLAs in these MgII systems is constant at 22\% and that their HI
column density is constant at $ 1.16 \times 10^{21}$ cm$^{-2}$. The
errors are therefore indicative of statistical errors in the MgII
sample alone. The open star at $z=0.06$ is similarly derived from the
HST MgII sample of Churchill (2001).  Including errors in the DLA
fraction and $N(HI)$ will systematically move the data up or down by
$\sim 0.1$ dex. We see, for the first time, possible evidence of a
decline in  $\Omega_{DLA}$ from $z\approx 0.5$ to $z=0$. }
\end{figure}
We have shown that a survey of MgII systems with \Wmi $\ge 0.6$ \AA\
is a reliable tracer of DLAs, and can be used to determine DLA
statistics.  The two $n_{DLA}$ data points at low redshift shown in
Figure 14 are  20 ($\pm$ 5)\% and 24 ($\pm$ 6)\%  of the corresponding
MgII redshift number density values for \Wmi $\ge 0.6$ \AA\  derived
by Nestor (2004) and NTR05, at $z=0.6$ and $z=1.2$ respectively. By
assuming that the DLA fraction in a  \Wmi $\ge 0.6$ \AA\ sample is
constant over the entire redshift interval $0.1<z<1.65$,  $n_{DLA}$
can be estimated in much smaller redshift bins from the statistics of
any MgII survey sample without a UV survey for DLAs.  The systematic
uncertainty in  $n_{DLA}$ will then be primarily  limited only by the
precision  to which the DLA fraction is known. Similarly, the
systematic uncertainty in  $\Omega_{DLA}$ will primarily be  limited
by how accurately the DLA HI column density  is known. For \Wmi $\ge
0.6$ \AA, we found the mean DLA column density  to be
$\left<N(HI)\right>=(1.16 \pm 0.20)\times 10^{21}$ cm$^{-2}$. Recall
that all the DLAs in our sample have  \Wmi $\ge 0.6$ \AA. Assuming a
constant DLA fraction of 22\% and a constant DLA HI column density of
$1.16\times 10^{21}$ cm$^{-2}$, we can estimate $n_{DLA}$ and
$\Omega_{DLA}$ from the MgII \Wmi $\ge 0.6$ \AA\ redshift number
density as follows  (see Equations 1 and 4):
\begin{equation}
n_{DLA}(z)=0.22\times n_{MgII}(z)
\end{equation}
and
\begin{equation}
\Omega_{DLA}(z)= \frac{\mu m_H H_0}{c \rho_c} 0.22\times n_{MgII}(z)
\times 1.16 \times 10^{21} \frac{E(z)}{(1+z)^2}.
\end{equation}
These data points are shown in Figures 19 and 20 as open
triangles. Only the error  in $n_{MgII}$ is propagated through to show
the statistical uncertainty in these data. The errors associated with
the DLA fraction and DLA column density  are systematic and will
affect all of these data points equally, moving  them uniformly up or
down by $\sim 25$\% in the case of $n_{DLA}$ and $\sim 0.1$ dex for
$\log \Omega_{DLA}$.  We also show $n_{DLA}$ and $\Omega_{DLA}$
inferred using the redshift number density derived by Churchill (2001)
for MgII systems detected in HST spectra, $n_{MgII}(z=0.06) =
0.22^{+0.12}_{-0.09}$; again, only the error in $n_{MgII}$ has been
propagated.  We now see for the first time that, under the assumption
of constant DLA fraction and HI column density suggested by our
current MgII-DLA sample, there may be evidence of a decreasing trend
in $\Omega_{DLA}$ from $z=0.5$ to $z=0$.   There also appears to be a
dip in $\Omega_{DLA}$ near $z=2$, albeit within $1\sigma$, that shows
up in both the high- and low-redshift data. It should be noted that
the highest of the low redshift data points comes from the red end of
SDSS spectra, and suffers from low signal-to-noise ratio due to the
presence  of atmospheric absorption. Similarly, the lowest of the high
redshift data points  comes from the blue end of SDSS spectra, and
also suffers from low signal-to-noise ratio. Whether this $1\sigma$
effect will persist with better quality data remains to be seen; but
if real, will be a challenge for galaxy formation models to explain.
\begin{figure}
\plotone{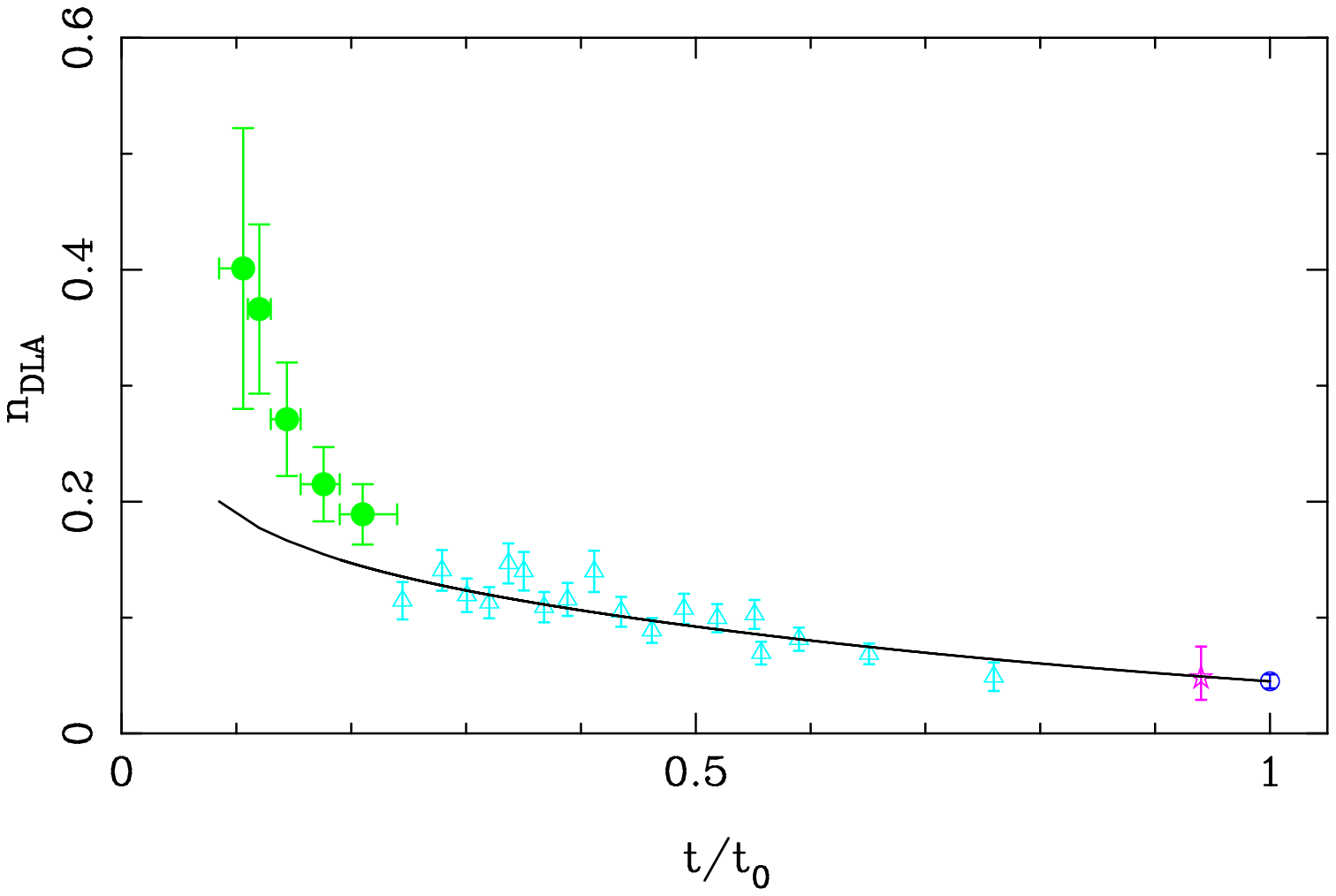}
\caption{Redshift number density of DLAs as a function
of cosmic time with $t_0$ being the current epoch. Symbols are the same
as in Figure 19. }
\end{figure}
\begin{figure}
\plotone{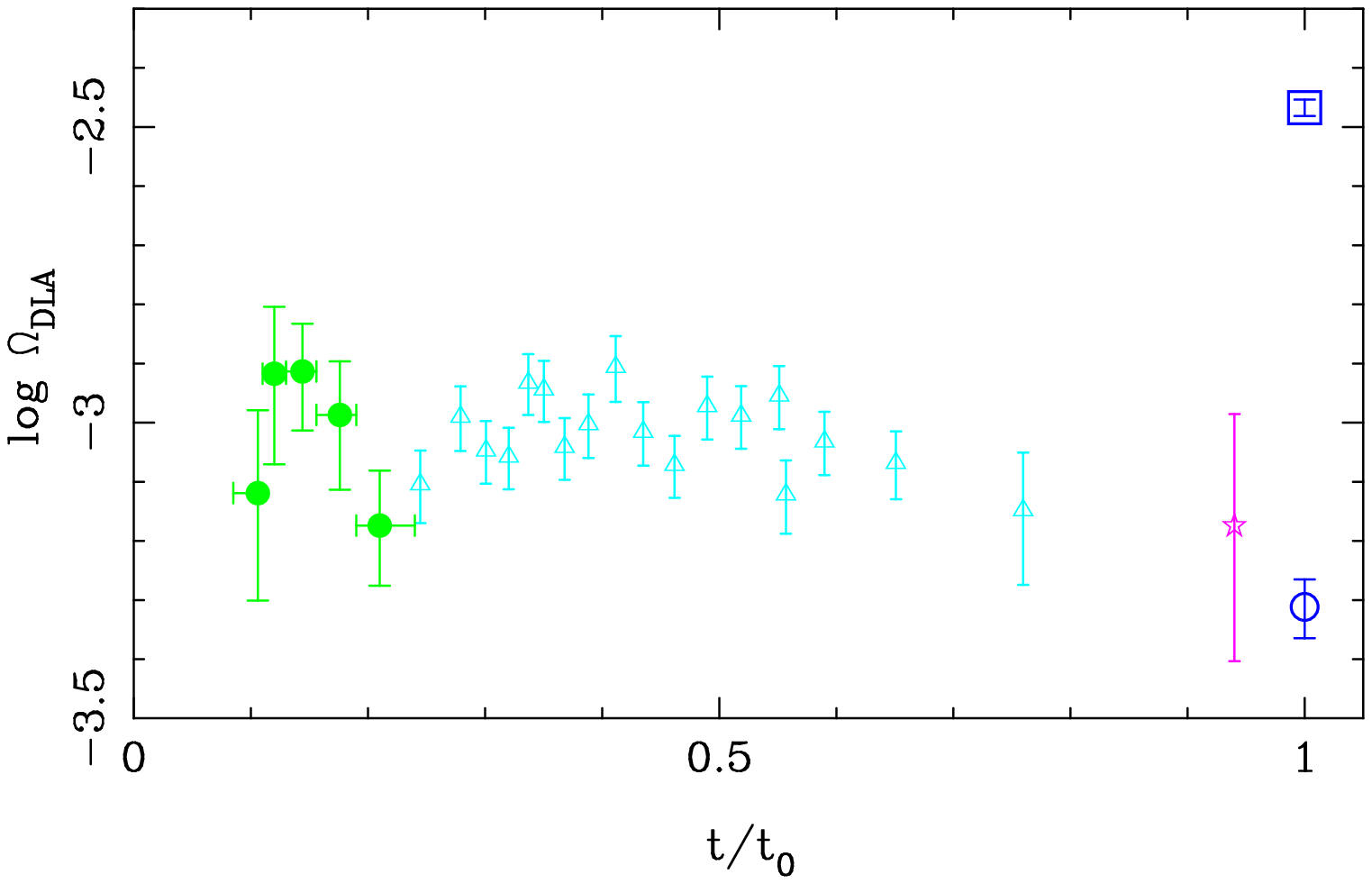}
\caption{Cosmological mass density of neutral gas in DLAs as a function
of cosmic time with $t_0$ being the current epoch. Symbols are the same
as in Figure 20.}
\end{figure}
These two figures illustrate our method for determining DLA
statistics. Using our sample of 197 MgII systems with follow-up UV
spectra, we have demonstrated that, to first order, the DLA fraction
in a  \Wmi\ $\ge 0.6$ \AA\ MgII sample and the DLA HI column density
are constant. Using this assumption, we have shown that details  in
the evolution of MgII systems can reveal details in the neutral gas
evolution. The data point at $z=0.28$ was derived from the MMT survey
for low redshift MgII systems (Nestor 2004; Nestor, Turnshek,  \& Rao,
in preparation), and is in a redshift regime inaccessible by the
SDSS. An even larger survey for MgII systems at $0.11<z<0.36$  is
clearly needed in order to understand the evolution of $\Omega_{DLA}$
in a redshift regime where most of the evolution appears to be  taking
place. Similarly, a high-redshift, near infrared survey for MgII
systems could extend this method into the optical regime, and any
evolution in the MgII-DLA relationship could be studied. In order to
underscore the importance of pursuing this work in the future, we show
$n_{DLA}$ and $\Omega_{DLA}$ as functions of cosmic time in Figures 21
and 22, respectively.

\section{Summary}

We have presented statistical results on UV surveys for low-redshift
($z<1.65$) DLAs with $N(HI) \ge 2\times10^{20}$ cm$^{-2}$ using
the largest sample of UV-detected DLAs ever assembled. The DLAs were
found by targeting QSOs with MgII systems identified optically
in the redshift range $0.11<z<1.65$. In total, UV observations
of the Ly$\alpha$ absorption line in 197 MgII systems with \Wmi\
$\ge 0.3$ \AA\ have been obtained. This is an efficient and
effective way to find DLAs because, in the absence of MgII, the
system evidently has no chance of being a DLA. This sample contains 
41 DLAs, all of which have \Wmi\ $\ge 0.6$ \AA. Our main findings can be
summarized as follows:

1. To a high level of completeness, DLAs can be studied through
follow-up observations of strong MgII absorbers.  In particular,
Figure 4 shows that for our sample the probability of a MgII system
being a DLA is $P \approx 0$ for \Wmi\ $< 0.6$ \AA\ and $P \approx
0.16 + 0.18($\Wmi$ - 0.6)$  for $0.6 \le$ \Wmi\ $< 3.3$ \AA. A MgII
absorber must generally have $1 <$ \Wmi/\Wf $< 2$ and \Wmiii $>0.1$\AA\
to be a DLA (see \S4.1).

2. UV spectroscopy, almost exclusively with HST, enabled us to
measure or place limits on $N(HI)$ for each of the 197 systems
studied. For MgII systems with $0.3$ \AA\ $\le$ \Wmi $< 0.6$ \AA,
$\left< N(HI) \right> = (9.7\pm2.2) \times 10^{18}$ cm$^{-2}$,
while for systems with \Wmi $\ge 0.6$ \AA, $\left< N(HI) \right> =
(3.5\pm0.7) \times 10^{20}$ cm$^{-2}$. This is basically
a step function (see Figure 4), with a factor of $\approx 36$ change 
in mean HI column density near \Wmi $\approx 0.6$ \AA. Since the MgII
absorption lines are saturated at \Wmi $> 0.6$ \AA, 
there is evidently a threshold in kinematic velocity
spread below which it is highly unlikely to encounter high column
density neutral DLA gas.

3. Above \Wmi $=0.6$ \AA, the mean HI column density of a sample of
MgII absorbers is found to be constant with increasing \Wmi. However,
owing to the increase in probability of finding a DLA with increasing
\Wmi, the mean HI column density of MgII absorbers that are DLAs is
found to decrease by about a factor of four with increasing \Wmi,
from \Wmi $\approx 0.6$ \AA\ to \Wmi $\approx 3.5$ \AA. 
Improved statistics are needed to study this effect owing to the
large scatter in the \Wmi versus $N(HI)$ plane for \Wmi $> 0.6$ \AA.

4. By combining results at all redshifts, including 
21 cm emission surveys at $z=0$, we find that the DLA incidence per unit
redshift can be parameterized as $n_{DLA}(z) = n_0 (1+z)^{\gamma}$ 
where $n_0=0.044\pm0.005$ and $\gamma=1.27\pm0.11$. In the standard
``737'' cosmology this indicates no evolution in the product of
neutral gas cross section times comoving number density at redshifts
$z\lesssim2$, but from $z\approx5$ to $z\approx2$ there is a decrease
of a factor of $\approx2$ in this quantity relative to the no evolution
prediction (Figure 19). This decline happens relatively rapidly,
in a time span that corresponds to $\lesssim1.5$ Gyrs. 

5. The cosmological mass density of neutral gas due to DLAs, $\Omega_{DLA}$,
follows a completely different evolutionary pattern. It remains
relatively constant in the redshift interval $0.5<z<5.0$, with
$\Omega_{DLA} \approx 10^{-3}$, but then it declines by a factor
of $\approx2$ between $z\approx0.5$ and $z=0$ (Figure 20).
This drop in neutral gas takes place during the last $\approx5$ Gyrs
of the history of the Universe. However, due to possible selection effects 
which are biased against finding regions with very high column densities
because the product of their gas cross section and comoving number
density is small, it is important to realize that the neutral gas
component as traced by the DLAs may not include all of the neutral and
molecular gas involved in star formation (Hopkins et al. 2005; Rao 2005;
Turnshek et al. 2005). 

6. Consistent with the $n_{DLA}(z)$ and $\Omega_{DLA}(z)$ results,
the HI CDD at $\left<z\right> \approx 1$ shows a higher 
incidence of high column density systems than at
$\left<z\right> \approx 3$. This presumably represents a build up of
neutral mass concentrations. By $z=0$
the higher incidence of high $N(HI)$ systems seen at 
$\left<z\right> \approx 1$ has disappeared, presumably due to
the depletion of gas during star formation. 

7. In the absence of future QSO absorption-line surveys that aim to 
identify DLAs and measure their $N(HI)$ in UV spectra, more detailed 
studies that lead to a better understanding of the strong
MgII systems may hold promise for reaching a better 
determination of the properties of the neutral gas phase of the 
universe at $z<1.65$ (e.g., Figures $19-22$).

\acknowledgments
 This work was funded by grants from NASA-STScI, NASA-LTSA, and
NSF. HST-UV spectroscopy made the $N(HI)$ determinations possible.  We
thank members of the SDSS collaboration who made the SDSS project a
success. Funding for creation and distribution of the SDSS Archive has
been provided by the Alfred P. Sloan Foundation, Participating
Institutions, NASA, NSF, DOE, the Japanese Monbukagakusho, and the Max
Planck Society. The SDSS Web site is www.sdss.org. The SDSS is managed
by the Astrophysical Research Consortium for the Participating
Institutions: University of Chicago, Fermilab, Institute for Advanced
Study, the Japan Participation Group, Johns Hopkins University, Los
Alamos National Laboratory, the Max-Planck-Institute for Astronomy
(MPIA), the Max-Planck-Institute for Astrophysics (MPA), New Mexico
State University, University of Pittsburgh, Princeton University, the
United States Naval Observatory, and University of Washington.

\clearpage
\LongTables
\begin{landscape}
\begin{deluxetable}{cccccccccccc}
\tablecolumns{12}
\tablewidth{0pc}
\tablenum{1}
\tablecaption{The \ion{Mg}{2} Sample\tablenotemark{a}}
\tablehead{
\colhead{QSO} &
\colhead{$m_{\rm V}$} &
\colhead{$z_{\rm em}$} &
\colhead{$z_{\rm abs}$} &
\colhead{$W^{\lambda2600}_{0}$ (\AA)} &
\colhead{$W^{\lambda2796}_{0}$ (\AA)} &
\colhead{$W^{\lambda2803}_{0}$ (\AA)} &
\colhead{$W^{\lambda2852}_{0}$ (\AA)} &
\colhead{Refs} &
\colhead{$\log$ N(HI)\tablenotemark{b}} &
\colhead{Sel\tablenotemark{c}} &
\colhead{UV sp\tablenotemark{d}}  \\[.2ex]
\colhead{} &
\colhead{} &
\colhead{} &
\colhead{MgII} &
\colhead{FeII} &
\colhead{MgII} &
\colhead{MgII} &
\colhead{MgI} &
\colhead{} &
\colhead{} &
\colhead{Crit} &
\colhead{source}
}
\startdata
0002$-$422 & 17.2 & 2.758 & 1.5413 & $<$0.1 & 0.48 $\pm$ 0.04 & 0.32 $\pm$ 0.04 & 0.27 $\pm$ 0.06 & 1 & $18.86^{+0.02}_{-0.02}$ & 1 & 06 \\ 
0002+051 & 16.2 & 1.899 & 0.8514 & 0.42 $\pm$ 0.02 & 1.09 $\pm$ 0.02 & 0.84 $\pm$ 0.03 & 0.17 $\pm$ 0.03 & 2 & $19.08^{+0.03}_{-0.04}$ & 1 & AR \\ 
0009$-$016 & 17.6 & 1.998 & 1.3862 & 0.46 $\pm$ 0.04 & 0.88 $\pm$ 0.08 & 0.74 $\pm$ 0.09 & \nodata & 2 & $20.26^{+0.02}_{-0.02}$ & 2 & 09 \\ 
0017+154 & 18.2 & 2.018 & 1.6261 & 0.83 & 1.42 & 1.30 & \nodata &  3 & $19.41^{+0.05}_{-0.05}$ & 3 & 09 \\ 
0021+0043 & 17.7 & 1.245 & 0.5203 & 0.280 $\pm$ 0.072 & 0.533 $\pm$ 0.036 & 0.342 $\pm$ 0.033 & 0.048 $\pm$ 0.040 & 4 & $19.54^{+0.02}_{-0.03}$ & 1 & 11  \\ 
\nodata & \nodata & \nodata & 0.9420 & 0.942 $\pm$ 0.036 & 1.777 $\pm$ 0.035 & 1.735 $\pm$ 0.037 & 0.576 $\pm$ 0.044 & 4 & $19.38^{+0.10}_{-0.15}$ & 4 & 11 \\ 
0021+0104 & 18.2 & 1.829 & 1.3259 & 2.251 $\pm$ 0.090 & 2.656 $\pm$ 0.076 & 2.414 $\pm$ 0.071 & 0.535 $\pm$ 0.069 & 4 & $20.04^{+0.07}_{-0.14}$ & 4 & 11 \\ 
\nodata & \nodata & \nodata & 1.5756 & 1.802 $\pm$ 0.060 & 3.264 $\pm$ 0.084 & 2.595 $\pm$ 0.080 & \nodata &  4 & $20.48^{+0.12}_{-0.18}$ & 4 & 11 \\ 
0041$-$266 & 17.8 & 3.053 & 0.8626 & \nodata & 0.67 $\pm$ 0.06 & 0.38 $\pm$ 0.06 & $<$0.1 &  5 & $<18.00$ & 2 & 09 \\ 
0058+019 & 17.2 & 1.959 & 0.6127 & 1.27 $\pm$ 0.04 & 1.63 $\pm$ 0.01 & 1.51 $\pm$ 0.03 & 0.30 $\pm$ 0.06 & 6,7 & $20.04^{+0.10}_{-0.09}$ & 1 & 06 \\ 
0106+0105 & 19.0 & 1.611 & 1.3002 & 1.181 $\pm$ 0.122 & 2.050 $\pm$ 0.088 & 1.747 $\pm$ 0.074 & 0.429 $\pm$ 0.110 & 4 & $20.95^{+0.05}_{-0.11}$ & 4 & 11 \\ 
\nodata & \nodata & \nodata & 1.3556 & 0.164 $\pm$ 0.118 & 1.312 $\pm$ 0.094 & 0.716 $\pm$ 0.079 & $-$0.066 $\pm$ 0.100 & 4 & $19.65^{+0.09}_{-0.11}$ & 1 & 11  \\ 
0107$-$0019 & 18.3 & 0.738 & 0.5260 & \nodata & 0.784 $\pm$ 0.080 & 0.488 $\pm$ 0.065 & $-$0.073 $\pm$ 0.065 & 4 & $18.48^{+0.30}_{-0.63}$ & 1 & 11  \\ 
0116$-$0043 & 18.7 & 1.282 & 0.9127 & 0.904 $\pm$ 0.104 & 1.379 $\pm$ 0.096 & 1.115 $\pm$ 0.095 & $-$0.001 $\pm$ 0.114 & 4 & $19.95^{+0.05}_{-0.11}$ & 4 & 11 \\ 
0117+213 & 16.1 & 1.491 & 0.5764 & 0.88 $\pm$ 0.04 & 0.91 $\pm$ 0.04 & 0.93 $\pm$ 0.04 & 0.23 $\pm$ 0.04 & 2 & $19.15^{+0.06}_{-0.07}$ & 1 & AR \\ 
\nodata & \nodata & \nodata & 1.0480 & 0.07 $\pm$ 0.01 & 0.42 $\pm$ 0.01 & 0.26 $\pm$ 0.03 & 0.03 $\pm$ 0.01 & 2,7 & $18.86^{+0.02}_{-0.02}$ & 1 & AR \\ 
0119$-$046 & 16.9 & 1.953 & 0.6577 & $<$0.1 & 0.30 $\pm$ 0.04 & 0.22 $\pm$ 0.04 & $<$0.1 & 8 & $18.76^{+0.09}_{-0.07}$ & 1 & 06 \\ 
0123$-$0058 & 18.6 & 1.551 & 0.8686 & 0.231 $\pm$ 0.071 & 0.757 $\pm$ 0.098 & 0.746 $\pm$ 0.116 & 0.077 $\pm$ 0.093 &  4 & $<18.62$ & 1 & 11  \\ 
\nodata & \nodata & \nodata & 1.4094 & 1.503 $\pm$ 0.062 & 1.894 $\pm$ 0.054 & 1.795 $\pm$ 0.054 & 0.610 $\pm$ 0.073 & 4 & $20.08^{+0.10}_{-0.08}$ & 4 & 11 \\ 
0126$-$0105 & 18.4 & 1.609 & 1.1916 & 1.374 $\pm$ 0.072 & 1.983 $\pm$ 0.055 & 1.798 $\pm$ 0.060 & 0.518 $\pm$ 0.061 &  4 & $20.60^{+0.02}_{-0.06}$ & 4 & 11 \\ 
0132+0116 & 18.9 & 1.786 & 1.2712 & 1.848 $\pm$ 0.144 & 2.739 $\pm$ 0.108 & 2.509 $\pm$ 0.144 & 0.931 $\pm$ 0.115 & 4 & $19.70^{+0.08}_{-0.10}$ & 4 & 11 \\ 
0138$-$0005 & 18.7 & 1.340 & 0.7821 & 1.103 $\pm$ 0.110 & 1.208 $\pm$ 0.096 & 1.384 $\pm$ 0.102 & 0.278 $\pm$ 0.094 &  4 & $19.81^{+0.06}_{-0.11}$ & 4 & 11 \\ 
0139$-$0023 & 19.0 & 1.384 & 0.6828 & 1.067 $\pm$ 0.098 & 1.243 $\pm$ 0.102 & 1.123 $\pm$ 0.121 & 0.592 $\pm$ 0.117 &  4 & $20.60^{+0.05}_{-0.12}$ & 4 & 11 \\ 
0141+339 & 17.6 & 1.450 & 0.4709 & $<$0.7 & 0.78 $\pm$ 0.07 & 0.65 $\pm$ 0.07 & $<$0.3 &  2 & $18.88^{+0.08}_{-0.10}$ & 1 & 06 \\ 
0143$-$015 & 17.7 & 3.141 & 1.0383 & $<$0.1 & 0.64 $\pm$ 0.06 & 0.53 $\pm$ 0.05 & $<$0.1 & 9 & $19.15^{+0.06}_{-0.10}$ & 1 & 06 \\ 
\nodata & \nodata & \nodata & 1.2853 & $<$0.1 & 0.56 $\pm$ 0.05 & 0.33 $\pm$ 0.04 & $<$0.1 & 9 & $18.83^{+0.03}_{-0.03}$ & 1 & 06 \\ 
0150$-$202 & 17.4 & 2.147 & 0.7800 & $<$0.3 & 0.36 $\pm$ 0.04 & 0.21 $\pm$ 0.04 & \nodata & 6 & $18.87^{+0.11}_{-0.14}$ & 1 & 06 \\ 
0152+0023 & 17.7 & 0.589 & 0.4818 & 0.884 $\pm$ 0.061 & 1.340 $\pm$ 0.057 & 0.949 $\pm$ 0.050 & 0.624 $\pm$ 0.067 & 4 & $19.78^{+0.07}_{-0.08}$ & 4 & 11 \\ 
0153+0009 & 17.8 & 0.837 & 0.7714 & 1.217 $\pm$ 0.061 & 2.960 $\pm$ 0.051 & 2.431 $\pm$ 0.058 & 0.755 $\pm$ 0.066 & 4 & $19.70^{+0.08}_{-0.10}$ & 4 & 11 \\ 
0153+0052 & 19.0 & 1.162 & 1.0599 & 1.219 $\pm$ 0.078 & 1.618 $\pm$ 0.096 & 1.375 $\pm$ 0.117 & 0.305 $\pm$ 0.126 & 4 & $20.43^{+0.10}_{-0.11}$ & 4 & 11 \\ 
0157$-$0048 & 17.9 & 1.548 & 1.4157 & 0.812 $\pm$ 0.050 & 1.292 $\pm$ 0.050 & 1.068 $\pm$ 0.044 & 0.445 $\pm$ 0.058 &  4 & $19.90^{+0.07}_{-0.06}$ & 4 & 11 \\ 
0215+015 & 16.0 & 1.715 & 1.3447 & 1.36 $\pm$ 0.02 & 1.93 $\pm$ 0.02 & 1.64 $\pm$ 0.02 & 0.32 $\pm$ 0.02 & 10 & $19.89^{+0.08}_{-0.09}$ & 1 & AR \\ 
0239$-$154 & 18.4 & 2.786 & 1.3035 & $<$0.1 & 1.04 $\pm$ 0.07 & 0.89 $\pm$ 0.07 & \nodata & 9 & $<18.70$ & 2 & 09 \\ 
0248+430 & 17.6 & 1.310 & 0.3939 & 1.03 $\pm$ 0.09 & 1.86 $\pm$ 0.09 & 1.42 $\pm$ 0.09 & 0.70 $\pm$ 0.08 & 2 & $21.59^{+0.06}_{-0.07}$ & 1 & 06 \\ 
\nodata & \nodata & \nodata & 0.4515 & $<$0.1 & 0.34 $\pm$ 0.07 & 0.29 $\pm$ 0.07 & $<$0.5 & 11 & $<19.51$ & 1 & 06 \\ 
0253+0107 & 18.8 & 1.035 & 0.6317 & 2.205 $\pm$ 0.182 & 2.571 $\pm$ 0.166 & 2.581 $\pm$ 0.161 & 1.326 $\pm$ 0.157 & 4 & $20.78^{+0.12}_{-0.08}$ & 4 & 11 \\ 
0254$-$334 & 16.0 & 1.849 & 0.2125 & \nodata & 2.23 & 1.73 & \nodata & 12 & $19.41^{+0.09}_{-0.14}$ & 2 & 09 \\ 
0256+0110 & 18.8 & 1.349 & 0.7254 & 2.467 $\pm$ 0.094 & 3.104 $\pm$ 0.115 & 2.861 $\pm$ 0.116 & 1.021 $\pm$ 0.146 & 4 & $20.70^{+0.11}_{-0.22}$ & 4 & 11 \\ 
0302$-$223 & 16.4 & 1.409 & 1.0096 & 0.63 $\pm$ 0.02 & 1.16 $\pm$ 0.04 & 0.96 $\pm$ 0.04 & 0.18 $\pm$ 0.03 & 13 & $20.36^{+0.04}_{-0.04}$ & 1 & AR \\ 
0316$-$203 & 19.5 & 2.869 & 1.4026 & $<$0.1 & 0.92 $\pm$ 0.06 & 0.52 $\pm$ 0.05 & \nodata & 9 & $18.68^{+0.14}_{-0.22}$ & 2 & 09 \\ 
0333+321 & 17.5 & 1.259 & 0.9531 & $<$0.2 & 0.47 $\pm$ 0.05 & 0.33 $\pm$ 0.04 & $<$0.3 &  2 & $<18.00$ & 1 & AR \\ 
0352$-$275 & 17.9 & 2.823 & 1.4051 & 1.68 $\pm$ 0.04 & 2.75 $\pm$ 0.07 & 2.54 $\pm$ 0.07 & \nodata & 9 & $20.18^{+0.12}_{-0.18}$ & 3 & 09 \\ 
0420$-$014 & 17.0 & 0.915 & 0.6331 & \nodata & 1.02 $\pm$ 0.10 & 0.86 $\pm$ 0.10 & \nodata & 14 & $18.54^{+0.07}_{-0.10}$ & 2 & 09 \\ 
0421+019 & 17.0 & 2.055 & 1.3918 & $<$0.1 & 0.34 $\pm$ 0.07 & 0.31 $\pm$ 0.07 & $<$0.3 &  2 & $<18.48$ & 1 & 06 \\ 
\nodata & \nodata & \nodata & 1.6379 & $<$0.4 & 0.34 $\pm$ 0.04 & 0.28 $\pm$ 0.05 & $<$0.2 &  2 & $18.88^{+0.03}_{-0.04}$ & 1 & 06 \\ 
0424$-$131 & 17.5 & 2.166 & 1.4080 & 0.44 $\pm$ 0.05 & 0.55 $\pm$ 0.07 & 0.35 $\pm$ 0.07 & $<$0.3 & 2 & $19.04^{+0.04}_{-0.04}$ & 1 & 06 \\ 
\nodata & \nodata & \nodata & 1.5623 & $<$0.2 & 0.38 $\pm$ 0.05 & 0.39 $\pm$ 0.05 & $<$0.2 & 2 & $18.90^{+0.04}_{-0.04}$ & 1 & 06 \\ 
0449$-$168 & 18.0 & 2.679 & 1.0072 & 1.88 $\pm$ 0.05 & 2.14 $\pm$ 0.06 & 2.13 $\pm$ 0.07 & 0.43 $\pm$ 0.05 & 6 & $20.98^{+0.06}_{-0.07}$ & 3 & 09 \\ 
0454+039 & 16.5 & 1.343 & 0.8596 & 1.23 $\pm$ 0.01 & 1.45 $\pm$ 0.01 & 1.40 $\pm$ 0.06 & 0.31 $\pm$ 0.02 & 2,7 & $20.67^{+0.03}_{-0.03}$ & 1 & AR \\ 
\nodata & \nodata & \nodata & 1.1532 & 0.08 $\pm$ 0.02 & 0.43 $\pm$ 0.01 & 0.36 $\pm$ 0.04 & 0.03 $\pm$ 0.01 & 2,7 & $18.59^{+0.03}_{-0.02}$ & 1 & AR \\ 
0454$-$220 & 16.1 & 0.534 & 0.4744 & 0.98 $\pm$ 0.03 & 1.38 $\pm$ 0.01 & 1.31 $\pm$ 0.04 & 0.33 $\pm$ 0.01 &  15,7 & $19.45^{+0.02}_{-0.03}$ & 1 & AR \\ 
\nodata & \nodata & \nodata & 0.4833 & 0.16 $\pm$ 0.04 & 0.43 $\pm$ 0.01 & 0.27 $\pm$ 0.03 & 0.07 $\pm$ 0.01 & 15 & $18.65^{+0.02}_{-0.02}$ & 1 & AR \\ 
0710+119 & 16.6 & 0.768 & 0.4629 & $<$0.4 & 0.62 $\pm$ 0.06 & 0.29 $\pm$ 0.05 & 0.24 $\pm$ 0.08 & 14 & $<18.30$ & 1 & AR \\ 
0729+818 & 17.5 & 1.024 & 0.7068 & 0.52 $\pm$ 0.04 & 1.27 $\pm$ 0.14 & 0.97 $\pm$ 0.14 & \nodata & 2 & $18.67^{+0.06}_{-0.07}$ & 3 & 09 \\ 
0735+178 & 14.9 & \nodata & 0.4240 & 0.87 $\pm$ 0.18 & 1.32 $\pm$ 0.03 & 1.03 $\pm$ 0.03 & 0.18 $\pm$ 0.03 & 16 & $<19.00$ & 1 & AR \\ 
0738+313 & 16.1 & 0.630 & 0.2213 & $<$0.6 & 0.61 $\pm$ 0.04 & 0.38 $\pm$ 0.02 & 0.24 $\pm$ 0.04 & 17 & $20.90^{+0.07}_{-0.08}$ & 1 & 06 \\ 
0742+318 & 15.6 & 0.462 & 0.1920 & \nodata & 0.33 $\pm$ 0.04 & 0.23 $\pm$ 0.04 & $<$0.2 &  18 & $<18.30$ & 1 & AR \\
0823$-$223 & 16.2 & 0.000 & 0.9110 & 0.42 $\pm$ 0.03 & 1.28 $\pm$ 0.02 & 0.68 $\pm$ 0.10 & 0.22 $\pm$ 0.03 &  19,7 & $19.04^{+0.04}_{-0.04}$ & 1 & 06 \\ 
0827+243 & 17.3 & 0.941 & 0.5247 & 1.90 & 2.90 & 2.20 & \nodata & 20,21 & $20.30^{+0.04}_{-0.05}$ & 1 & 06 \\ 
0843+136 & 17.8 & 1.877 & 0.6064 & $<$0.7 & 1.08 $\pm$ 0.08 & 0.67 $\pm$ 0.09 & $<$0.1 & 22 & $19.56^{+0.11}_{-0.14}$ & 1 & 06 \\ 
0933+732 & 17.3 & 2.525 & 1.4789 & 0.76 $\pm$ 0.08 & 0.95 $\pm$ 0.08 & 1.15 $\pm$ 0.08 & $<$0.3 &  2 & $21.62^{+0.08}_{-0.09}$ & 1 & 06 \\ 
\nodata & \nodata & \nodata & 1.4973 & 1.15 $\pm$ 0.09 & 1.71 $\pm$ 0.09 & 1.98 $\pm$ 0.09 & 0.67 $\pm$ 0.06 & 2 & $20.00^{+0.18}_{-0.30}$ & 1 & 06 \\ 
0952+179 & 17.2 & 1.478 & 0.2377 & \nodata & 0.63 $\pm$ 0.11 & 0.79 $\pm$ 0.11 & $<$0.4 &  2 & $21.32^{+0.05}_{-0.06}$ & 1 & 06 \\ 
0953$-$0038 & 18.4 & 1.383 & 0.6381 & 1.029 $\pm$ 0.139 & 1.668 $\pm$ 0.080 & 1.195 $\pm$ 0.083 & 0.143 $\pm$ 0.135 &  4 & $19.90^{+0.07}_{-0.09}$ & 4 & 11 \\ 
0957+003 & 17.6 & 0.907 & 0.6720 & \nodata & 1.77 & 1.32 & \nodata & 23 & $19.59^{+0.03}_{-0.03}$ & 2 & 09 \\ 
0957+561A & 17.0 & 1.414 & 1.3911 & 1.67 $\pm$ 0.21 & 2.12 $\pm$ 0.03 & 1.97 $\pm$ 0.03 & 0.19 $\pm$ 0.04 & 24,25 & $20.32^{+0.09}_{-0.12}$ & 1 & AR \\ 
0958+551 & 16.0 & 1.760 & 0.2413 & \nodata & 0.55 $\pm$ 0.07 & 0.38 $\pm$ 0.07 & $<$0.2 &  6 & $19.80^{+0.08}_{-0.09}$ & 1 & 06 \\ 
0959$-$0035 & 18.9 & 1.871 & 1.5985 & 1.548 $\pm$ 0.064 & 2.310 $\pm$ 0.092 & 1.770 $\pm$ 0.104 & 0.188 $\pm$ 0.076 &  4 & $20.54^{+0.11}_{-0.10}$ & 4 & 11 \\ 
1007+0042 & 19.1 & 1.681 & 0.9321 & 0.127 $\pm$ 0.113 & 0.896 $\pm$ 0.148 & 0.528 $\pm$ 0.143 & $-$0.360 $\pm$ 0.146 & 4 & $18.60^{+0.16}_{-0.26}$ & 1 & 11  \\ 
\nodata & \nodata & \nodata & 1.0373 & 1.990 $\pm$ 0.184 & 2.980 $\pm$ 0.233 & 3.275 $\pm$ 0.260 & 0.648 $\pm$ 0.255 & 4 & $21.15^{+0.15}_{-0.24}$ & 4 & 11 \\ 
1009$-$0026 & 17.4 & 1.244 & 0.8426 & 0.403 $\pm$ 0.043 & 0.713 $\pm$ 0.038 & 0.629 $\pm$ 0.038 & 0.135 $\pm$ 0.044 &  4 & $20.20^{+0.05}_{-0.06}$ & 1 & 11  \\ 
\nodata & \nodata & \nodata & 0.8866 & 1.068 $\pm$ 0.039 & 1.900 $\pm$ 0.039 & 1.525 $\pm$ 0.039 & 0.326 $\pm$ 0.046 & 4 & $19.48^{+0.01}_{-0.08}$ & 4 & 11 \\ 
1009+0036 & 19.0 & 1.699 & 0.9714 & 1.081 $\pm$ 0.078 & 1.093 $\pm$ 0.111 & 1.361 $\pm$ 0.131 & 0.226 $\pm$ 0.129 & 4 & $20.00^{+0.11}_{-0.05}$ & 4 & 11 \\ 
1010+0003 & 18.2 & 1.399 & 1.2651 & 0.973 $\pm$ 0.110 & 1.122 $\pm$ 0.069 & 1.118 $\pm$ 0.058 & 0.420 $\pm$ 0.068 & 4 & $21.52^{+0.06}_{-0.07}$ & 4 & 11 \\ 
1010$-$0047 & 18.0 & 1.671 & 1.0719 & 0.172 $\pm$ 0.109 & 0.571 $\pm$ 0.083 & 0.449 $\pm$ 0.078 & $-$0.034 $\pm$ 0.077 & 4 & $18.90^{+0.03}_{-0.03}$ & 1 & 11  \\ 
\nodata & \nodata & \nodata & 1.3270 & 1.339 $\pm$ 0.060 & 2.059 $\pm$ 0.051 & 1.859 $\pm$ 0.047 & 0.298 $\pm$ 0.048 & 4 & $19.81^{+0.03}_{-0.07}$ & 4 & 11 \\ 
1019+309 & 17.5 & 1.319 & 0.3461 & \nodata & 0.70 $\pm$ 0.05 & 0.68 $\pm$ 0.05 & \nodata &  2 & $18.18^{+0.08}_{-0.10}$ & 2 & 09 \\ 
1022+0101 & 18.9 & 1.563 & 1.4240 & 0.414 $\pm$ 0.146 & 0.862 $\pm$ 0.074 & 0.383 $\pm$ 0.070 & 0.095 $\pm$ 0.088 & 4 & $18.97^{+0.07}_{-0.03}$ & 1 & 11  \\ 
1028$-$0100 & 18.2 & 1.531 & 0.6322 & 1.140 $\pm$ 0.101 & 1.579 $\pm$ 0.087 & 1.216 $\pm$ 0.093 & 0.568 $\pm$ 0.091 &  4 & $19.95^{+0.05}_{-0.08}$ & 4 & 11 \\ 
\nodata & \nodata & \nodata & 0.7087 & 0.888 $\pm$ 0.091 & 1.210 $\pm$ 0.066 & 1.008 $\pm$ 0.068 & 0.718 $\pm$ 0.088 & 4 & $20.04^{+0.07}_{-0.04}$ & 4 & 11 \\ 
1032+0003 & 18.9 & 1.193 & 1.0168 & 0.697 $\pm$ 0.119 & 1.919 $\pm$ 0.128 & 1.650 $\pm$ 0.130 & $-$0.083 $\pm$ 0.160 & 4 & $19.00^{+0.04}_{-0.10}$ & 4 & 11 \\ 
1035$-$276 & 19.0 & 2.168 & 0.8242 & 0.52 $\pm$ 0.02 & 1.08 $\pm$ 0.02 & 0.87 $\pm$ 0.02 & $<$0.6 & 26 & $18.81^{+0.15}_{-0.24}$ & 1 & 06 \\ 
1037+0028 & 18.4 & 1.733 & 1.4244 & 1.767 $\pm$ 0.081 & 2.563 $\pm$ 0.031 & 2.171 $\pm$ 0.031 & 0.303 $\pm$ 0.042 & 4 & $20.04^{+0.10}_{-0.14}$ & 4 & 11 \\ 
1038+064 & 16.7 & 1.265 & 0.4416 & $<$0.2 & 0.66 $\pm$ 0.05 & 0.57 $\pm$ 0.04 & $<$0.2 &  2 & $18.30^{+0.18}_{-0.30}$ & 1 & AR \\ 
1040+123 & 17.3 & 1.028 & 0.6591 & $<$0.2 & 0.58 $\pm$ 0.10 & 0.42 $\pm$ 0.10 & $<$0.2 &  14 & $18.38^{+0.05}_{-0.08}$ & 1 & AR \\ 
1047$-$0047 & 18.4 & 0.740 & 0.5727 & 0.765 $\pm$ 0.130 & 1.063 $\pm$ 0.117 & 0.697 $\pm$ 0.094 & 0.055 $\pm$ 0.096 &  4 & $19.36^{+0.17}_{-0.19}$ & 4 & 11 \\ 
1048+0032 & 18.6 & 1.649 & 0.7203 & 1.252 $\pm$ 0.077 & 1.878 $\pm$ 0.063 & 1.636 $\pm$ 0.070 & 0.417 $\pm$ 0.085 & 4 & $18.78^{+0.18}_{-0.48}$ & 4 & 11 \\ 
1049+616 & 16.5 & 0.421 & 0.2255 & $<$0.4 & 0.51 $\pm$ 0.03 & 0.56 $\pm$ 0.03 & $<$0.1 & 18 & $<18.00$ & 1 & 06 \\
\nodata & \nodata & \nodata & 0.3937 & $<$0.1 & 0.34 $\pm$ 0.03 & 0.29 $\pm$ 0.03 & \nodata & 18 & $<18.00$ & 1 & 06 \\ 
1054$-$0020 & 18.3 & 1.021 & 0.8301 & 0.860 $\pm$ 0.049 & 1.156 $\pm$ 0.041 & 1.102 $\pm$ 0.041 & 0.333 $\pm$ 0.048 &  4 & $18.95^{+0.09}_{-0.26}$ & 4 & 11 \\*
\nodata & \nodata & \nodata & 0.9514 & 0.408 $\pm$ 0.047 & 0.834 $\pm$ 0.047 & 0.506 $\pm$ 0.049 & 0.263 $\pm$ 0.053 & 4 & $19.28^{+0.02}_{-0.02}$ & 1 & 11  \\ 
1100$-$264 & 16.0 & 2.148 & 1.1872 & 0.04 $\pm$ 0.01 & 0.51 $\pm$ 0.01 & 0.28 $\pm$ 0.01 & $<$0.2 &  13,27 & $18.51^{+0.04}_{-0.03}$ & 1 & AR \\ 
\nodata & \nodata & \nodata & 1.2028 & $<$0.2 & 0.54 $\pm$ 0.02 & 0.43 $\pm$ 0.02 & 0.27 $\pm$ 0.05 & 13,27,28 & $18.40^{+0.05}_{-0.04}$ & 1 & AR \\ 
1107+0003 & 18.6 & 1.740 & 0.9545 & 0.868 $\pm$ 0.048 & 1.356 $\pm$ 0.066 & 1.187 $\pm$ 0.059 & $-$0.044 $\pm$ 0.036 & 4 & $20.26^{+0.09}_{-0.14}$ & 4 & 11 \\ 
1107+0048 & 17.5 & 1.392 & 0.7404 & 2.375 $\pm$ 0.020 & 2.952 $\pm$ 0.025 & 2.809 $\pm$ 0.025 & 0.913 $\pm$ 0.032 & 4 & $21.00^{+0.02}_{-0.05}$ & 4 & 11 \\ 
\nodata & \nodata & \nodata & 1.0703 & 0.035 $\pm$ 0.038 & 0.532 $\pm$ 0.033 & 0.260 $\pm$ 0.031 & 0.026 $\pm$ 0.032 & 4 & $18.60^{+0.03}_{-0.03}$ & 1 & 11  \\ 
1109+0051 & 18.7 & 0.957 & 0.4181 & \nodata & 1.361 $\pm$ 0.105 & 1.083 $\pm$ 0.110 & 0.198 $\pm$ 0.110 & 4 & $19.08^{+0.22}_{-0.38}$ & 1 & 11  \\ 
\nodata & \nodata & \nodata & 0.5520 & 0.934 $\pm$ 0.092 & 1.417 $\pm$ 0.085 & 1.193 $\pm$ 0.075 & 0.482 $\pm$ 0.094 & 4 & $19.60^{+0.10}_{-0.12}$ & 4 & 11 \\ 
1110+0048 & 18.6 & 0.761 & 0.5604 & 2.132 $\pm$ 0.117 & 2.273 $\pm$ 0.089 & 2.623 $\pm$ 0.083 & 0.436 $\pm$ 0.083 & 4 & $20.20^{+0.10}_{-0.09}$ & 4 & 11 \\ 
1112+0013 & 18.8 & 1.433 & 1.2420 & 0.667 $\pm$ 0.106 & 2.158 $\pm$ 0.083 & 1.752 $\pm$ 0.079 & 0.029 $\pm$ 0.054 & 4 & $19.30^{+0.18}_{-0.15}$ & 4 & 11 \\ 
1115+080 & 17.0 & 1.732 & 1.0431 & $<$0.1 & 0.31 $\pm$ 0.03 & 0.18 $\pm$ 0.03 & $<$0.2 &  2 & $18.62^{+0.02}_{-0.02}$ & 1 & AR \\ 
1127$-$145 & 16.9 & 1.187 & 0.3130 & 1.14 $\pm$ 0.27 & 2.21 $\pm$ 0.12 & 1.90 $\pm$ 0.12 & 1.14 $\pm$ 0.12 & 29 & $21.71^{+0.07}_{-0.08}$ & 1 & 06 \\ 
1137+660 & 16.3 & 0.652 & 0.1164 & $<$0.2 & 0.50 $\pm$ 0.14 & 0.53 $\pm$ 0.12 & $<$0.2 &  30 & $18.60^{+0.10}_{-0.12}$ & 1 & AR \\ 
1148+386 & 17.0 & 1.304 & 0.5533 & $<$0.2 & 0.92 $\pm$ 0.05 & 0.99 $\pm$ 0.05 & $<$0.2 &  2 & $<18.00$ & 1 & 06 \\ 
1206+459 & 15.5 & 1.155 & 0.9276 & 0.08 $\pm$ 0.02 & 0.88 $\pm$ 0.02 & 0.79 $\pm$ 0.04 & 0.04 $\pm$ 0.02 & 2,7 & $19.04^{+0.04}_{-0.04}$ & 1 & AR \\ 
1209+107 & 17.8 & 2.193 & 0.3930 & $<$0.4 & 1.00 $\pm$ 0.07 & 0.54 $\pm$ 0.09 & $<$0.2 & 31 & $19.46^{+0.08}_{-0.08}$ & 1 & AR \\ 
\nodata & \nodata & \nodata & 0.6295 & 1.50 $\pm$ 0.20 & 2.92 $\pm$ 0.23 & 2.05 $\pm$ 0.18 & $<$0.4 & 31 & $20.30^{+0.18}_{-0.30}$ & 1 & AR \\ 
1213$-$002 & 17.0 & 2.691 & 1.5543 & 1.11 $\pm$ 0.09 & 2.09 $\pm$ 0.06 & 1.65 $\pm$ 0.06 & $<$0.2 & 2 & $19.56^{+0.02}_{-0.02}$ & 1 & 06 \\ 
1220$-$0040 & 18.5 & 1.411 & 0.9746 & 1.005 $\pm$ 0.098 & 1.952 $\pm$ 0.145 & 1.793 $\pm$ 0.125 & 0.143 $\pm$ 0.101 &  4 & $20.20^{+0.05}_{-0.09}$ & 4 & 11 \\ 
1222+228 & 16.6 & 2.048 & 0.6681 & $<$0.1 & 0.43 $\pm$ 0.04 & 0.41 $\pm$ 0.04 & $<$0.1 &  6 & $18.59^{+0.03}_{-0.03}$ & 1 & 06 \\ 
1224+0037 & 18.7 & 1.482 & 1.2346 & 0.950 $\pm$ 0.091 & 1.093 $\pm$ 0.072 & 1.028 $\pm$ 0.069 & 0.119 $\pm$ 0.063 & 4 & $20.88^{+0.04}_{-0.06}$ & 4 & 11 \\ 
\nodata & \nodata & \nodata & 1.2665 & 1.634 $\pm$ 0.183 & 2.094 $\pm$ 0.062 & 1.985 $\pm$ 0.073 & 0.330 $\pm$ 0.077 & 4 & $20.00^{+0.08}_{-0.05}$ & 4 & 11 \\ 
1225+0035 & 18.9 & 1.226 & 0.7730 & 1.316 $\pm$ 0.141 & 1.744 $\pm$ 0.138 & 1.447 $\pm$ 0.150 & 0.929 $\pm$ 0.135 & 4 & $21.38^{+0.11}_{-0.12}$ & 4 & 11 \\ 
1226+105 & 18.5 & 2.305 & 0.9376 & 0.86 & 1.36 & 1.20 & \nodata & 3 & $19.41^{+0.12}_{-0.18}$ & 3 & 09 \\ 
1229$-$021 & 16.8 & 1.038 & 0.7571 & \nodata & 0.52 $\pm$ 0.07 & 0.48 $\pm$ 0.07 & $<$0.1 & 14 & $18.36^{+0.09}_{-0.08}$ & 1 & AR \\ 
1241+176 & 15.9 & 1.282 & 0.5505 & 0.24 $\pm$ 0.05 & 0.48 $\pm$ 0.02 & 0.37 $\pm$ 0.05 & 0.10 $\pm$ 0.03 &  2,7 & $18.90^{+0.07}_{-0.09}$ & 1 & AR \\ 
1246$-$057 & 16.7 & 2.224 & 1.2015 & 0.40 $\pm$ 0.03 & 0.90 $\pm$ 0.04 & 0.75 $\pm$ 0.04 & $<$0.4 & 2 & $18.91^{+0.03}_{-0.03}$ & 1 & AR \\ 
\nodata & \nodata & \nodata & 1.6453 & 0.38 $\pm$ 0.06 & 0.52 $\pm$ 0.04 & 0.58 $\pm$ 0.04 & $<$0.3 &  2 & $18.89^{+0.02}_{-0.02}$ & 1 & AR \\ 
1247+267 & 15.6 & 2.043 & 1.2232 & \nodata & 0.48 $\pm$ 0.03 & 0.38 $\pm$ 0.02 & 0.11 $\pm$ 0.02 & 2 & $19.87^{+0.01}_{-0.01}$ & 1 & AR \\ 
1248+401 & 16.1 & 1.032 & 0.7730 & 0.25 $\pm$ 0.02 & 0.69 $\pm$ 0.01 & 0.50 $\pm$ 0.08 & 0.07 $\pm$ 0.02 & 2,7 & $18.60^{+0.02}_{-0.02}$ & 1 & AR \\ 
1254+047 & 16.4 & 1.018 & 0.5193 & 0.40 $\pm$ 0.05 & 0.46 $\pm$ 0.04 & 0.38 $\pm$ 0.04 & $<$0.2 &  2 & $18.86^{+0.06}_{-0.06}$ & 1 & AR \\ 
1317+277 & 16.0 & 1.014 & 0.2887 & \nodata & 0.33 $\pm$ 0.04 & 0.31 $\pm$ 0.04 & $<$0.2 &  2 & $<18.30$ & 1 & AR \\ 
\nodata & \nodata & \nodata & 0.6601 & 0.13 $\pm$ 0.02 & 0.34 $\pm$ 0.01 & 0.33 $\pm$ 0.04 & 0.03 $\pm$ 0.01 & 2,7 & $18.57^{+0.02}_{-0.02}$ & 1 & AR \\ 
1323$-$0021 & 18.2 & 1.390 & 0.7160 & 1.452 $\pm$ 0.077 & 2.229 $\pm$ 0.071 & 1.864 $\pm$ 0.066 & 0.940 $\pm$ 0.069 &  4 & $20.54^{+0.15}_{-0.15}$ & 4 & 11 \\ 
1323+655 & 17.5 & 1.624 & 1.5181 & \nodata & 0.57 & 0.53 & $<$0.1 &  3 & $18.56^{+0.02}_{-0.02}$ & 1 & 06 \\ 
\nodata & \nodata & \nodata & 1.6101 & 0.88 & 2.20 & 1.85 & 0.16 & 3 & high $b$ & 1 & 06 \\ 
1327$-$206 & 17.0 & 1.165 & 0.8526 & 0.76 $\pm$ 0.27 & 2.11 & 1.48 & $<$0.2 &  32,33 & $19.40^{+0.02}_{-0.02}$ & 1 & AR \\ 
1329+412 & 17.2 & 1.937 & 1.2820 & \nodata & 0.49 $\pm$ 0.05 & 0.31 $\pm$ 0.05 & $<$0.3 & 2 & $19.46^{+0.09}_{-0.10}$ & 1 & 06 \\ 
\nodata & \nodata & \nodata & 1.6011 & $<$0.2 & 0.70 $\pm$ 0.04 & 0.35 $\pm$ 0.04 & $<$0.2 &  2 & $19.04^{+0.04}_{-0.04}$ & 1 & 06 \\ 
1338+416 & 16.1 & 1.204 & 0.6213 & $<$0.1 & 0.31 $\pm$ 0.05 & 0.17 $\pm$ 0.04 & $<$0.3 & 2 & $19.08^{+0.03}_{-0.04}$ & 1 & AR \\ 
1341+0059 & 18.8 & 1.714 & 1.1176 & 0.929 $\pm$ 0.119 & 1.711 $\pm$ 0.154 & 1.145 $\pm$ 0.133 & 0.414 $\pm$ 0.088 & 4 & $19.93^{+0.11}_{-0.08}$ & 4 & 11 \\ 
1342$-$0035 & 18.2 & 0.787 & 0.5380 & 1.453 $\pm$ 0.090 & 2.256 $\pm$ 0.068 & 1.882 $\pm$ 0.083 & 0.780 $\pm$ 0.088 &  4 & $19.78^{+0.12}_{-0.14}$ & 4 & 11 \\ 
1345$-$0023 & 17.6 & 1.095 & 0.6057 & 0.635 $\pm$ 0.049 & 1.177 $\pm$ 0.049 & 1.219 $\pm$ 0.052 & 0.155 $\pm$ 0.043 &  4 & $18.85^{+0.15}_{-0.24}$ & 4 & 11 \\ 
1354+195 & 16.0 & 0.719 & 0.4566 & 0.32 $\pm$ 0.04 & 0.89 $\pm$ 0.04 & 0.82 $\pm$ 0.04 & 0.16 $\pm$ 0.03 & 2 & $18.54^{+0.04}_{-0.04}$ & 1 & AR \\ 
1354+258 & 18.0 & 2.006 & 0.8585 & $<$0.2 & 1.00 & 0.86 & $<$0.1 & 3 & $18.57^{+0.07}_{-0.08}$ & 1 & 06 \\*
\nodata & \nodata & \nodata & 0.8856 & $<$0.2 & 0.81 & 0.57 & $<$0.2 &  3 & $18.76^{+0.10}_{-0.13}$ & 1 & 06 \\* 
\nodata & \nodata & \nodata & 1.4205 & 0.55 & 0.61 & 0.50 & 0.20 & 3 & $21.51^{+0.03}_{-0.03}$ & 1 & 06 \\ 
1419$-$0036 & 18.3 & 0.969 & 0.6238 & 0.075 $\pm$ 0.083 & 0.597 $\pm$ 0.069 & 0.476 $\pm$ 0.071 & 0.106 $\pm$ 0.073 &  4 & $19.04^{+0.07}_{-0.14}$ & 1 & 11  \\ 
\nodata & \nodata & \nodata & 0.8206 & 0.848 $\pm$ 0.073 & 1.145 $\pm$ 0.057 & 0.963 $\pm$ 0.064 & 0.167 $\pm$ 0.067 & 4 & $18.78^{+0.26}_{-0.23}$ & 4 & 11 \\ 
1420$-$0054 & 18.9 & 1.458 & 1.3475 & 1.079 $\pm$ 0.097 & 1.532 $\pm$ 0.108 & 1.227 $\pm$ 0.104 & 0.365 $\pm$ 0.105 &  4 & $20.90^{+0.05}_{-0.06}$ & 4 & 11 \\ 
1421+330 & 16.7 & 1.906 & 0.9026 & \nodata & 1.08 $\pm$ 0.03 & 0.85 $\pm$ 0.03 & \nodata &  2 & $18.85^{+0.01}_{-0.01}$ & 2 & 09 \\ 
\nodata & \nodata & \nodata & 1.1725 & $<$0.1 & 0.54 $\pm$ 0.03 & 0.40 $\pm$ 0.03 & $<$0.1 & 2 & $18.79^{+0.01}_{-0.01}$ & 1 & 09 \\ 
1426+0051 & 18.8 & 1.333 & 0.7352 & 0.275 $\pm$ 0.072 & 0.857 $\pm$ 0.080 & 0.751 $\pm$ 0.085 & 0.047 $\pm$ 0.086 & 4 & $18.85^{+0.03}_{-0.03}$ & 1 & 11  \\ 
\nodata & \nodata & \nodata & 0.8424 & 1.081 $\pm$ 0.109 & 2.618 $\pm$ 0.125 & 1.972 $\pm$ 0.109 & 0.454 $\pm$ 0.131 & 4 & $19.65^{+0.09}_{-0.07}$ & 4 & 11 \\ 
1431$-$0050 & 18.1 & 1.190 & 0.6085 & 1.239 $\pm$ 0.074 & 1.886 $\pm$ 0.076 & 1.581 $\pm$ 0.079 & 0.203 $\pm$ 0.060 &  4 & $19.18^{+0.30}_{-0.27}$ & 4 & 11 \\ 
\nodata & \nodata & \nodata & 0.6868 & 0.123 $\pm$ 0.059 & 0.613 $\pm$ 0.066 & 0.283 $\pm$ 0.057 & 0.049 $\pm$ 0.059 & 4 & $18.40^{+0.06}_{-0.08}$ & 1 & 11  \\ 
1436$-$0051 & 18.5 & 1.275 & 0.7377 & 1.051 $\pm$ 0.095 & 1.142 $\pm$ 0.084 & 1.146 $\pm$ 0.089 & 0.625 $\pm$ 0.081 &  4 & $20.08^{+0.10}_{-0.12}$ & 4 & 11 \\ 
\nodata & \nodata & \nodata & 0.9281 & 0.662 $\pm$ 0.080 & 1.174 $\pm$ 0.065 & 0.971 $\pm$ 0.063 & 0.132 $\pm$ 0.072 & 4 & $<18.82$ & 4 & 11 \\ 
1437+624 & 19.0 & 1.090 & 0.8723 & \nodata & 0.71 $\pm$ 0.09 & 0.64 $\pm$ 0.09 & \nodata & 14 & $<18.00$ & 2 & 09 \\ 
1455$-$0045 & 18.0 & 1.378 & 1.0929 & 1.273 $\pm$ 0.043 & 1.625 $\pm$ 0.056 & 1.577 $\pm$ 0.061 & 0.373 $\pm$ 0.055 &  4 & $20.08^{+0.03}_{-0.08}$ & 4 & 11 \\ 
1501+0019 & 18.1 & 1.930 & 1.4832 & 1.481 $\pm$ 0.042 & 2.168 $\pm$ 0.052 & 1.853 $\pm$ 0.052 & 0.952 $\pm$ 0.054 & 4 & $20.85^{+0.11}_{-0.15}$ & 4 & 11 \\ 
1517+239 & 16.4 & 1.903 & 0.7382 & $<$0.1 & 0.30 $\pm$ 0.04 & 0.34 $\pm$ 0.06 & \nodata & 6 & $18.72^{+0.03}_{-0.03}$ & 1 & AR \\ 
1521$-$0009 & 19.0 & 1.318 & 0.9590 & 1.610 $\pm$ 0.090 & 1.848 $\pm$ 0.096 & 1.588 $\pm$ 0.093 & 0.944 $\pm$ 0.107 &  4 & $19.40^{+0.08}_{-0.14}$ & 4 & 11 \\ 
1525+0026 & 17.0 & 0.801 & 0.5674 & 1.140 $\pm$ 0.053 & 1.852 $\pm$ 0.035 & 1.525 $\pm$ 0.035 & 0.359 $\pm$ 0.042 & 4 & $19.78^{+0.07}_{-0.08}$ & 4 & 11 \\ 
1537+0021 & 19.1 & 1.754 & 1.1782 & 2.087 $\pm$ 0.113 & 2.502 $\pm$ 0.102 & 2.732 $\pm$ 0.104 & 0.517 $\pm$ 0.115 & 4 & $20.18^{+0.08}_{-0.10}$ & 4 & 11 \\ 
\nodata & \nodata & \nodata & 1.6455 & 1.498 $\pm$ 0.074 & 2.272 $\pm$ 0.111 & 2.172 $\pm$ 0.087 & 0.553 $\pm$ 0.079 & 4 & $20.48^{+0.22}_{-0.18}$ & 4 & 11 \\ 
1554$-$203 & 19.2 & 1.947 & 0.7869 & \nodata & 0.73 & 0.76 & \nodata &  3 & $<19.00$ & 2 & 09 \\ 
1622+239 & 17.5 & 0.927 & 0.6561 & 1.02 $\pm$ 0.05 & 1.45 $\pm$ 0.03 & $<$1.7 & 0.29 $\pm$ 0.03 & 2,7 & $20.36^{+0.07}_{-0.08}$ & 1 & AR \\ 
\nodata & \nodata & \nodata & 0.8913 & 1.02 $\pm$ 0.15 & 1.55 $\pm$ 0.08 & 1.27 $\pm$ 0.08 & 0.31 $\pm$ 0.03 & 2 & $19.23^{+0.02}_{-0.03}$ & 1 & AR \\ 
1623+269 & 16.0 & 2.521 & 0.8881 & 0.21 $\pm$ 0.02 & 0.93 $\pm$ 0.03 & 0.75 $\pm$ 0.04 & 0.14 $\pm$ 0.02 & 6 & $18.66^{+0.02}_{-0.03}$ & 1 & 06 \\ 
1629+120 & 18.4 & 1.792 & 0.5313 & 0.71 $\pm$ 0.10 & 1.40 $\pm$ 0.07 & 1.35 $\pm$ 0.07 & 0.31 $\pm$ 0.08 & 14 & $20.70^{+0.08}_{-0.10}$ & 3 & 09 \\ 
\nodata & \nodata & \nodata & 0.9004 & 0.63 $\pm$ 0.10 & 1.20 $\pm$ 0.09 & 0.69 $\pm$ 0.09 & \nodata & 3,14 & $19.70^{+0.03}_{-0.04}$ & 3 & 09 \\ 
1634+706 & 14.9 & 1.335 & 0.9902 & 0.13 $\pm$ 0.01 & 0.56 $\pm$ 0.01 & 0.42 $\pm$ 0.03 & 0.06 $\pm$ 0.01 & 2,7 & $18.34^{+0.02}_{-0.02}$ & 1 & AR \\ 
1704+608 & 15.3 & 0.371 & 0.2220 & \nodata & 0.55 $\pm$ 0.03 & 0.33 $\pm$ 0.03 & $<$0.2 & 18 & $18.23^{+0.05}_{-0.05}$ & 1 & AR \\ 
1712+5559 & 18.7 & 1.358 & 1.1584 & 0.605 $\pm$ 0.053 & 0.889 $\pm$ 0.058 & 0.847 $\pm$ 0.061 & 0.227 $\pm$ 0.063 & 4 & $19.54^{+0.06}_{-0.15}$ & 1 & 11  \\ 
 \nodata & \nodata & \nodata& 1.2093 & 1.173 $\pm$ 0.054 & 1.742 $\pm$ 0.058 & 1.388 $\pm$ 0.058 & \nodata &  4 & $20.72^{+0.05}_{-0.05}$ & 4 & 11 \\ 
1714+5757 & 18.6 & 1.252 & 0.7481 & 0.880 $\pm$ 0.108 & 1.099 $\pm$ 0.084 & 0.864 $\pm$ 0.085 & 0.071 $\pm$ 0.099 & 4 & $19.23^{+0.17}_{-0.33}$ & 4 & 11 \\ 
1715+5747 & 18.3 & 0.697 & 0.5579 & 0.532 $\pm$ 0.095 & 1.001 $\pm$ 0.067 & 0.813 $\pm$ 0.075 & 0.295 $\pm$ 0.073 & 4 & $19.18^{+0.15}_{-0.18}$ & 4 & 11 \\ 
1716+5654 & 19.0 & 0.937 & 0.5301 & 1.404 $\pm$ 0.161 & 1.822 $\pm$ 0.130 & 1.329 $\pm$ 0.108 & 0.081 $\pm$ 0.122 & 4 & $19.98^{+0.20}_{-0.28}$ & 4 & 11 \\ 
1722+5442 & 18.8 & 1.215 & 0.6338 & 0.588 $\pm$ 0.096 & 1.535 $\pm$ 0.098 & 1.350 $\pm$ 0.093 & 0.203 $\pm$ 0.099 & 4 & $19.00^{+0.30}_{-0.22}$ & 4 & 11 \\ 
1727+5302 & 18.3 & 1.444 & 0.9448 & 2.188 $\pm$ 0.116 & 2.832 $\pm$ 0.070 & 2.511 $\pm$ 0.071 & 0.988 $\pm$ 0.067 & 4 & $21.16^{+0.04}_{-0.05}$ & 4 & 11 \\ 
\nodata & \nodata & \nodata & 1.0312 & 0.755 $\pm$ 0.114 & 0.922 $\pm$ 0.057 & 1.181 $\pm$ 0.079 & 0.331 $\pm$ 0.095 & 4 & $21.41^{+0.03}_{-0.03}$ & 4 & 11 \\ 
1729+5758 & 17.5 & 1.342 & 0.5541 & 1.015 $\pm$ 0.058 & 1.836 $\pm$ 0.046 & 1.608 $\pm$ 0.046 & 0.121 $\pm$ 0.044 & 4 & $18.60^{+0.18}_{-0.43}$ & 4 & 11 \\ 
1733+5533 & 18.0 & 1.072 & 0.9981 & 1.344 $\pm$ 0.056 & 2.173 $\pm$ 0.069 & 2.004 $\pm$ 0.071 & 0.362 $\pm$ 0.080 & 4 & $20.70^{+0.04}_{-0.03}$ & 4 & 11 \\ 
1736+5938 & 18.8 & 1.410 & 1.0664 & 1.439 $\pm$ 0.092 & 2.177 $\pm$ 0.104 & 2.039 $\pm$ 0.110 & \nodata & 4 & $20.00^{+0.08}_{-0.10}$ & 4 & 11 \\ 
1821+107 & 17.3 & 1.364 & 1.2528 & $<$0.3 & 0.71 $\pm$ 0.04 & 0.48 $\pm$ 0.03 & 0.09 $\pm$ 0.02 & 2 & high $b$ & 1 & 06 \\ 
1857+566 & 17.3 & 1.578 & 0.7151 & \nodata & 0.65 & 0.64 & \nodata & 3 & $18.56^{+0.05}_{-0.06}$ & 2 & 09 \\ 
\nodata & \nodata & \nodata & 1.2345 & 0.55 & 0.82 & 0.70 & \nodata &  3 & $18.46^{+0.04}_{-0.06}$ & 3 & 09 \\ 
1901+319 & 17.5 & 0.635 & 0.3901 & 00.0 & 0.45 $\pm$ 0.04 & 0.15 $\pm$ 0.04 & $<$0.1 &  14 & $<18.00$ & 1 & 06 \\ 
2003$-$025 & 19.0 & 1.457 & 1.2116 & 1.27 $\pm$ 0.14 & 2.65 $\pm$ 0.14 & 2.17 $\pm$ 0.14 & \nodata & 14 & $19.32^{+0.06}_{-0.07}$ & 3 & 09 \\ 
\nodata & \nodata & \nodata & 1.4106 & 0.34 $\pm$ 0.08 & 0.74 $\pm$ 0.07 & 0.62 $\pm$ 0.07 & \nodata &  14 & $20.54^{+0.15}_{-0.24}$ & 2 & 09 \\ 
2048+196 & 18.5 & 2.367 & 1.1157 & 1.31 & 1.52 & 1.28 & \nodata & 3 & $19.26^{+0.05}_{-0.08}$ & 3 & 09 \\ 
2128$-$123 & 16.1 & 0.501 & 0.4297 & 0.27 $\pm$ 0.05 & 0.41 $\pm$ 0.01 & 0.37 $\pm$ 0.05 & 0.16 $\pm$ 0.03 &  13,7 & $19.18^{+0.03}_{-0.03}$ & 1 & AR \\ 
2145+067 & 16.5 & 0.999 & 0.7908 & 0.04 $\pm$ 0.01 & 0.48 $\pm$ 0.02 & 0.41 $\pm$ 0.06 & $<$0.04 & 2,7 & $18.43^{+0.03}_{-0.03}$ & 1 & AR \\ 
2149+212 & 19.0 & 1.538 & 0.9114 & 0.95 & 0.72 & 0.62 & \nodata & 3 & $20.70^{+0.08}_{-0.10}$ & 3 & 09 \\ 
\nodata & \nodata & \nodata & 1.0023 & 1.00 & 2.46 & 1.70 & \nodata & 3 & $19.30^{+0.02}_{-0.05}$ & 3 & 09 \\ 
2212$-$299 & 17.4 & 2.706 & 0.6329 & \nodata & 1.26 & 1.00 & \nodata &  3 & $19.75^{+0.03}_{-0.03}$ & 2 & 09 \\ 
2223$-$052 & 18.4 & 1.404 & 0.8472 & $<$0.4 & 0.65 & 0.42 & $<$0.2 &  34,35 & $18.48^{+0.41}_{-0.88}$ & 1 & AR \\ 
2326$-$477 & 16.8 & 1.306 & 1.2608 & $<$0.1 & 0.50 $\pm$ 0.04 & 0.38 $\pm$ 0.04 & $<$0.1 & 13 & $18.36^{+0.41}_{-0.76}$ & 1 & AR \\ 
2328+0022 & 17.9 & 1.308 & 0.6519 & 1.258 $\pm$ 0.065 & 1.896 $\pm$ 0.077 & 1.484 $\pm$ 0.073 & 0.550 $\pm$ 0.079 & 4 & $20.32^{+0.06}_{-0.07}$ & 4 & 11 \\ 
2331+0038 & 17.8 & 1.486 & 1.1414 & 1.437 $\pm$ 0.064 & 3.007 $\pm$ 0.071 & 2.417 $\pm$ 0.066 & 0.579 $\pm$ 0.060 & 4 & $20.00^{+0.04}_{-0.05}$ & 4 & 11 \\ 
2334+0052 & 18.2 & 1.040 & 0.4713 & 1.113 $\pm$ 0.128 & 1.226 $\pm$ 0.107 & 1.281 $\pm$ 0.091 & 0.301 $\pm$ 0.089 & 4 & $20.65^{+0.12}_{-0.18}$ & 4 & 11 \\ 
2339$-$0029 & 18.6 & 1.340 & 0.9664 & 1.715 $\pm$ 0.072 & 2.932 $\pm$ 0.074 & 2.530 $\pm$ 0.076 & 0.882 $\pm$ 0.090 &  4 & $20.48^{+0.07}_{-0.10}$ & 4 & 11 \\ 
2352$-$0028 & 18.2 & 1.628 & 0.8730 & 0.122 $\pm$ 0.080 & 1.254 $\pm$ 0.095 & 0.816 $\pm$ 0.075 & $-$0.048 $\pm$ 0.091 & 4 & $19.18^{+0.08}_{-0.10}$ & 1 & 11  \\ 
\nodata & \nodata & \nodata & 1.0318 & 1.494 $\pm$ 0.074 & 2.160 $\pm$ 0.102 & 1.714 $\pm$ 0.110 & 0.128 $\pm$ 0.119 & 4 & $19.81^{+0.14}_{-0.11}$ & 4 & 11 \\ 
\nodata & \nodata & \nodata & 1.2467 & 0.896 $\pm$ 0.132 & 2.926 $\pm$ 0.088 & 2.261 $\pm$ 0.120 & 0.087 $\pm$ 0.067 & 4 & $19.60^{+0.18}_{-0.30}$ & 4 & 11 \\ 
2353$-$0028 & 17.9 & 0.765 & 0.6044 & 1.024 $\pm$ 0.104 & 1.601 $\pm$ 0.082 & 1.292 $\pm$ 0.083 & 0.606 $\pm$ 0.080 &  4 & $21.54^{+0.15}_{-0.15}$ & 4 & 11 
\enddata 
\tablenotetext{a}{\
For systems other than those from Nestor (2004),
upper limits are either given by the authors or estimated by us from
the published spectrum. If the line is part of a blend, then the rest
equivalent width entered is an upper limit. The upper limit is a $1\sigma$
upper limit if estimated by us from the published spectrum. In some cases,
the absorption line ($\lambda2600$ or $\lambda2852$) was clearly visible in 
the spectrum but was not identified by the authors since the rest equivalent
width of the line did not meet their detection criterion, usually $5\sigma$.
In these cases, we estimated the equivalent width of the line from the
spectrum and tabulated the measurement as an upper limit. The trends that
we have established are evident from the
measurements of rest equivalent widths by the authors of the published
spectra; our upper limit measurements only serve to provide a certain
degree of completeness to our sample and, in fact, do not indicate or
contribute to any trends by themselves. We note that the quasar 0151+045
was erroneously included in Table 4 of RT00. The MgII system towards this
quasar was detected after the galaxy-quasar pair was known and is, therefore,
a biased system. In addition,
the $z_{abs}=0.213$ system towards 1148+386 and the $z_{abs}=0.1634$ system
towards 1704+608 were flagged as doubtful systems
by Boiss\'e et al. (1992). Finally,
the nature of the $z_{abs}=1.3284$ system towards 1331+170 was deemed 
inconclusive from the low signal-to-noise ratio IUE spectrum. Therefore,
these four were eliminated from our current MgII sample. }
\tablenotetext{b}{\
In two cases (the $z_{abs}=1.6101$ system towards 1329+412 and the 
$z_{abs}=1.2528$ system towards 1821+107), the HI column density could 
not be determined because the
Ly$\alpha$ and higher order Lyman lines could not be fit with a unique
value of $N(HI)$ under the assumption that the lines are damped. A large 
$b$ value (greater than 100 km s$^{-1}$)
was necessary in both cases. Thus, it is clear that these systems are not
DLAs. These are flagged as ``high b'' systems.}
\tablenotetext{c}{\
Flag for $W^{\lambda2796}_{0}$ selection criterion used in the determination 
of $n_{DLA}$. See \S 4.}
\tablenotetext{d}{\
Source for UV spectrum from which Ly$\alpha$ information was obtained. 
AR: HST or IUE archive, 06: {\it HST}-cycle 6 program 6577, P.I. Rao,
09: {\it HST}-cycle 9 program 8569, P.I. Rao, 11: {\it HST}-cycle 11 program
9382, P.I. Rao}
\tablerefs{
(1) Lanzetta, Turnshek, \& Wolfe 87.
(2) SS92.
(3) Barthel, Tytler, \& Thomson 1990.
(4) Nestor 2004.
(5) Steidel 1990.
(6) Sargent, Boksenberg, \& Steidel 1988.
(7) Churchill et al. 2000.
(8) Sargent, Young, \& Boksenberg 1982.
(9) Sargent, Steidel, \& Boksenberg 1989.
(10) Bergeron \& D'Odorico 1986.
(11) Womble et al. 1990.
(12) Wright et al. 1982.
(13) Petijean \& Bergeron 1990.
(14) Aldcroft, Bechtold, \& Elvis 1994.
(15) Tytler et al. 1987.
(16) Boksenberg, Carswell, \& Sargent 1979.
(17) Khare et al. 2004.
(18) Boiss\'e et al. 1992.
(19) Falomo 1990.
(20) Wills 1978.
(21) Ulrich \& Owen 1977.
(22) Foltz et al. 1986.
(23) Bergeron \& Boiss\'e 1984.
(24) Caulet 1989.
(25) Wills \& Wills 1980.
(26) Dinshaw \& Impey 1996.
(27) Jannuzi et al. 1998.
(28) Boiss\'e \& Bergeron, 1985.
(29) Bergeron \& Boiss\'e 1991.
(30) Bahcall et al. 1993.
(31) Young, Sargent, \& Boksenberg 1982.
(32) Kunth \& Bergeron 1984.
(33) Bergeron, D'Odorico, \& Kunth 1987.
(34) Le Brun et al. 1993.
(35) Miller \& French 1978.
}
\end{deluxetable}
\clearpage
\end{landscape}



\end{document}